\documentclass[12pt, preprint, trackchanges]{aastex62}
\usepackage{epsfig} 
\usepackage{amsmath}
\usepackage{multirow}

\newcommand{\shk}{$S_{\rm HK}$}

\newcommand{\ms}{\mbox{m s$^{-1}$}}

\newcommand{\msun}{M$_{\sun}$}
\newcommand{\rsun}{R$_{\sun}$}
\newcommand{\lsun}{L$_{\sun}$}
\newcommand{\mjup}{M$_{\rm JUP}$}

\newcommand{\msini}{$M \sin i$}

\newcommand{\chisq}{$\chi_{\nu}^2$}

\newcommand{\sn}{$\rm S/N$}

\newcommand{\feh}{\ensuremath{[\mbox{Fe}/\mbox{H}]}}
\newcommand{\rphk}{\ensuremath{R'_{\mbox{\scriptsize HK}}}}
\newcommand{\lrphk}{\ensuremath{\log{\rphk}}}

\received{February 15, 2018}
\revised{August 23, 2018}
\accepted{August 29, 2018}
\submitjournal{AJ}

\shorttitle{The N2K Project}
\shortauthors{Ment}

\begin{document}

\title{Radial velocities from the N2K Project: 6 new cold gas giant planets orbiting HD 55696, HD 98736, HD 148164, HD 203473, and HD 211810}

\author[0000-0001-5847-9147]{Kristo Ment}
\affiliation{Department of Astronomy, Yale University, 52 Hillhouse Avenue, New Haven, CT 06511, USA}
\affiliation{Harvard Smithsonian Center for Astrophysics, 60 Garden Street, Cambridge, MA 02138, USA}

\author[0000-0003-2221-0861]{Debra A. Fischer}
\affiliation{Department of Astronomy, Yale University, 52 Hillhouse Avenue, New Haven, CT 06511, USA}

\author[0000-0001-7204-6727]{Gaspar Bakos}
\affiliation{Department of Astrophysical Sciences, Princeton University, NJ 08544, USA}

\author[0000-0001-8638-0320]{Andrew W. Howard}
\affiliation{Department of Astronomy, California Institute of Technology, Pasadena, CA 91125, USA}

\author[0000-0002-0531-1073]{Howard Isaacson}
\affiliation{Department of Astronomy, University of California, Berkeley, CA 94720, USA}

\correspondingauthor{Kristo Ment}
\email{kristo.ment@cfa.harvard.edu}

\begin{abstract}
The N2K planet search program was designed to exploit the planet-metallicity correlation by searching for gas giant planets orbiting metal-rich stars. Here, we present the radial velocity measurements for 378 N2K target stars that were observed with the HIRES spectrograph at Keck Observatory between 2004 and 2017. With this data set, we announce the discovery of six new gas giant exoplanets: a double-planet system orbiting HD~148164 (\msini\ of 1.23 and 5.16 \mjup) and single planet detections around HD~55696 (\msini\ = 3.87 \mjup), HD~98736 (\msini = 2.33 \mjup), HD~203473 (\msini\ = 7.8 \mjup), and HD~211810 (\msini\ = 0.67 \mjup). These gas giant companions have orbital semi-major axes between 1.0 and 6.2 AU and eccentricities ranging from 0.13 to 0.71. We also report evidence for three gravitationally bound companions with \msini\ between 20 to 30 \mjup, placing them in the mass range of brown dwarfs, around HD~148284, HD~214823, and HD~217850, and four low mass stellar companions orbiting HD~3404, HD~24505, HD~98630, and HD~103459. In addition, we present updated orbital parameters for 42 previously announced planets. We also report a nondetection of the putative companion HD 73256 b. Finally, we highlight the most promising candidates for direct imaging and astrometric detection, and find that many hot Jupiters from our sample could be detectable by state-of-the-art telescopes such as Gaia.
\end{abstract}


\section{Introduction}
Before exoplanet searches began in earnest, radial velocity programs were used to search for low mass binary star and brown dwarf companions \citep{Marcy1987, Latham1989, DuquennoyMayor1991, FischerMarcy1992}. As precision improved to a few meters per second \citep{MarcyButler1992, Butler1996}, the Doppler technique was used to survey a few thousand stars with the goal of detecting exoplanets around nearby stars \citep{MayorQueloz1995}. Although most of the initial discoveries were larger amplitude gas giant planets, the state of the art measurement precision for radial velocities is now about 1 \ms\ \citep{Fischer2016}, enabling the detection of lower mass planets with velocity amplitudes as small as about 2 \ms.

A turning point in the exoplanet field occurred in 1999 when the short-period Doppler-detected planet, HD~209458b, was found to transit its host star \citep{Henry2000a, Charbonneau2000}. The combination of the planet mass (from Doppler measurements) and planet radius (from transit observations) allowed for a derivation of the bulk density of HD~209458b, showing that this planet was akin to Saturn and Jupiter in our solar system. Following this discovery, photometric transit surveys intensified \citep{Horne2003} and Doppler measurements were used to support new ground-based transit search programs, including the Hungarian-made Automated Telescope (HAT) system \citep{Bakos2004}, the WASP project \citep{Pollacco2006}, and the XO project \citep{McCullough2005}. The close-in gas giant planets, called ``hot Jupiters,'' exhibited some unexpected differences in comparison to solar system planets. For example, HD~209458b was found to have an inflated radius that was almost 1.4 times the radius of Jupiter \citep{Henry2000a, Charbonneau2000}. The powerful combination of exoplanet mass and radius offered a deeper understanding of the nature of these other worlds and transiting planets with mass measurements from radial velocities became the holy grail of exoplanets in the ensuing years. 

As the first few gas giant planets were being identified with radial velocity measurements, \citet{Gonzalez1997} noticed that the host stars tended to have super-solar metallicity. This observation was supported by subsequent spectroscopic analyses of planet-bearing stars \citep{Fuhrmann1997, Gonzalez1998, Gonzalez1999, Santos2000, Santos2004, Sadakane2002, Laws2003, FischerValenti2005}, and ultimately confirmed a statistically significant planet-metallicity correlation that provided important clues about conditions in the protoplanetary disk where hot Jupiters formed. 

The ``Next 2000" or N2K consortium \citep{Fischer2005} was started in 2003, and exploited the planet-metallicity correlation with the goal of detecting additional hot Jupiters for transit-search programs. Here, we publish the data collected at Keck Observatory as part of this program. We provide an overview of the N2K project and our target selection in Section \ref{sec:n2k}. Our Keplerian fitting methodology is described in Section \ref{sec:method}. The modeling results are divided into three sections: Section \ref{sec:results1} presents a series of new discoveries, Section \ref{sec:results2} lists systems with interesting or unusual characteristics, and Section \ref{sec:results3} provides updated orbital parameters for the rest of the known companions in the data sample. Section \ref{sec:discussion} provides a discussion of a few relevant results, and Section \ref{sec:summary} concludes the article.
	
\section{N2K project \label{sec:n2k}}
The N2K consortium built on collaboration to expedite the discovery of gas giant exoplanets with a high transit probability \citep{Fischer2005}. The N2K partners included U.S. astronomers, who used the HIRES spectrograph \citep{Vogt1994} at Keck; Japanese astronomers, who secured time on the High Dispersion Spectrometer \citep[HDS]{Noguchi2002} at Subaru; and Chilean astronomers who collected data with the Magellan Inamori Kyocera Echelle \citep[MIKE]{Bernstein2003} spectrograph at the Las Campanas Observatory. 

An initial list of more than 14,000 main sequence and subgiant stars brighter than tenth magnitude and closer than 100 parsecs was culled from the Hipparcos catalog \citep{Perryman1997}. A training set of stars with well-determined spectroscopic metallicities \citep{ValentiFischer2005} was used to calibrate the available broadband photometric indices \citep{Ammons2006}, reducing the initial sample to about 2000 stars that were likely to have super-solar metallicity. As telescope time was scheduled at each of the consortium telescopes, a subset of about 100 stars was distributed to the N2K partners and a series of three observations were made on nearly consecutive nights to quickly flag prospective hot Jupiters. One additional radial velocity (RV) measurement was scheduled a few weeks later to look for slightly longer period planets and an assessment of the data set was based on the rms RV scatter. Stars with constant radial velocities were temporarily retired from active observations and stars that showed more than 2$\sigma$ RV variability were retained as active targets for follow-up measurements. 

Altogether, the N2K program detected 42 exoplanets at the Keck, Subaru and Magellan telescopes using this RV screening strategy. Two of these exoplanets turned out to be extremely valuable transiting systems: HD~149026 b, the most compact transiting planet at the time \citep{Sato2005}, and HD~17156 b, the transiting planet with the longest orbital period of the time \citep{Fischer2007}. Table \ref{t_N2K} summarizes the N2K exoplanet discoveries, listing the orbital period, \msini, and a reference to the original publication. The orbital parameters have been updated for all systems where longer time baseline of Keck RV measurements are available and are refitted with the algorithms described in this paper.

\startlongtable
\begin{deluxetable}{lccl}
	\tablewidth{0pt}
	\tabletypesize{\scriptsize}
	\tablecaption{N2K substellar discoveries to date \label{t_N2K}}
	\tablehead{
		\colhead{Star}           & \colhead{Period}      &
		\colhead{$M \sin i$}          & \colhead{Reference}  \\
		& \colhead{(days)}  & \colhead{(\mjup)} &
	}
	\startdata
	HD 86081 b & 2.14 & 1.48 & \citet{Johnson2006} \\
	HD 149026 b & 2.88 & 0.33\tablenotemark{a} & \citet{Sato2005} \\
	HD 88133 b & 3.41 & 0.28 & \citet{Fischer2005} \\
	HD 149143 b & 4.07 & 1.33 & \citet{Fischer2006} \\
	HD 125612 c & 4.16 & 0.05 & \citet{Fischer2007} \\
	HD 109749 b & 5.24 & 0.27 & \citet{Fischer2006} \\
	HIP 14810 b & 6.67 & 3.90 & \citet{Wright2007} \\
	HD 179079 b & 14.5 & 0.08 & \citet{Valenti2009} \\
	HD 33283 b & 18.2 & 0.33 & \citet{Johnson2006} \\
	HD 17156 b & 21.2 & 3.16\tablenotemark{a} & \citet{Fischer2007} \\
	HD 224693 b & 26.7 & 0.70 & \citet{Johnson2006} \\
	HD 163607 b & 75.2 & 0.79 & \citet{Giguere2012} \\
	HD 231701 b & 142 & 1.13 & \citet{Fischer2007} \\
	HIP 14810 c & 148 & 1.31 & \citet{Wright2007} \\
	HD 154672 b & 164\tablenotemark{b} & 4.96\tablenotemark{b} & \citet{LopezMorales2008} \\
	HD 11506 c & 223 & 0.41 & \citet{Giguere2015} \\
	HD 164509 b & 280 & 0.44 & \citet{Giguere2012} \\
	HD 205739 b & 280\tablenotemark{b} & 1.37\tablenotemark{b} & \citet{LopezMorales2008} \\
	HD 148164 c & 329 & 1.23 & \textbf{This work} \\
	HD 148284 B & 339 & 33.7 & \textbf{This work} \\
	HD 75784 b & 342 & 1.08 & \citet{Giguere2015} \\
	HD 75898 b & 423 & 2.71 & \citet{Robinson2007} \\
	HD 16760 b & 466 & 15.0 & \citet{Sato2009} \\
	HD 96167 b & 498 & 0.71 & \citet{Peek2009} \\
	HD 125612 b & 557 & 3.10 & \citet{Fischer2007} \\
	HD 5319 b & 639 & 1.56 & \citet{Robinson2007} \\
	HD 38801 b & 687 & 9.97 & \citet{Harakawa2010} \\
	HD 5319 c & 877 & 1.02 & \citet{Giguere2015} \\
	HD 98736 b & 969 & 2.33 & \textbf{This work} \\
	HIP 14810 d & 982 & 0.59 & \citet{Wright2009} \\
	HD 16175 b & 990\tablenotemark{b} & 4.40\tablenotemark{b} & \citet{Peek2009} \\
	HD 10442 b & 1053 & 1.53 & \citet{Giguere2015} \\
	HD 163607 c & 1267 & 2.16 & \citet{Giguere2012} \\
	HD 203473 b & 1553 & 7.84 & \textbf{This work} \\
	HD 211810 b & 1558 & 0.67 & \textbf{This work} \\
	HD 11506 b & 1622 & 4.83 & \citet{Fischer2007} \\
	HD 73534 b & 1721 & 1.01 & \citet{Valenti2009} \\
	HD 55696 b & 1827 & 3.87 & \textbf{This work} \\
	HD 214823 B & 1854 & 20.3 & \textbf{This work} \\
	HD 217850 B & 3501 & 21.6 & \textbf{This work} \\
	HD 75784 c & 3878 & 4.50 & \citet{Giguere2015} \\
	HD 148164 b & 5062 & 5.16 & \textbf{This work} \\
	\enddata
	\tablenotetext{a}{This planet has a confirmed transit. The value reported here is the planetary mass $M$ since the inclination $i$ is known.}
	\tablenotetext{b}{This planet's host is not in our dataset. The value reported here is taken from the original reference.}
\end{deluxetable}

\subsection{Doppler Measurements}
The data presented here were obtained at the 10-m Keck Observatory atop Maunakea in Hawai'i, using the HIRES spectrograph. The spectra have a fairly consistent \sn\ of about 200 since the exposures are automatically terminated with an exposure meter that picks off a small fraction of light behind the slit. The B5 decker was used, providing a spectral resolution of about 50,000. Typical exposure times range from a few to fifteen minutes.  

Our Doppler analysis makes use of a glass cell that contains laboratory-grade iodine to provide wavelength calibration. The intrinsically narrow iodine lines also permit modeling of the spectral line spread function \citep{MarcyButler1992, Butler1996} in the extracted data. The iodine cell is positioned in front of the entrance slit of the spectrograph and wrapped in a heating blanket to prevent the iodine from condensing onto the walls of the cell. As the starlight passes through the cell, a forest of iodine absorption lines are imprinted on the stellar spectrum.

Forward modeling is driven by a Levenberg-Marquardt chi-squared minimization to derive the Doppler shift in about 700 2-Angstrom chunks over the wavelength range of about 510 to 620 nm. The model ingredients include a high-S/N, high-resolution ``template" spectrum of the star, which is constructed from observations of the star without iodine, and a Fourier Transform Spectrograph (FTS) scan of the iodine cell obtained with resolution approaching one million and signal-to-noise of about one thousand. In our model of the program observations with the iodine cell, the template spectrum of the star is multiplied by the FTS iodine spectrum and the product is convolved with a model of the line spread function (LSF). The model for each 2-Angstrom chunk contains 21 free parameters: the wavelength zero point, the wavelength dispersion, the Doppler shift, a continuum shift, and 17 amplitudes of the Gaussian components that model the LSF \citep{Valenti1995}. Once a Doppler shift is measured, it is converted to a radial velocity measurement using the non-relativistic Doppler equation. The flux-weighted centroid of the observation is then used to calculate and correct for the barycentric velocity.   

\begin{deluxetable}{lrrrr}
	\tablewidth{0pt}
	\tabletypesize{\scriptsize}
	\tablecaption{Radial velocities for N2K stars \label{t_rv_stub}}
	\tablehead{
		\colhead{Star} & \colhead{JD-2440000} & \colhead{Velocity} & \colhead{Error} & \colhead{\shk} \\
		 & day & \ms & \ms & 
	}
	\startdata
	HAT-P-1	&	13927.068	&	-50.478726	&	1.60722	&	0.156 \\
	HAT-P-1	&	13927.966	&	-45.501405	&	1.69255	&	0.158 \\
	HAT-P-1	&	13931.037	&	-24.217690	&	1.72525	&	0.157 \\
	HAT-P-1	&	13931.941	&	-56.605246	&	2.08687	&	0.158 \\
	HAT-P-1	&	13932.036	&	-58.051372	&	1.87104	&	0.157 \\
	HAT-P-1	&	13933.000	&	-12.589293	&	1.84088	&	0.162 \\
	\enddata
	\tablecomments{Table \ref{t_rv_stub} is published in its entirety in the machine-readable format.
      A portion is shown here for guidance regarding its form and content.}
\end{deluxetable}

\subsection{Stellar sample}

\begin{figure}
	\plotone{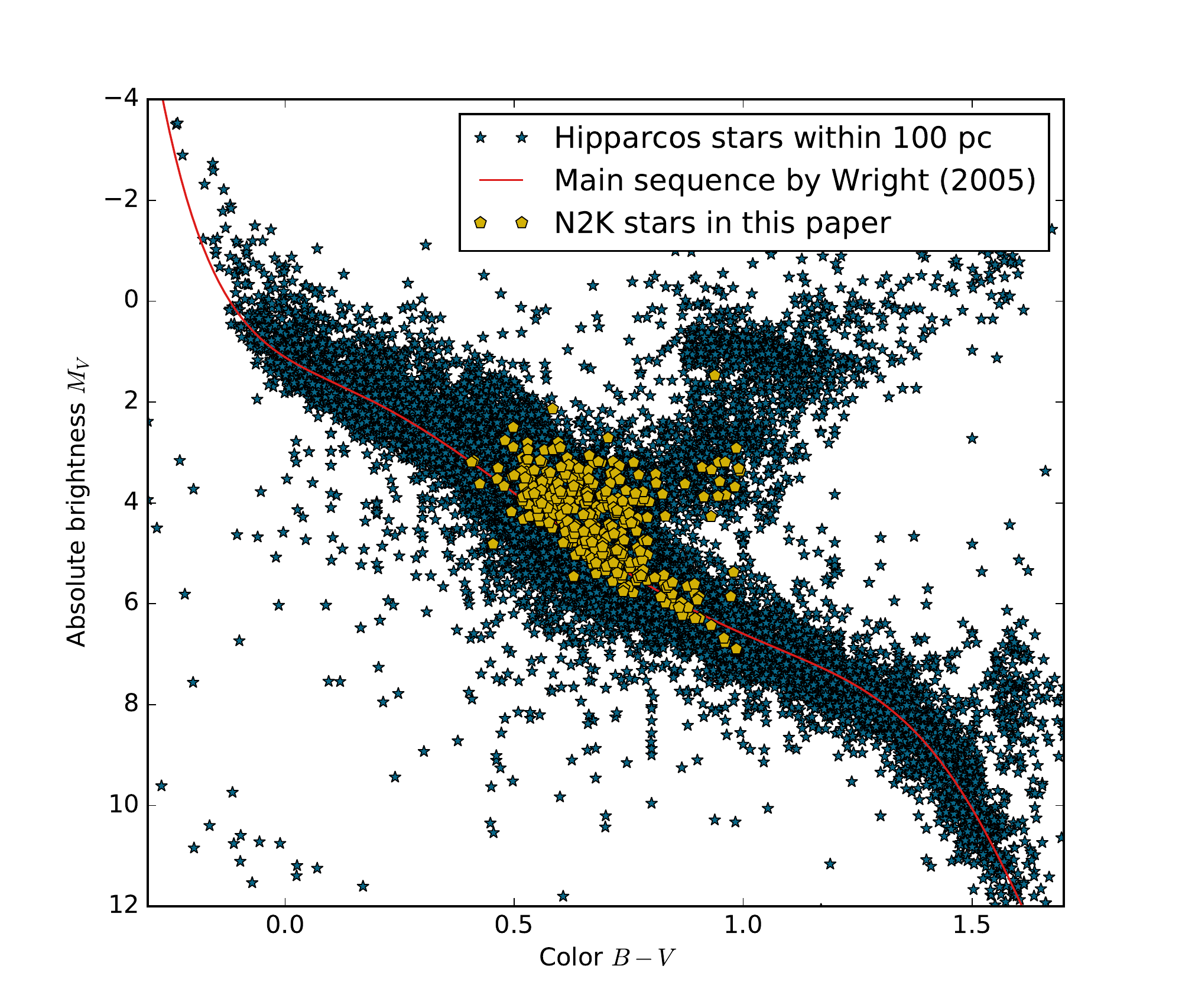}
	\caption{A Hertzsprung-Russell diagram of all the stars in this sample, overplotted on the stars in the Hipparcos catalog within 100 pc of the Sun. A model for the main sequence (solid red line) is given by \citet{Wright2005}. As can be seen in the figure, the N2K sample is primarily composed of main sequence stars and subgiants in the G, F, and K spectral types.\label{fig:hrplot}}
\end{figure}

\begin{figure}
	\plotone{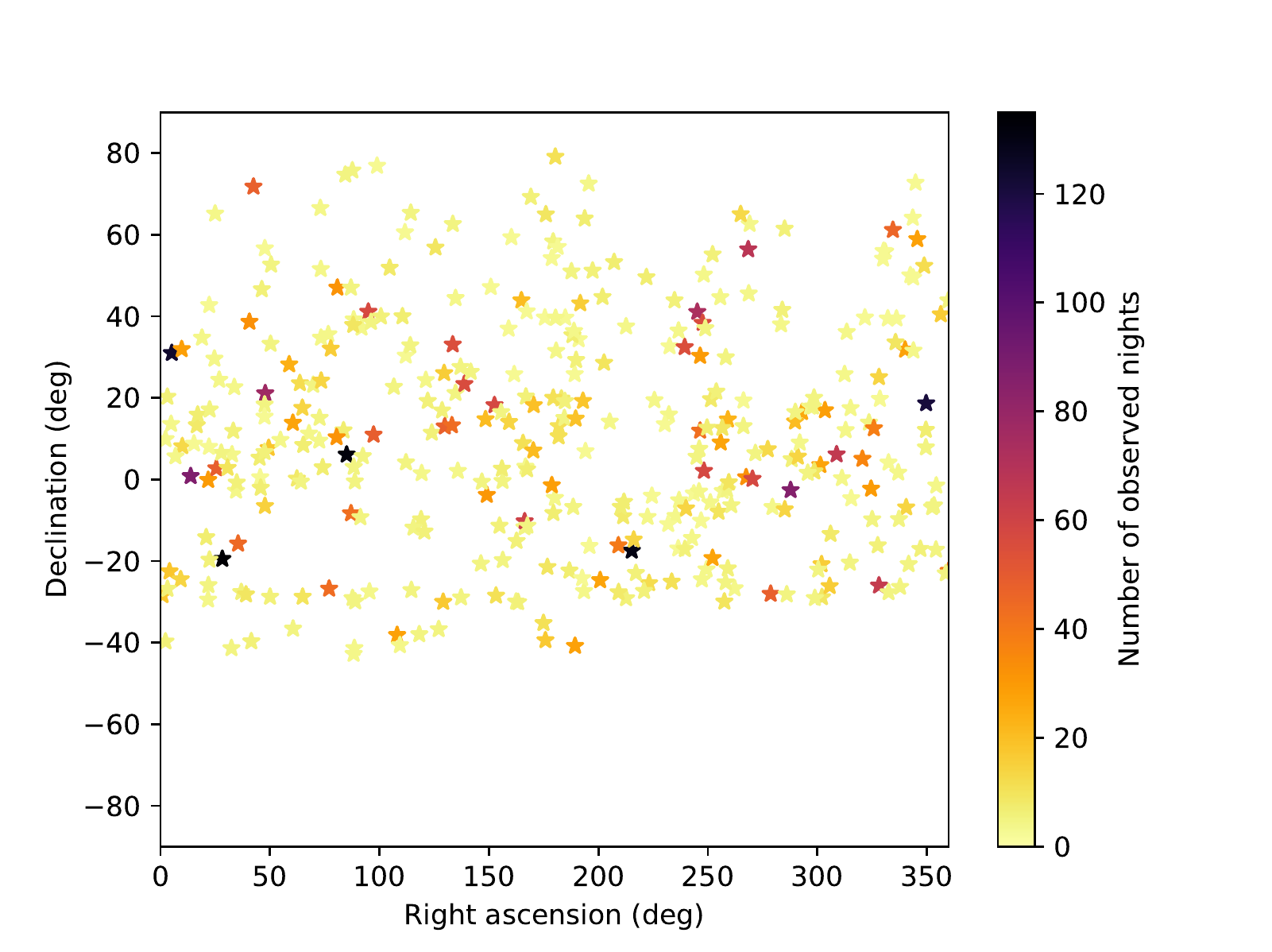}
	\caption{Coordinates of the N2K targets. Darker colors indicate a greater number of visits. The targets cover the sky relatively uniformly between declination angles of -30 and +70 degrees. The positions were downloaded from SIMBAD.\label{fig:coords}}
\end{figure}

\begin{figure}
	\plotone{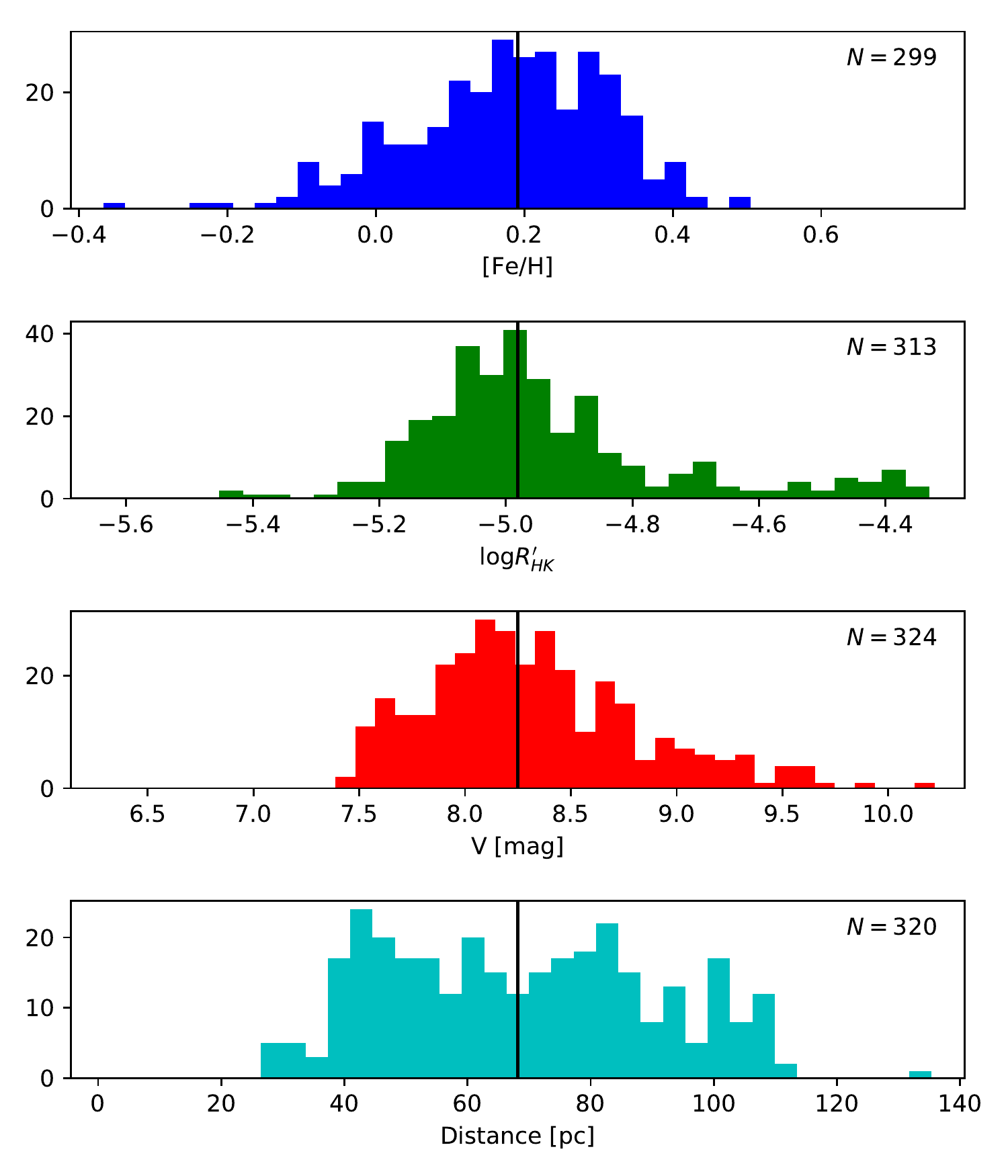}
	\caption{Histograms of the stellar sample. \textit{Top to bottom:} Histograms of metallicity [Fe/H], stellar activity index \lrphk, brightness $V$, and distance from the Sun in parsecs. The median value of each distribution is given as a vertical solid black line. The total number of stars used in each histogram is given as $N$ whereas the vertical axis displays the number of stars in each histogram bin.\label{fig:starhist}}
\end{figure}

In this paper, we list radial velocities for a total of 378 stars that have been observed at least three times by the N2K Consortium with the Keck HIRES spectrograph in Table \ref{t_rv_stub}. These stars are predominantly main-sequence stars and subgiants in the G, F, and occasionally K, spectral types. An HR diagram of the stellar sample can be seen in Figure \ref{fig:hrplot}, overplotted on all stars in the Hipparcos catalog within 100 pc of the Sun. Figure \ref{fig:hrplot} also includes an empirically fitted main sequence model by \citet{Wright2005}, which we employed to estimate the height above the main sequence $M_{V,MS}$ as described in Section \ref{sec:stellarpars}. Due to the location of the Keck Observatory at 19$^\circ$ 49' N, our sample spans mainly the declinations between -30 and +70 degrees with no significant variations in coverage. The positions of the N2K targets from SIMBAD are plotted as a function of right ascension and declination in Figure \ref{fig:coords}. Figure \ref{fig:starhist} gives the distributions as well as the median values of iron abundance [Fe/H], activity index \lrphk, brightness in the V-band, and the distance from the Sun for most stars in the N2K sample. The sample is intentionally biased towards metal-rich stars: the median iron abundance is $\rm [Fe/H] = 0.19$, and nearly 90\% of the N2K stars have a higher [Fe/H] value compared to the Sun. Most stars also have a low chromospheric activity index, with the median at \lrphk $= -4.98$. The V-band brightness varies between 7.4 and 10.2 magnitudes, and the median distance of an N2K star from Earth is around 68 parsecs.

A comprehensive overview of the parameters of all N2K stars is given in Table \ref{t_starswithplanets}. The version printed here only lists stars with at least one known companion (based on the results of this paper). However, we also provide a complete version with all 378 stars in the machine-readable format.

\begin{longrotatetable}
\begin{deluxetable}{lrrcrrrrrrrrrrrr}
	\tablewidth{0pt}
	\tabletypesize{\scriptsize}
	\tablecaption{Properties of N2K Stars \label{t_starswithplanets}}
	\tablehead{
		\colhead{Star}           & \colhead{V}      &
		\colhead{B$-$V}          & \colhead{Type}  &
		\colhead{T$_{\rm eff}$}  &
		\colhead{[Fe/H]}         & \colhead{M}    &
		\colhead{log L} & \colhead{R} &
		\colhead{Age}			 & \colhead{S$_{\rm HK}$} &
		\colhead{log R$_{\rm HK}^{'}$}   & \colhead{Jitter} &
		\colhead{N$_{\rm RV}$} & \colhead{N$_{\rm fit}$\tablenotemark{a}} &
		\colhead{$\sigma_{\rm resid}$\tablenotemark{b}}\\
		& & & & 
		\colhead{(K)}  & &
		\colhead{(\msun)} &
		\colhead{(\lsun)} & \colhead{(\rsun)} &
		\colhead{(Gyr)} & & & \colhead{(\ms)} &
		& & \colhead{(\ms)}
	}
	\startdata
	HAT-P-1& 9.87& 0.95& G0V& 6067& \nodata& 1.12(0.09)
	& 0.18(0.16)& 1.15(0.09)& 3.6(0.0)& 0.158(0.002)& -5.119(0.008)& \nodata& 19& 1& 3.70 \\
	HAT-P-12& 12.84& -0.24& K4& 4559& -0.30& 0.73(0.02)
	& -0.68(0.04)& 0.70(0.01)& 2.5(2.0)& 0.257(0.006)& nan(nan)& \nodata& 21& 1& 7.31 \\
	HAT-P-3& 11.58& 0.67& K& 5084& 0.27& 0.92(0.03)
	& -0.36(0.05)& 0.80(0.04)& 1.6(2.1)& 0.205(0.005)& -4.788(0.020)& \nodata& 10& 1& 3.16 \\
	HAT-P-4& 11.22& 0.71& F& 5907& 0.33& 1.26(0.10)
	& 0.43(0.16)& 1.59(0.07)& 4.2(1.6)& 0.145(0.004)& -5.111(0.035)& \nodata& 24& 1& 9.92 \\
	HD 103459& 7.60& 0.69& G5& 5721& 0.24& 1.18(0.20)
	& 0.44(0.05)& 1.69(0.10)& 6.03(0.65)& 0.152(0.003)& -5.059(0.021)& 2.31& 29& 1& 5.35 \\
	HD 10442& 7.84& 0.93& K0IV& 4912& 0.10& 1.01(0.81)
	& 0.31(0.35)& 1.97(0.79)& 10.65(3.86)& 0.139(0.002)& -5.181(0.009)& 4.22& 49& 1& 5.48 \\
	HD 109749& 8.08& 0.714& G3V& 5824& 0.24& 1.13(0.22)
	& 0.19(0.06)& 1.22(0.09)& 3.30(1.2)& 0.166(0.003)& -4.983(0.016)& 2.19& 28& 1& 2.75 \\
	HD 11506& 7.51& 0.607& G0V& 6030& 0.34& 1.24(0.18)
	& 0.33(0.03)& 1.34(0.05)& 2.30(0.58)& 0.159(0.002)& -4.986(0.014)& 2.52& 135& 2& 5.67 \\
	HD 125612& 8.31& 0.628& G3V& 5841& 0.22& 1.11(0.20)
	& 0.06(0.06)& 1.05(0.07)& 1.77(1.24)& 0.181(0.004)& -4.872(0.018)& 2.85& 130& 3& 4.73 \\
	HD 147506& 8.72& 0.463& F8& 6380& 0.18& 1.33(0.03)
	& 0.53(0.08)& 1.39(0.09)& 1.44(0.47)& 0.190(0.003)& -4.779(0.015)& 3.03& 73& 1& 44.72 \\
	HD 148164& 8.23& 0.589& F8& 6032& 0.24& 1.21(0.24)
	& 0.33(0.07)& 1.34(0.10)& 2.41(0.81)& 0.154(0.002)& -5.016(0.014)& 2.37& 43& 2& 5.62 \\
	HD 148284& 9.01& 0.751& K0& 5572& 0.16& 1.07(0.26)
	& 0.28(0.09)& 1.48(0.16)& 8.66(1.17)& 0.143(0.001)& -5.127(0.008)& 2.15& 30& 1& 2.97 \\
	HD 149026& 8.15& 0.611& G0& 6084& 0.36& 1.30(0.22)
	& 0.44(0.05)& 1.50(0.09)& 2.22(0.44)& 0.153(0.002)& -5.030(0.012)& 2.35& 56& 1& 5.78 \\
	HD 149143& 7.89& 0.68& G0& 5856& 0.29& 1.20(0.20)
	& 0.35(0.05)& 1.44(0.08)& 4.28(0.68)& 0.170(0.005)& -4.950(0.030)& 2.81& 58& 1& 11.80 \\
	HD 1605& 7.52& 0.961& K1IV& 4915& 0.22& 1.33(0.09)
	& 0.81(0.05)& 3.49(0.22)& 4.36(0.97)& 0.126(0.001)& -5.258(0.005)& 4.21& 124& 2& 6.68 \\
	HD 163607& 8.00& 0.777& G5& 5522& 0.19& 1.12(0.16)
	& 0.42(0.03)& 1.76(0.07)& 7.75(0.67)& 0.164(0.002)& -5.013(0.010)& 4.25& 68& 2& 2.92 \\
	HD 164509& 8.10& 0.665& G5& 5859& 0.20& 1.12(0.19)
	& 0.12(0.05)& 1.11(0.06)& 2.32(1.27)& 0.183(0.005)& -4.879(0.023)& 2.82& 58& 1& 5.33 \\
	HD 16760& 8.70& 0.715& G5& 5593& -0.03& 0.96(0.24)
	& -0.22(0.09)& 0.82(0.09)& 3.56(2.92)& 0.177(0.001)& -4.928(0.006)& 2.24& 33& 1& 4.02 \\
	HD 171238& 8.61& 0.767& G8V& 5440& 0.20& 0.96(0.17)
	& -0.09(0.05)& 1.01(0.07)& 7.56(2.96)& 0.315(0.008)& -4.576(0.015)& 2.84& 48& 1& 11.59 \\
	HD 17156& 8.17& 0.643& G5& 5969& 0.22& 1.23(0.21)
	& 0.39(0.05)& 1.47(0.09)& 3.46(0.71)& 0.155(0.002)& -5.027(0.016)& 2.49& 48& 1& 3.62 \\
	HD 179079& 7.95& 0.744& G5IV& 5646& 0.24& 1.14(0.19)
	& 0.38(0.05)& 1.63(0.09)& 6.99(0.71)& 0.161(0.002)& -5.019(0.011)& 4.24& 86& 1& 4.14 \\
	HD 203473& 8.23& 0.663& G5& 5780& 0.19& 1.12(0.21)
	& 0.25(0.06)& 1.33(0.10)& 5.22(0.98)& 0.155(0.002)& -5.031(0.014)& 2.3& 36& 1& 3.34 \\
	HD 207832& 8.78& 0.691& G5V& 5726& 0.14& 1.05(0.17)
	& -0.11(0.05)& 0.89(0.05)& 1.36(1.26)& 0.243(0.009)& -4.681(0.025)& 4.03& 64& 1& 30.04 \\
	HD 211810& 8.59& 0.732& G5& 5652& 0.17& 1.03(0.02)
	& 0.07(0.43)& 1.13(1.13)& 0(0.0)& 0.150(0.001)& -5.081(0.009)& 2.19& 46& 1& 2.55 \\
	HD 214823& 8.06& 0.631& G0& 5933& 0.14& 1.31(0.24)
	& 0.67(0.06)& 2.04(0.15)& 4.34(0.5)& 0.149(0.001)& -5.066(0.011)& 2.33& 28& 1& 4.23 \\
	HD 217850& 8.52& 0.791& G8V& 5544& 0.26& 1.03(0.16)
	& 0.10(0.04)& 1.21(0.06)& 7.59(1.29)& 0.157(0.001)& -5.051(0.006)& 2.13& 28& 1& 2.88 \\
	HD 219828& 8.04& 0.654& G0IV& 5807& 0.18& 1.18(0.20)
	& 0.42(0.05)& 1.61(0.10)& 5.69(0.71)& 0.168(0.001)& -4.950(0.007)& 2.66& 121& 2& 2.99 \\
	HD 224693& 8.23& 0.639& G2V& 5894& 0.27& 1.31(0.29)
	& 0.61(0.08)& 1.93(0.18)& 4.12(0.67)& 0.147(0.004)& -5.083(0.032)& 2.46& 40& 1& 5.73 \\
	HD 231701& 8.97& 0.539& F8V& 6101& 0.05& 1.21(0.34)
	& 0.47(0.11)& 1.53(0.20)& 3.43(0.71)& 0.169(0.002)& -4.901(0.012)& 2.77& 26& 1& 4.60 \\
	HD 24505& 8.05& 0.737& G5III& 5688& 0.07& 1.11(0.20)
	& 0.40(0.06)& 1.64(0.11)& 7.33(0.77)& 0.141(0.001)& -5.141(0.004)& 4.23& 24& 1& 3.02 \\
	HD 33283& 8.05& 0.641& G3/G5V& 5935& 0.35& 1.38(0.24)
	& 0.64(0.05)& 1.97(0.13)& 3.63(0.48)& 0.136(0.001)& -5.179(0.014)& 4.26& 44& 1& 3.47 \\
	HD 3404& 7.94& 0.824& G2V& 5339& 0.20& 1.17(0.22)
	& 0.49(0.06)& 2.05(0.15)& 6.73(1.29)& 0.160(0.003)& -5.046(0.014)& 4.27& 14& 1& 2.02 \\
	HD 37605& 8.67& 0.827& K0& 5329& 0.27& 0.94(0.15)
	& -0.22(0.05)& 0.91(0.05)& 5.65(3.42)& 0.167(0.004)& -5.017(0.018)& 2.17& 132& 2& 6.44 \\
	HD 38801& 8.26& 0.873& K0& 5207& 0.32& 1.29(0.26)
	& 0.58(0.07)& 2.41(0.19)& 4.87(0.96)& 0.195(0.010)& -4.941(0.030)& 4.32& 43& 1& 7.71 \\
	HD 43691& 8.03& 0.596& G0& 6093& 0.33& 1.32(0.25)
	& 0.50(0.06)& 1.60(0.11)& 2.39(0.37)& 0.180(0.004)& -4.860(0.021)& 2.88& 57& 1& 12.49 \\
	HD 45652& 8.10& 0.846& G8/K0& 5294& 0.26& 0.92(0.13)
	& -0.20(0.03)& 0.94(0.04)& 8.40(3.24)& 0.205(0.009)& -4.891(0.027)& 2.3& 49& 1& 18.16 \\
	HD 5319& 8.05& 0.985& G5& 4871& 0.17& 1.27(0.30)
	& 0.92(0.09)& 4.06(0.42)& 5.04(1.69)& 0.123(0.002)& -5.285(0.007)& 4.24& 88& 2& 6.84 \\
	HD 55696& 7.95& 0.604& G0V& 6012& 0.37& 1.29(0.20)
	& 0.43(0.04)& 1.52(0.07)& 2.64(0.5)& 0.164(0.004)& -4.955(0.023)& 2.53& 28& 1& 7.18 \\
	HD 73534& 8.23& 0.962& G5& 4917& 0.28& 1.16(0.21)
	& 0.54(0.06)& 2.58(0.17)& 7.04(1.37)& 0.130(0.001)& -5.239(0.006)& 4.21& 48& 1& 4.20 \\
	HD 75784& 7.84& 0.99& G5& 4867& 0.28& 1.26(0.25)
	& 0.77(0.07)& 3.40(0.26)& 5.36(1.39)& 0.133(0.002)& -5.244(0.010)& 4.2& 44& 2& 3.79 \\
	HD 75898& 8.03& 0.626& G0& 5963& 0.28& 1.26(0.23)
	& 0.46(0.06)& 1.58(0.11)& 3.56(0.65)& 0.149(0.004)& -5.065(0.031)& 2.33& 55& 2& 5.82 \\
	HD 79498& 8.05& 0.693& G5& 5748& 0.18& 1.08(0.15)
	& 0.03(0.03)& 1.05(0.04)& 2.71(1.45)& 0.139(0.001)& -5.153(0.006)& 2.49& 56& 1& 5.57 \\
	HD 86081& 8.73& 0.664& F8& 5939& 0.22& 1.21(0.28)
	& 0.38(0.08)& 1.46(0.14)& 3.61(0.86)& 0.164(0.004)& -4.980(0.022)& 2.47& 31& 1& 3.44 \\
	HD 88133& 8.01& 0.81& G5& 5392& 0.32& 1.26(0.25)
	& 0.57(0.07)& 2.20(0.17)& 5.37(0.91)& 0.134(0.001)& -5.186(0.010)& 4.21& 58& 1& 4.41 \\
	HD 96167& 8.09& 0.731& G5& 5733& 0.34& 1.27(0.26)
	& 0.57(0.07)& 1.94(0.16)& 4.85(0.63)& 0.138(0.002)& -5.163(0.021)& 4.23& 61& 1& 4.27 \\
	HD 98630& 8.10& 0.601& G0& 6002& 0.29& 1.44(0.07)
	& ()& 1.94(0.19)& (0.0)& 0.154(0.001)& -5.016(0.008)& 2.5& 21& 1& 7.30 \\
	HD 98736& 7.93& 0.894& K6+...& 5271& 0.36& 0.92(0.13)
	& -0.22(0.03)& 0.93(0.04)& 7.39(3.05)& 0.210(0.008)& -4.914(0.021)& 2.2& 20& 1& 3.08 \\
	HIP 14810& 8.52& 0.777& G5& 5544& 0.24& 1.01(0.19)
	& -0.01(0.06)& 1.07(0.08)& 5.84(2.42)& 0.154(0.002)& -5.064(0.012)& 2.11& 77& 3& 3.29 \\
	XO-5& 12.13& 0.84& G8V& 5470& 0.05& 0.88(0.03)
	& -0.06(0.04)& 1.08(0.04)& 14.8(2.0)& 0.150(0.005)& -5.098(0.024)& \nodata& 34& 1& 16.60 \\
	\enddata
	\tablenotetext{a}{Number of fitted companions.}
	\tablenotetext{b}{RMS of the RV residuals after subtracting the fitted companions.}
	\tablecomments{Table \ref{t_starswithplanets} is published in its entirety in the machine-readable format. The portion shown here omits N2K stars with no found companions.}
\end{deluxetable}
\end{longrotatetable}

\section{Methodology \label{sec:method}}

\subsection{Keplerian fitting \label{sec:fitting}}

We developed our own Keplerian fitting code in C++ specifically for the analysis presented here. The radial velocity contribution $V_i(t)$ of a single planet $i$ at an observation epoch $t$ is modeled through the following five free parameters: radial velocity semi-amplitude $K_i$, orbital period $P_i$, eccentricity $e_i$, longitude of periastron $\omega_i$, and time of periastron passage $T_{P,i}$. These parameters combine to give a value for the velocity via the following equations:
\begin{eqnarray}
V_i(t) = K_i [\cos(\theta(t) + \omega_i) + e_i \cos(\omega_i)] \label{eq:vit} \\
\theta(t) = 2 \arctan \left[ \sqrt{\frac{1 + e_i}{1 - e_i}} \tan \frac{E(t)}{2} \right] \label{eq:thetat} \\
M(t) = E(t) - e_i \sin E(t) = 2 \pi \frac{t - T_{P,i}}{P_i} \label{eq:Kepler}
\end{eqnarray}

For each planet with parameters $\{K_i,P_i,e_i,\omega_i,T_{P,i}\}$ in the model, we start out by solving Kepler's Equation (\ref{eq:Kepler}) for the eccentric anomaly $E(t)$ which we then plug into Equation \ref{eq:thetat} to obtain a true anomaly $\theta(t)$. Finally, we can use Equation \ref{eq:vit} to get the velocity contribution $V_i(t)$ of the planet. The total radial velocity $V(t)$ for $n$ planets at time $t$ is then given by
\begin{equation}\label{eq:vt}
V(t) = \sum_{i=1}^n V_i(t) + \gamma_{\rm Keck} + \dot{\gamma} (t - t_0) + \frac{1}{2} \ddot{\gamma} (t - t_0)^2
\end{equation}

In Equation \ref{eq:vt}, we included an arbitrary radial velocity offset $\gamma_{\rm Keck}$ due to the instrument, and possible contributions from a linear radial velocity trend $\dot{\gamma}$ as well as a curvature term $\ddot{\gamma}$ which are often set to zero. $t_0$ is a reference time for curvature measurements which we set to the time of the earliest observation $t_0 = \min t_j$ in each data set since that is the earliest time $t$ for which our model is evaluated. Given a series of radial velocity measurements $\{v_j\}$ with uncertainties $\{\sigma_j\}$ at times $\{t_j\}$, the goodness of the fit can then be estimated via the $\chi^2$ parameter:
\begin{equation}\label{eq:chi2}
\chi^2 = \sum_{j} \left( \frac{v_j - V(t_j)}{\sigma_j} \right)^2
\end{equation}

The value of $\chi^2$ is often normalized to the number of free parameters $n_{\rm dof}$ used in the model, and the ``reduced chi-squared'' $\chi_{\nu}^2 = \chi^2 / n_{\rm dof}$ is reported instead. For some stars we include additional observations with other instruments from the literature - in these cases, we use a separate radial velocity offset $\gamma$ for each instrument when calculating $\chi^2$. Finally, instead of looking for a single solution, we set up a Monte Carlo routine with many walkers as described in the following section.

\subsection{Affine-invariant MCMC}

First, we obtain three best guesses for the orbital period of a planet by using a generalized Lomb-Scargle periodogram from the astroML\footnote{Introduction to astroML: Machine learning for astrophysics, Vanderplas et al, proc. of CIDU, pp. 47-54, 2012} package. We then distribute walkers around those guesses by drawing from a normal distribution centered at the guessed periods. We also draw from normal and uniform distributions to obtain values for all the other free parameters $K_i$, $e_i$, $\omega_i$, $T_{P,i}$, $\gamma$, $\dot{\gamma}$, and $\ddot{\gamma}$. We often set $\ddot{\gamma} = 0$ and test models with $\dot{\gamma} = 0$ as well. Additional constraints may be implemented based on the characteristics of each individual system, e.g. we may generate models where the period has been fixed to a previously known value from transit ephemeris. Such constraints will be noted in the results section.

Second, we run an affine-invariant Markov Chain Monte Carlo based on the method described by \citet{GoodmanWeare2010}. At every step, we propose a move for each walker $X_k$ by randomly selecting another walker $X_j$ from the ensemble and picking a location on the line connecting the two walkers. This is called a stretch move and can be described by the equation
\begin{eqnarray}
X_k \rightarrow Y = X_j + Z (X_k - X_j) \nonumber
\end{eqnarray}
where the scaling variable Z is drawn from a probability distribution $g(z) \propto 1 / \sqrt{z}$ for $z \in [\frac{1}{2},2]$ and $g(z)=0$ otherwise. The proposed position Y is then accepted with probability
\begin{eqnarray}\label{eq:pacc}
\Pr (X_k \rightarrow Y) = \min \left( 1, Z^{n-1} \frac{\pi(Y)}{\pi(X_k)} \right)
\end{eqnarray}
where $n$ is the number of dimensions and $\pi(X)$ is the unnormalized posterior probability
\begin{equation}\label{eq:ppost}
\pi(X) \equiv p(X) \exp \left( -\frac{\chi^2}{2} \right)
\end{equation}

In the formula above, $p(X)$ is the eccentricity prior discussed in Section \ref{sec:prior} and $\chi^2$  is given by Equation \ref{eq:chi2}. If the proposed stretch move is not accepted then we just set $X_k \rightarrow X_k$ as usual. The above formulation assures that all the moves are symmetric in probability:
\begin{eqnarray}
\Pr (X_k \rightarrow Y) = \Pr (Y \rightarrow X_k) \nonumber
\end{eqnarray}

Subsequently, the resulting number density of the walker distribution is proportional to the real posterior probability density in the usual Monte Carlo way. The affine-invariant method has been shown to have significantly shorter burn-in and autocorrelation times compared to the basic Metropolis-Hastings method (\citet{GoodmanWeare2010}), and it minimizes human input as the routine dynamically adjusts the scaling of each parameter based on the posterior probability distributions.

Third, we take the best set of parameters $X_b$ from the step above (the one that gives the highest posterior probability $\pi(X_b))$ and use a Levenberg-Marquardt algorithm to find the local posterior probability maximum. This is done by adopting the C++ code from the publicly available ALGLIB package (www.alglib.net) developed by Sergey Bochkanov.

Finally, we take the best result $X_B$ in terms of maximizing $\pi(X_B)$ from all three steps above and report it as our “best fit” for the current model. We then report errors on every free parameter by using 68\% confidence intervals - that is, for every free parameter $x_B$ in the set $X_B = \{x_B \pm \sigma_B\}$, at least 68\% of the marginalized probability is contained within $x_B - \sigma_B \leq x \leq x_B + \sigma_B$.

\subsection{Model selection \label{sec:modelselection}}

The fitting for multiple planets is done sequentially. We start out by fitting the observed radial velocities for a single planet as described above. We then proceed by subtracting the single-planet model from the data, and by repeating the analysis for the velocity residuals. While fitting for additional planets, the walker positions for previously fitted planets are allowed to vary as well, and they are initially drawn from normal distributions centered at the result of the previous model. We generate models both with and without a linear trend $\dot{\gamma}$. We generally set $\ddot{\gamma} = 0$ unless there is strong evidence for the existence of a long-period planet whose orbital period cannot be restricted with a sufficient precision based on the current time baseline. Therefore, we end up with a series of different models for each star.

Since the $\chi^2$ parameter depends on radial velocity residuals, introducing additional parameters into the model generally reduces its value. Thus, we generally prefer to use the ``reduced chi-squared'' $\chi_{\nu}^2$ instead for model comparison purposes since it also takes into account the number of free parameters. However, relying solely on $\chi_{\nu}^2$ might not sufficiently penalize the risks and additional (e.g. dynamical) constraints introduced by adding another companion into the system. We also want to make sure that the reduction in $\chi_{\nu}^2$ is statistically significant to minimize the number of false positives due to uncertainties in the data. Therefore, we test a $\chi^2$-difference test for nested models. We calculate the difference $\Delta \chi^2$ between the two models as well as the difference in the number of degrees of freedom $\Delta n_{\rm dof}$. The latter parameter is generally 5 for an additional companion and 1 when introducing a linear trend. We then plug these into a chi-square distribution with $k = \Delta n_{\rm dof}$ degrees of freedom, described by the following probability density function (pdf):
\begin{equation}
f_k(x) = \frac{x^{\frac{k}{2} - 1} e^{-\frac{x}{2}}}{2^{\frac{k}{2}} \Gamma\left(\frac{k}{2}\right)}
\end{equation}
Subsequently, we obtain the significance of the improvement via
\begin{equation}\label{eq:pvalue}
p = \int_{\Delta \chi^2}^{\infty} f_k(x) dx
\end{equation}
Equation \ref{eq:pvalue} is the $p$-value of observing a difference in $\chi^2$ at least as extreme as $\Delta \chi^2$. We typically discard the more complex model if $p > 0.001$.

Finally, we inspect any system with known or prospective companions by eye and conduct additional tests if necessary to rule out additional sources of false positives. To identify temporal window functions in the data, we equip the Lomb-Scargle periodograms with false alarm probability (FAP) estimates. These are obtained by bootstrapping the observed radial velocities with replacement while keeping the observation times and uncertainties fixed, generating Lomb-Scargle periodograms for the bootstrap samples, and counting the percentage of samples with periodogram peaks at least as high as the original data sample. In addition, we calculate the Pearson correlation coefficient between the radial velocities and the respective $S_{\rm HK}$ values characterizing the emission flux in the Ca II H\&K lines. The ``S-values'' are well-established indices for estimating chromospheric activity \citep{Noyes1984, Duncan1991} and stellar rotation periods. Therefore, we can use them to check if the data displays any stellar activity masquerading as a periodic Keplerian signal. If the derived correlation is significant, the radial velocities can be decorrelated and the analysis can be repeated.

Our best models for all of the planetary and stellar companions are summarized in Table \ref{t_planets}. Many of these results are described in greater detail in the sections below.


\begin{longrotatetable}
\begin{deluxetable}{lrrrrrrrrrrr}
	\rotate
	\tablewidth{0pt}
	\tabletypesize{\scriptsize}
	\tablecaption{Best Fits for Stellar Companions \label{t_planets}}
	\tablehead{
		\colhead{Planet}         &
		\colhead{P} & \colhead{K}          &
		\colhead{e}  & \colhead{$\omega$}  &
		\colhead{T$_0$}         & \colhead{M sin i}    &
		\colhead{a} & \colhead{Trend} &
		\colhead{$\chi_{\nu}^2$} \\
		& \colhead{(day)} & \colhead{(m/s)} & &
		\colhead{(deg)} & \colhead{(JD-2450000)} &
		\colhead{(\mjup)} & \colhead{(AU)} &
		\colhead{(m/s/yr)} & 
	}
	\startdata
	HAT-P-1 b& 4.4652934(fixed)& 61.2(2.2)& 0(fixed)& 253.1(2.0)& 13897.11172(fixed)& 0.534(0.048)& 0.0551(0.0015)& \nodata& 0.94 \\
	HAT-P-12 b& 3.2130589(fixed)& 35.5(1.4)& 0(fixed)& 354.3(2.1)& 14187.027(fixed)& 0.208(0.012)& 0.03837(0.00035)& \nodata& 3.91 \\
	HAT-P-3 b& 2.899736(fixed)& 89.2(2.3)& 0(fixed)& 107.7(1.7)& 14187.008(fixed)& 0.591(0.028)& 0.03871(0.00042)& \nodata& 1.66 \\
	HAT-P-4 b& 3.056328(0.000028)& 79.0(1.1)& 0.084(0.014)& 97.4(9.7)& 14187.86(0.081)& 0.655(0.045)& 0.0445(0.0012)& 7.83(0.41)& 13.45 \\
	HD 103459 b& 1831.91(0.87)& 3013(60)& 0.6993(0.0046)& 182.585(0.068)& 13924.4(1.9)& 140(20)& 3.21(0.16)& \nodata& 4.51 \\
	HD 10442 b& 1053.2(3.4)& 30.53(0.66)& 0.09(0.019)& 217(13)& 13972(44)& 1.53(0.86)& 2.03(0.55)& \nodata& 1.79 \\
	HD 109749 b& 5.239891(0.000099)& 29.2(1.1)& 0(fixed)& 109.8(2.1)& 13015.166(fixed)& 0.27(0.045)& 0.0615(0.004)& \nodata& 1.54 \\
	HD 11506 b& 1622.1(2.1)& 78.17(0.57)& 0.3743(0.0053)& 220.16(0.97)& 13391.8(4.7)& 4.83(0.52)& 2.9(0.14)& -7.133(0.097)& 3.48 \\
	HD 11506 c& 223.41(0.32)& 12.1(0.41)& 0.193(0.038)& 259(16)& 13230(11)& 0.408(0.057)& 0.774(0.038)& -7.133(0.097)& 3.48 \\
	HD 125612 b& 557.04(0.35)& 80.46(0.53)& 0.4553(0.0055)& 42.0(1.0)& 13221.8(1.7)& 3.1(0.4)& 1.372(0.083)& \nodata& 2.31 \\
	HD 125612 c& 4.15514(0.00026)& 6.46(0.44)& 0.049(0.038)& 123(147)& 13057.6(1.7)& 0.055(0.01)& 0.0524(0.0031)& \nodata& 2.31 \\
	HD 125612 d& 2835.0(7.9)& 98.37(0.59)& 0.1172(0.0056)& 313.2(2.9)& 14508(28)& 7.28(0.93)& 4.06(0.25)& \nodata& 2.31 \\
	HD 147506 b& 5.6335158(0.0000036)& 953.3(3.6)& 0.5172(0.0019)& 188.01(0.2)& 13982.4915(0.0024)& 8.62(0.17)& 0.06814(0.00051)& -47.2(1.2)\tablenotemark{a}& 21.73 \\
	HD 148164 b& 5062(114)& 54.28(0.89)& 0.125(0.017)& 152(11)& 14193(155)& 5.16(0.82)& 6.15(0.5)& \nodata& 3.73 \\
	HD 148164 c& 328.55(0.41)& 39.6(1.7)& 0.587(0.026)& 141.5(2.7)& 13472.7(4.7)& 1.23(0.25)& 0.993(0.066)& \nodata& 3.73 \\
	HD 148284 b& 339.331(0.018)& 1022.0(1.2)& 0.38926(0.00089)& 35.56(0.14)& 13750.96(0.21)& 33.7(5.5)& 0.974(0.079)& \nodata& 1.31 \\
	HD 149026 b& 2.8758911(fixed)& 39.22(0.68)& 0.051(0.019)& 109(21)& 13208.8(0.17)& 0.326(0.043)& 0.0432(0.0024)& \nodata& 4.42 \\
	HD 149143 b& 4.07182(0.00001)& 150.3(0.65)& 0.0167(0.004)& 217(17)& 13199.12(0.2)& 1.33(0.15)& 0.053(0.0029)& \nodata& 4.74 \\
	HD 1605 b& 2149(16)& 47.47(0.71)& 0.099(0.011)& 233.0(9.2)& 14691(79)& 3.62(0.23)& 3.584(0.099)& -8.63(0.2)& 4.38 \\
	HD 1605 c& 577.2(2.5)& 18.97(0.63)& 0.095(0.057)& 216(41)& 13760(67)& 0.934(0.079)& 1.492(0.038)& -8.63(0.2)& 4.38 \\
	HD 163607 b& 75.195(0.034)& 53.0(1.4)& 0.744(0.012)& 79.7(2.0)& 13584.7(0.74)& 0.79(0.11)& 0.362(0.017)& -2.7(1.0)\tablenotemark{b}& 0.54 \\
	HD 163607 c& 1267.4(7.2)& 37.72(0.95)& 0.076(0.023)& 287(20)& 13887(72)& 2.16(0.27)& 2.38(0.12)& -2.7(1.0)\tablenotemark{b}& 0.54 \\
	HD 164509 b& 280.17(0.82)& 13.15(0.77)& 0.238(0.062)& 326(13)& 13741(11)& 0.443(0.083)& 0.87(0.051)& -6.1(0.57)\tablenotemark{c}& 3.41 \\
	HD 16760 b& 466.048(0.057)& 407.16(0.71)& 0.0812(0.0018)& 241.9(1.4)& 13802.6(1.9)& 15.0(2.5)& 1.161(0.097)& \nodata& 1.15 \\
	HD 171238 b& 1532(12)& 50.7(2.6)& 0.234(0.028)& 97(11)& 13101(45)& 2.72(0.49)& 2.57(0.16)& \nodata& 4.46 \\
	HD 17156 b& 21.2167(0.0003)& 274.7(3.5)& 0.6753(0.0048)& 121.51(0.32)& 13759.802(0.039)& 3.16(0.42)& 0.1607(0.0091)& \nodata& 1.73 \\
	HD 179079 b& 14.479(0.01)& 6.22(0.78)& 0.049(0.087)& 308(77)& 13211.3(3.0)& 0.081(0.02)& 0.1214(0.0068)& \nodata& 0.93 \\
	HD 203473 b& 1552.9(3.4)& 133.6(2.4)& 0.289(0.01)& 18.0(1.1)& 13333.6(9.1)& 7.8(1.1)& 2.73(0.17)& -24.9(2.0)\tablenotemark{d}& 2.0 \\
	HD 207832 b& 160.07(0.23)& 20.73(0.91)& 0.197(0.053)& 1(16)& 13276.9(6.3)& 0.56(0.091)& 0.586(0.032)& \nodata& 1.82 \\
	HD 211810 b& 1558(22)& 15.6(7.2)& 0.68(0.14)& 98(14)& 14763(86)& 0.67(0.44)& 2.656(0.043)& \nodata& 1.2 \\
	HD 214823 b& 1854.4(1.1)& 285.47(0.96)& 0.1641(0.0026)& 124.0(1.2)& 13793.1(5.9)& 20.3(2.6)& 3.23(0.2)& \nodata& 2.61 \\
	HD 217850 b& 3501.3(2.1)& 439.0(5.8)& 0.7621(0.0019)& 165.95(0.22)& 14048.4(3.9)& 21.6(2.6)& 4.56(0.24)& \nodata& 1.81 \\
	HD 219828 b& 4682(99)& 270.7(7.2)& 0.8102(0.0051)& 145.68(0.44)& 14180(114)& 14.6(2.3)& 5.79(0.41)& \nodata& 0.68 \\
	HD 219828 c& 3.83492(0.00014)& 7.73(0.43)& 0.101(0.063)& 248(55)& 11450.84(0.64)& 0.066(0.012)& 0.0507(0.0029)& \nodata& 0.68 \\
	HD 224693 b& 26.6904(0.0019)& 39.96(0.68)& 0.104(0.017)& 358(10)& 13193.79(0.77)& 0.7(0.12)& 0.191(0.014)& \nodata& 4.31 \\
	HD 231701 b& 141.63(0.067)& 39.2(1.2)& 0.13(0.032)& 68(14)& 13330.6(5.3)& 1.13(0.25)& 0.567(0.053)& \nodata& 2.2 \\
	HD 24505 b& 11315(92)& 3294.0(2.8)& 0.798(0.0012)& 157.895(0.072)& 16995(65)& 222(30)& 10.82(0.61)& \nodata& 0.59 \\
	HD 33283 b& 18.1991(0.0017)& 22.4(1.6)& 0.399(0.056)& 155.5(7.1)& 13017.31(0.29)& 0.329(0.071)& 0.1508(0.0087)& \nodata& 0.66 \\
	HD 3404 b& 1540.8(1.9)& 3535(188)& 0.7381(0.0044)& 0.86(0.9)& 13455.3(9.0)& 145(28)& 2.86(0.16)& \nodata& 0.37 \\
	HD 37605 b& 55.01292(0.00062)& 203.47(0.75)& 0.6745(0.0019)& 220.78(0.27)& 13048.16(0.025)& 2.69(0.3)& 0.277(0.015)& \nodata& 1.38 \\
	HD 37605 c& 2720(15)& 48.51(0.57)& 0.03(0.012)& 227(30)& 14881(237)& 3.19(0.38)& 3.74(0.21)& \nodata& 1.38 \\
	HD 38801 b& 687.14(0.46)& 194.4(1.8)& 0.0572(0.0063)& 2(11)& 13976(20)& 10.0(1.4)& 1.66(0.11)& 5.07(0.32)& 3.41 \\
	HD 43691 b& 36.9987(0.0011)& 130.06(0.84)& 0.0796(0.0067)& 292.7(4.8)& 13048.04(0.49)& 2.55(0.34)& 0.238(0.015)& \nodata& 2.6 \\
	HD 45652 b& 44.073(0.0048)& 33.2(1.8)& 0.607(0.026)& 227.7(5.6)& 13720.92(0.45)& 0.433(0.076)& 0.237(0.011)& 2.83(0.21)& 8.23 \\
	HD 5319 b& 638.6(1.2)& 31.45(0.82)& 0.015(0.016)& 86(69)& 13066(123)& 1.56(0.29)& 1.57(0.13)& \nodata& 2.67 \\
	HD 5319 c& 877.0(4.9)& 18.53(0.91)& 0.109(0.067)& 245(42)& 13445(106)& 1.02(0.22)& 1.94(0.16)& \nodata& 2.67 \\
	HD 55696 b& 1827(10)& 76.7(3.9)& 0.705(0.022)& 137.0(2.4)& 13648(26)& 3.87(0.72)& 3.18(0.18)& 1.34(0.34)& 4.4 \\
	HD 73534 b& 1721(36)& 15.5(1.2)& 0(fixed)& 34(10)& 13014.922(fixed)& 1.01(0.21)& 2.95(0.22)& \nodata& 1.05 \\
	HD 75784 b& 3878(261)& 51.9(4.9)& 0.266(0.04)& 302(14)& 14623(187)& 4.5(1.2)& 5.22(0.58)& 3.1(1.3)& 1.05 \\
	HD 75784 c& 341.5(1.3)& 27.2(4.8)& 0.142(0.078)& 29(44)& 13046(46)& 1.08(0.35)& 1.033(0.071)& 3.1(1.3)& 1.05 \\
	HD 75898 b& 422.9(0.29)& 63.39(0.71)& 0.11(0.01)& 241.1(5.2)& 13299.0(5.9)& 2.71(0.36)& 1.191(0.073)& \nodata& 5.49 \\
	HD 75898 c& 6066(337)& 27.8(1.5)& 0(fixed)& 77.5(5.7)& 13014.946(fixed)& 2.9(0.57)& 7.03(0.69)& \nodata& 5.49 \\
	HD 79498 b& 1807(15)& 26.0(1.2)& 0.575(0.023)& 226.8(6.9)& 13380(32)& 1.34(0.21)& 2.98(0.15)& \nodata& 1.49 \\
	HD 86081 b& 2.1378431(0.0000031)& 205.53(0.78)& 0.0119(0.0047)& 3(23)& 13695.46(0.14)& 1.48(0.23)& 0.0346(0.0027)& -1.0(0.23)& 1.69 \\
	HD 88133 b& 3.414887(0.000045)& 32.7(1.0)& 0(fixed)& 205.3(3.3)& 13014.948(fixed)& 0.282(0.046)& 0.0479(0.0032)& \nodata& 1.11 \\
	HD 96167 b& 498.1(0.81)& 21.1(1.6)& 0.681(0.033)& 288.3(6.4)& 13060.3(4.5)& 0.71(0.18)& 1.332(0.092)& \nodata& 1.02 \\
	HD 98630 b& 13074(982)& 2613(24)& 0.059(0.032)& 71(14)& 14297(542)& 359(26)& 13.17(0.84)& \nodata& 6.25 \\
	HD 98736 b& 968.8(2.2)& 52(12)& 0.226(0.064)& 162(22)& 13541(67)& 2.33(0.78)& 1.864(0.091)& -3.08(0.17)& 2.05 \\
	HIP 14810 b& 6.673892(0.000008)& 423.34(0.4)& 0.14399(0.00087)& 158.83(0.38)& 13694.5879(0.0067)& 3.9(0.49)& 0.0696(0.0044)& -2.15(0.11)& 2.68 \\
	HIP 14810 c& 147.747(0.029)& 50.91(0.45)& 0.1566(0.0099)& 331.1(2.6)& 13786.4(1.2)& 1.31(0.18)& 0.549(0.034)& -2.15(0.11)& 2.68 \\
	HIP 14810 d& 981.8(6.9)& 12.17(0.46)& 0.185(0.035)& 247(12)& 14194(40)& 0.59(0.1)& 1.94(0.13)& -2.15(0.11)& 2.68 \\
	XO-5 b& 4.18776(fixed)& 143.6(1.3)& 0(fixed)& 331.01(0.72)& 14186.946(fixed)& 1.044(0.034)& 0.04872(0.00055)& \nodata& 4.77 \\
	\enddata
	\tablenotetext{a}{The model for HD 147506 includes a curvature term of 6.11(0.26) m/s/yr$^2$.}
	\tablenotetext{b}{The model for HD 163607 includes a curvature term of 0.96(0.18) m/s/yr$^2$.}
	\tablenotetext{c}{The model for HD 164509 includes a curvature term of 0.58(0.11) m/s/yr$^2$.}
	\tablenotetext{d}{The model for HD 203473 includes a curvature term of 3.91(0.30) m/s/yr$^2$.}
\end{deluxetable}
\end{longrotatetable}

\subsection{Eccentricity prior \label{sec:prior}}

\begin{figure}
    \plottwo{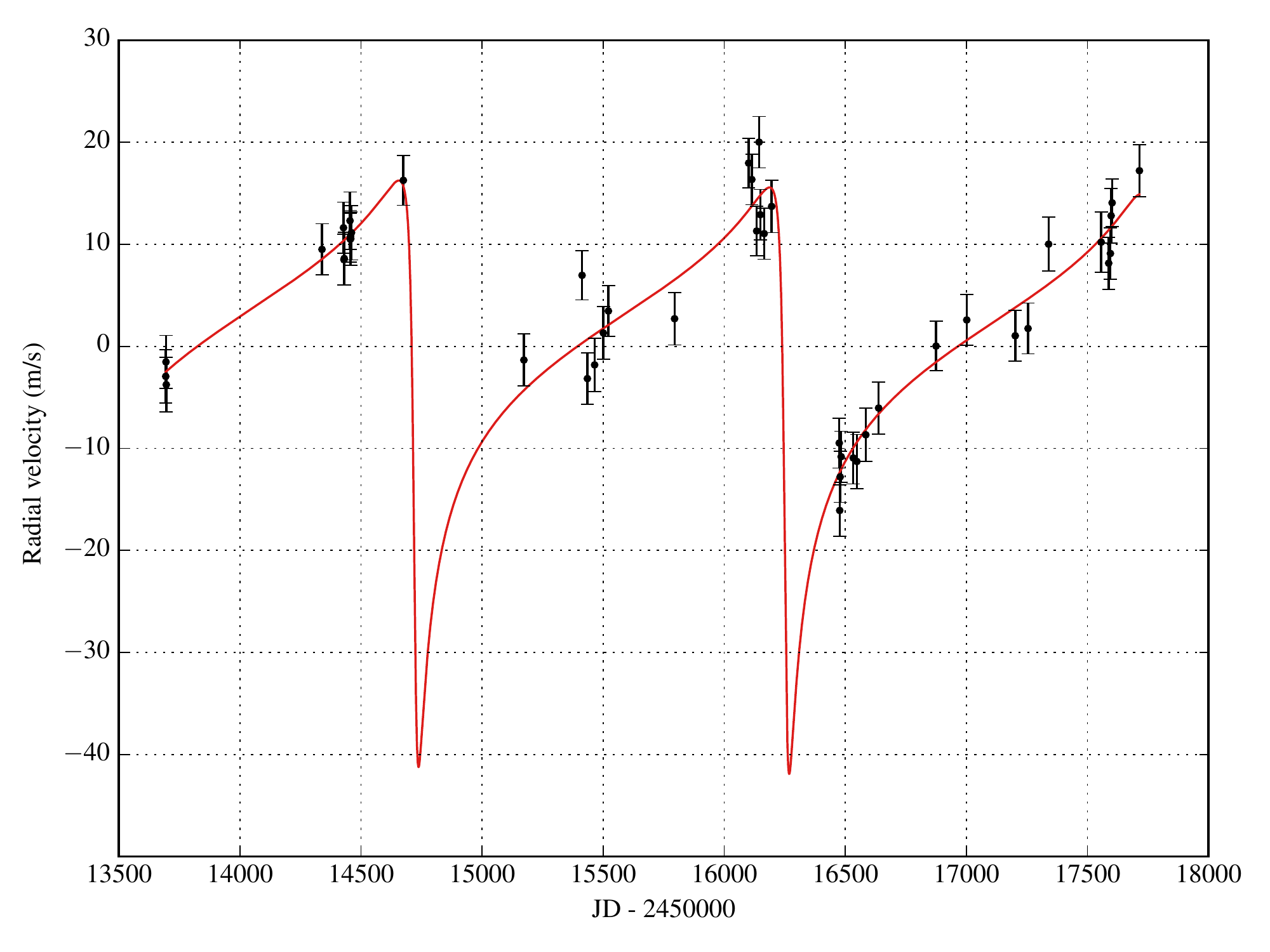}{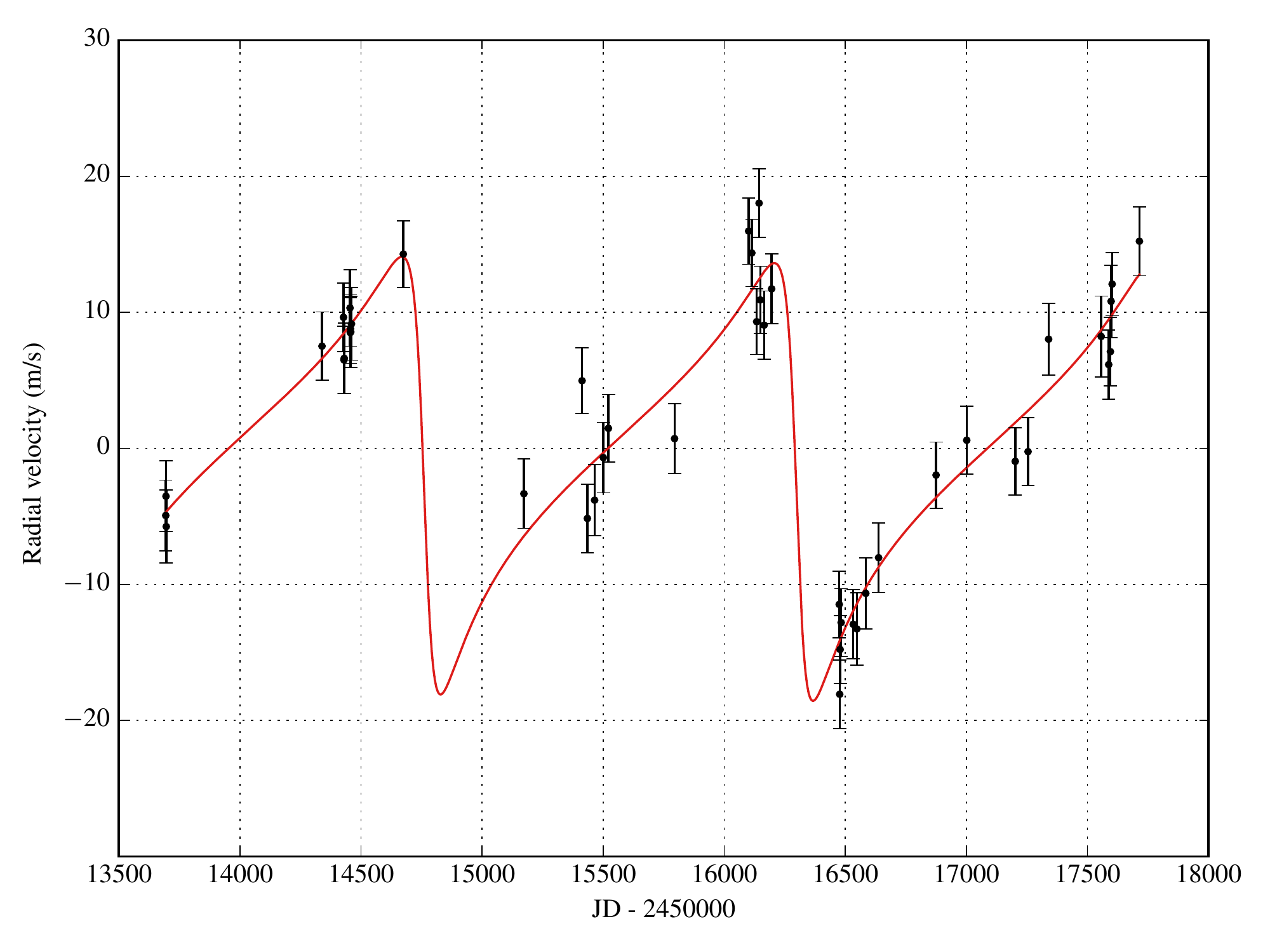}
    \caption{Two Keplerian models for a single planet orbiting HD 211810 based on 46 RV observations reported in this paper. On the left, the orbital eccentricity is allowed to vary freely, yielding a model with $e = 0.84 \pm 0.09$ and $\chi_\nu^2 = 1.21$. However, the model includes rapid RV variations during orbital phases with no data coverage. On the right, the same data has been fitted for using the eccentricity prior described in Section \ref{sec:prior}. We obtain an orbital eccentricity of $e = 0.68 \pm 0.16$ whereas the \chisq\ comes out as 1.22 - a statistically insignificant difference from the case with no prior. Importantly, the latter model displays fewer spurious features not backed up by the observed data. Both models include a small non-significant linear trend.}
    \label{figE10}
\end{figure}

\begin{figure}
	\plottwo{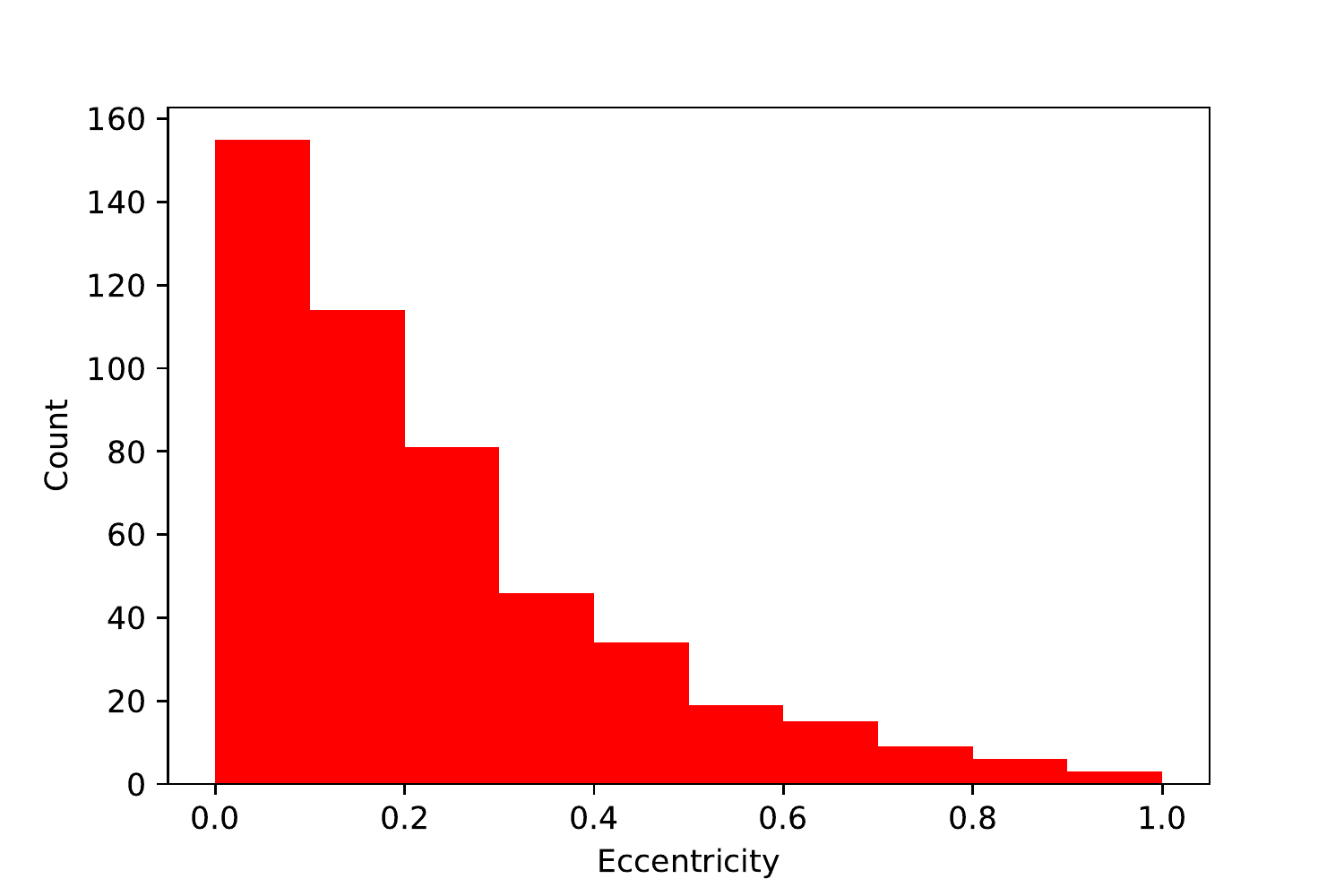}{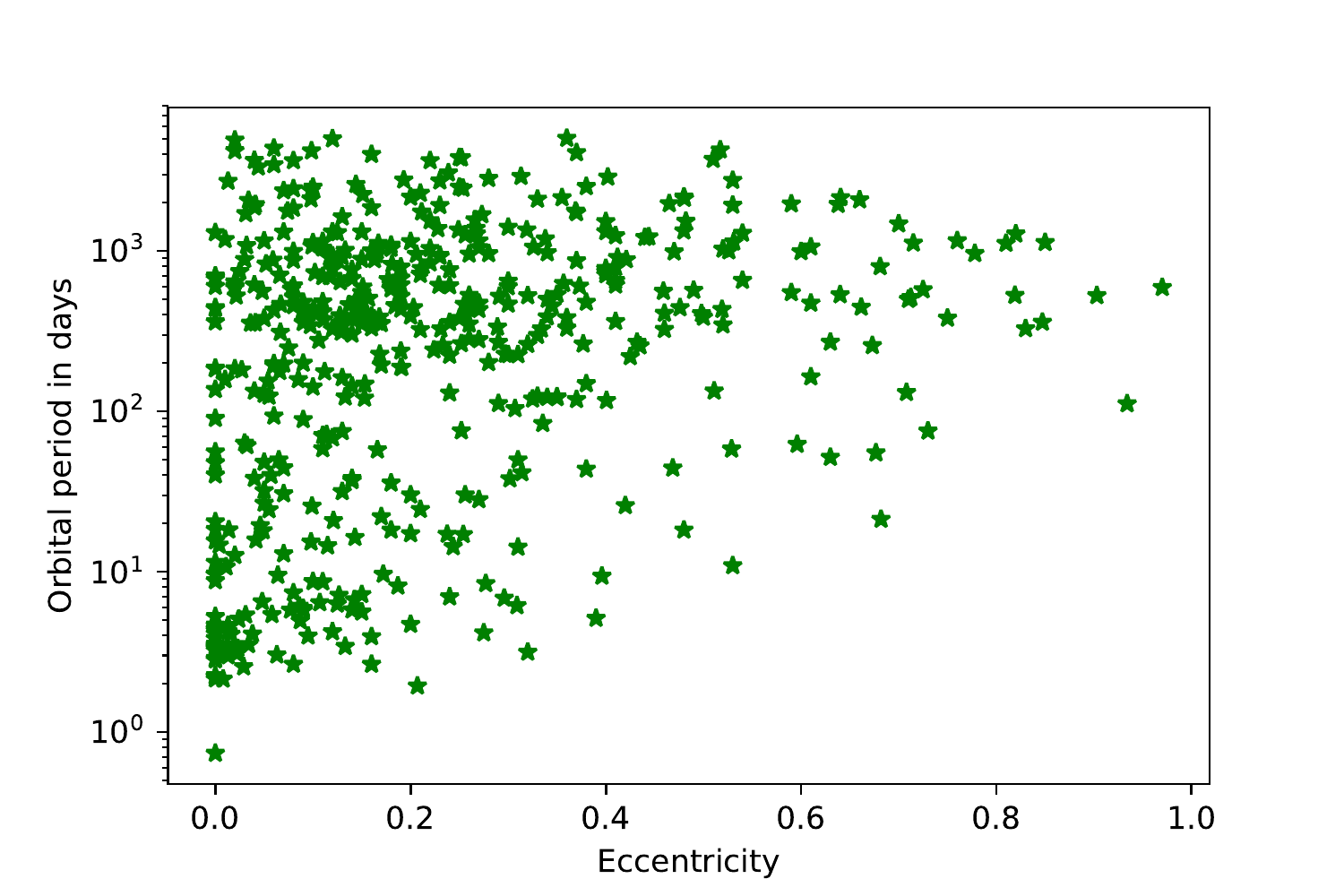}
	\caption{\textit{Left:} The eccentricity distribution of 482 known exoplanets discovered by the radial velocity technique. \textit{Right:} The eccentricity distribution of known exoplanets versus their orbital periods. Short-period orbits are much more likely to be circularized by tidal forces. Source: exoplanets.org\label{figE12}}
\end{figure}

Simple \chisq-minimization routines tend to yield orbital solutions with spuriously high eccentricities for the following reasons. To begin with, orbital eccentricites can never be negative. Therefore, any noise in the radial velocity measurements in the case of a circular orbit tends to drive the model away from the true eccentricity of 0. For example, \citet{Valenti2009} simulated $10^5$ RV data sets based on an $e = 0$ model only to recover a \chisq-minimizing eccentricity of $e = 0.115 \pm 0.087$. Furthermore, incomplete orbital phase coverage can drive the model towards higher eccentricities by allowing the RV signal to vary rapidly during the unobserved orbital phases. This is illustrated in Figure \ref{figE10} for the case of HD 211810 b, one of the discoveries of this paper. Allowing the orbital eccentricity of HD 211810 b to vary freely yields a non-significant decrease in the model \chisq\ from 1.22 to 1.21 at the cost of introducing a rapid RV variation during the orbital phases with no data coverage. By Occam's razor alone, we prefer the simpler model with no ambiguous features in the RV signal. This is in accordance with the eccentricity distribution of observed exoplanets which is heavily biased towards low-eccentricity orbits, as illustrated in Figure \ref{figE12}. The predisposition of truly circular orbits to appear spuriously eccentric was described in detail by \citet{LucySweeney1971} and is therefore known as the Lucy-Sweeney bias.

In order to alleviate this problem, we adjust the modeled likelihood by using a prior probability distribution on the orbital eccentricity $e$. A common choice for the eccentricity prior is a Rayleigh distribution which is predicted by some planet-planet scattering theories (\citet{JuricTremaine2008}, \citet{WangFord2011}). Others have developed their own models that are more specifically geared towards radial velocity surveys, e.g. \citet{ShenTurner2008}. We decided to develop a model for the prior probability $p(e)$ based on a linear combination of the Rayleigh and \citet{ShenTurner2008} models, given by Equation \ref{eq:ptE}.
\begin{equation}
    \tilde{p}(e) = Ae \exp(-Ce^2) + B \Big[(1 + e)^{-\gamma} - e \cdot 2^{-\gamma}\Big]
    \label{eq:ptE}
\end{equation}

While Equation \ref{eq:ptE} is a good starting point, we also noticed that the orbital eccentricities obtained from radial velocity surveys are visibly correlated with the orbital period, as can be seen on the right side of Figure \ref{figE12}. This is plausibly due to the fact that close-in orbits with shorter periods are much more likely to be tidally circularized. Thus, the conditional distributions $p(e | P)$ can vary noticeably depending on the value of the orbital period $P$. In order to incorporate this effect, we transformed our prior into a joint distribution $\tilde{p}(e,P)$ by expanding the coefficients ($A$, $B$, $C$, $\gamma$) in Equation \ref{eq:ptE} as quadratic functions of $\tilde{P} = \log_{10} P$ where $P$ is expressed in days. At the same time, we wanted to keep the marginal probabilities $p(P) = \int_0^1 p(e,P) de$ equal for every orbital period $P$ in order to avoid implementing a prior on $P$. Thus, we normalized our prior by dividing the unnormalized probability $\tilde{p}(e,P)$ with the marginal probability $F(P) \equiv \int_0^1 \tilde{p}(e,P) de$. This ensures that every orbital period $P$ has the same marginal probability.

As a result, our joint prior takes the following form:
\begin{equation}
    p(e, P) = \frac{ \tilde{p}(e,P) }{ \int_0^1 \tilde{p}(e,P) de} = \frac{ Ae \exp(-Ce^2) + B \Big[(1 + e)^{-\gamma} - e \cdot 2^{-\gamma}\Big] }{F}
    \label{eq:pE}
\end{equation}
where $A$, $B$, $C$, $\gamma$, and $F$ are all functions of $\tilde{P} = \log_{10} P$ only and can be calculated using Equations \ref{eq:pE:A} through \ref{eq:pE:F}. Notice that $p(P) = \int_0^1 p(e,P) de = 1$ so the prior in Equation \ref{eq:pE} is still properly normalized in the sense of probability (every orbital period has a marginal probability of 1). However, as can be inferred by combining Equations \ref{eq:pacc} and \ref{eq:ppost}, we are only interested in the ratios of the prior (and posterior) probabilities.
\begin{eqnarray}
A = 40.5 \tilde{P}^2 - 80.1 \tilde{P} + 152.6 \label{eq:pE:A} \\
B = 10.6 \tilde{P}^2 + 3.1 \tilde{P} + 24.3 \\
C = 7.0 \tilde{P}^2 - 43.8 \tilde{P} + 80.8 \\
\gamma = 0.7 \tilde{P}^2 - 4.8 \tilde{P} + 10.5 \label{eq:pE:g} \\
F = \frac{A}{2C} \big(1 - \exp(-C)\big) + \frac{B}{\gamma - 1} \Big(1 - \frac{\gamma + 3}{2^{\gamma + 1}}\Big) \label{eq:pE:F}
\end{eqnarray}

\begin{figure}
	\plotone{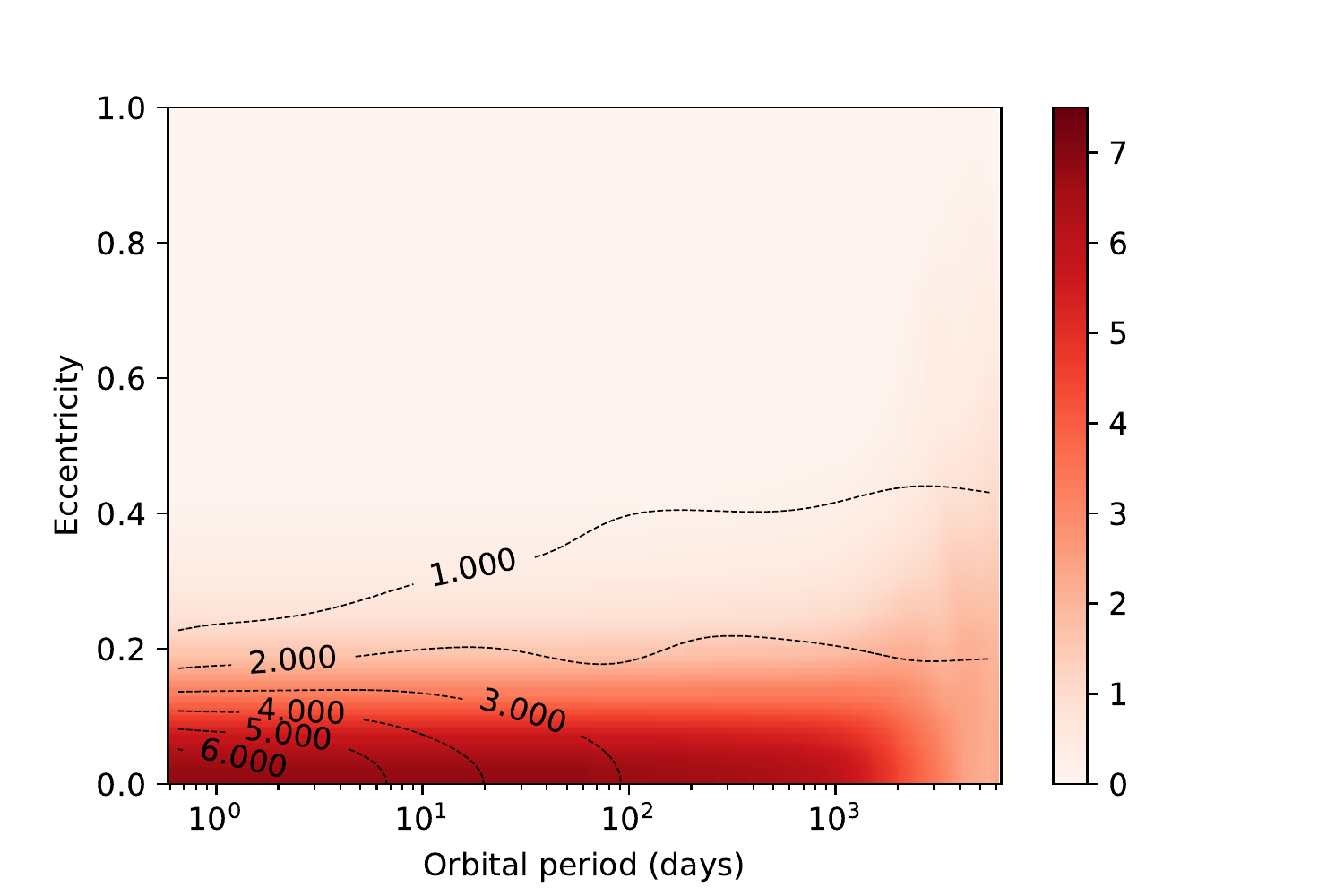}
	\caption{The eccentricity prior as a function of orbital period and eccentricity. The value of the prior is indicated by the color depth.}
	\label{fig:eccprior}
\end{figure}

\begin{figure}
	\plotone{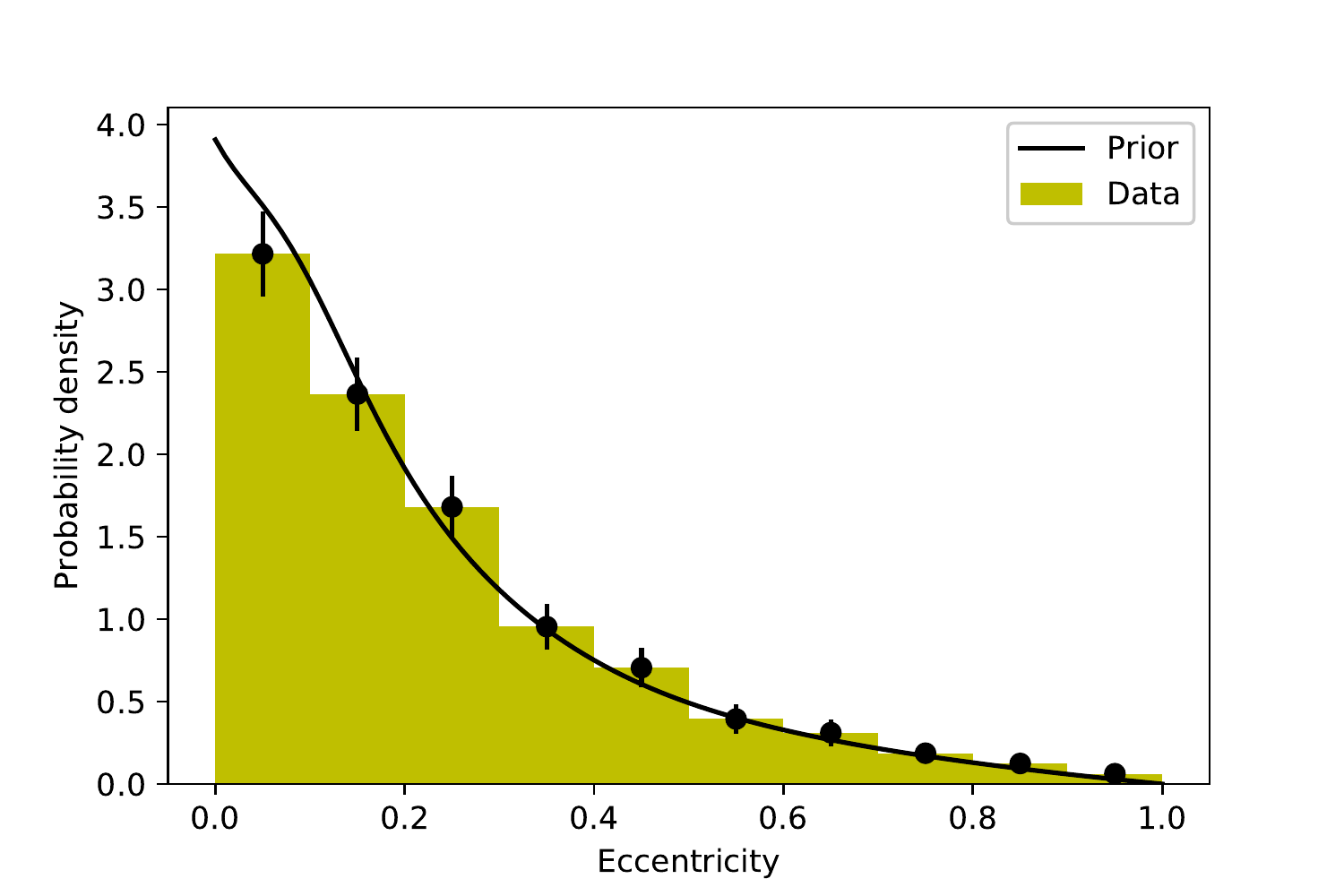}
	\caption{The marginalized eccentricity distribution from our model (solid line) compared with the eccentricity distribution of 482 known exoplanets discovered by the radial velocity method (source: exoplanets.org). The errors on the histogram columns are equal to $\sqrt{N}/N$ where $N$ is the number of planets in the bin. Our model is generally within 1$\sigma$ everywhere.}
	\label{figE14}
\end{figure}

The coefficients in Equations \ref{eq:pE:A}-\ref{eq:pE:g} were found by grid-sampling a Kernel density estimator (KDE) of the period-eccentricity distribution given in Figure \ref{figE12} to estimate the probability density at each sample point $(e,P)$, and then by using a least-squares optimizer to minimize the difference between the estimated and the modeled probability densities. We used a reflection method to correct the KDE for the boundary effects resulting from the fact that there are no data points with $e < 0$. The prior is displayed in Figure \ref{fig:eccprior}. We also calculated the marginal probability $p(e) = \int p(e,P) dP$ and compared it with the histogram of eccentricities in Figure \ref{figE12}. This comparison can be seen in Figure \ref{figE14}. Assuming Poisson errors for each histogram bin, our model is roughly within 1$\sigma$ everywhere.

The right hand side of Figure \ref{figE10} demonstrates the model with the highest posterior probability for the case of HD 211810 b mentioned above.  We tested the eccentricity prior on numerous simulated data sets, and found that it tends to disfavor models with higher eccentricities unless they are clearly backed up by data. Models with spuriously high eccentricities were mainly caused by a relative shortage of data points or significant gaps in orbital coverage. At the same time, it did not dampen the eccentricities by a significant amount where a highly eccentric model was warranted. This is also demonstrated by the fact that a large majority of our quoted orbital eccentricities for previously known planets are in agreement with the previously published values. Finally, we tested the significance of any eccentric model against a circular one by comparing the $\chi^2$-differences as described in Section \ref{sec:modelselection} (for $\Delta n_{\rm dof} = 2$), and preferred the simpler (circular) model whenever possible.

\subsection{Stellar parameters \label{sec:stellarpars}}

For each of the target stars, we used the following methodology to obtain the stellar parameters, unless otherwise stated. We adopt the spectral type from SIMBAD unless we have a reason to believe the reported type is inaccurate based on other stellar parameters. We then query the Hipparcos catalog to obtain values for the V-magnitude brightness $V$, B-V color, and the trigonometric parallax $p''$ in arcseconds. This allows us to calculate the absolute brightness as $M_V = V + 5 + 5 \log p''$. Furthermore, we predict the star’s height above the main sequence $\Delta M_V = M_{V,MS} (B-V) - M_V$, where the main sequence is modelled by \cite{Wright2005}:
\begin{eqnarray} \label{eq:sp1}
M_{V,MS} (B-V) = \sum_{i=0}^{9} a_i (B-V)^i \\
a = \{1.11255,5.79062,-16.76829,76.47777,-140.08488, \nonumber \\
	127.38044,-49.71805,-8.24265,14.07945,-3.43155\} \nonumber
\end{eqnarray}

We fetch estimates for the stellar mass, age, luminosity, and radius from \cite{Brewer2016} who determined these parameters by using the Yonsei-Yale isochrones. We also adopt their spectroscopically derived effective temperature, surface gravity ($\log g$), activity ($\log R_{HK}$), and metallicity ([Fe/H]). 

Chromospheric activity is measured by emission in the cores of Ca II H\&K lines as $S_{HK}$ and were determined for our stars to estimate contributions to the radial velocity from the stellar photosphere. We follow the procedure of \cite{IsaacsonFischer2010}, using the B-V color in the following formula to first calculate the baseline activity $S_{BL}$:
\begin{eqnarray}
S_{BL} = \sum_{i=0}^{7} b_i (B-V)^i \nonumber \\
b = \{2.7,-16.19,36.22,-27.54,-14.39,34.97,-18.71,3.17\} \nonumber
\end{eqnarray}

The excess activity is then defined as the difference $\Delta S \equiv \left| S_{HK} - S_{BL} \right|$. Subsequently, we use the relations in \cite{IsaacsonFischer2010} to estimate the radial velocity noise from the stellar photosphere (``jitter") in meters per second for main sequence stars (defined by $\Delta M_V < 1.5$) based on their spectral type:
\begin{subequations}
	\begin{align}
	\mathrm{Jitter\ (\ms)} & =  2.3 + 17.4 \cdot \Delta S \qquad \left(0.4<B-V<0.7\right) \label{eq:jit4a} \\
	\mathrm{Jitter\ (\ms)} & =  2.1 + 4.7 \cdot \Delta S \qquad \left(0.7<B-V<1.0\right) \label{eq:jit4b} \\
	\mathrm{Jitter\ (\ms)} & =  1.6 - 0.003 \cdot \Delta S \qquad \left(1.0<B-V<1.3\right) \\
	\mathrm{Jitter\ (\ms)} & =  2.1 + 2.7 \cdot \Delta S \qquad \left(1.3<B-V<1.6\right)
	\end{align}
\end{subequations}

For subgiants ($\Delta M_V \geq 1.5$), the following formulae are used instead:
\begin{eqnarray}
S_{BL} & = & 0.2 - 0.07 (B-V) \nonumber \\
\mathrm{Jitter\ (\ms)} & = & 4.2 + 3.8 \cdot \Delta S \label{eq:jit5}
\end{eqnarray}

We note that the method of \citet{IsaacsonFischer2010} provides a lower limit to the amount of RV jitter, and therefore our quoted jitter values likely underestimate the amount of real RV variations due to chromospheric activity in most cases. For most N2K stars, the RMS of the RV residuals (listed in Table \ref{t_starswithplanets}) after subtracting out any fitted linear trends and/or Keplerian models is significantly higher than the jitter estimate for the same star. However, these residuals are likely caused by a mix of chromospheric activity, instrumental noise, and possible unseen companions. Separating the contributions from each individual component is not straightforward, and therefore we decided not to make any assumptions about the unfitted RV residuals and merely report their RMS in Table \ref{t_starswithplanets}.

\subsection{Planetary mass and orbital semi-major axis \label{sec:pl}}

For any planet, the five orbital parameters $\{K,P,e,\omega,T_P\}$ given in Section \ref{sec:fitting} are related to the planetary mass $m$ and orbital semi-major axis $a$ via the following formulae:
\begin{eqnarray}
K = \left(\frac{2\pi G}{P}\right)^{1/3} \frac{m}{M} \left(m + M\right)^{1/3} \frac{\sin i}{\sqrt{1 - e^2}} \label{eq:pl1} \\
a^3 = \frac{G}{4\pi^2} (m + M) P^2 \label{eq:pl2}
\end{eqnarray}
where $M$ is the mass of the parent star and $i$ is the inclination angle between the orbital plane and the line-of-sight of the observer. For planets ($m < 0.013$ \msun) and brown dwarfs ($m < 0.08$ \msun) we can use the approximation $m + M \approx M$. In this case, we can express $m \sin i$ and $a$ as follows:
\begin{eqnarray}
m \sin i = \left(\frac{P}{2\pi G}\right)^{1/3} KM^{2/3} \sqrt{1 - e^2} \\
a = \left(\frac{GMP^2}{4\pi^2}\right)^{1/3}
\end{eqnarray}

However, our sample also includes a few stars with stellar companions (\msini $> 0.08$ \msun). In these cases, we can provide a lower limit on the companion mass by plugging $\sin i = 1$ into Equation \ref{eq:pl1}. We can then rewrite Equation \ref{eq:pl1} with the help of an additional variable $W \equiv -\frac{1}{2\pi G} K^3 M^3 P (1 - e^2)^{3/2}$ as a quartic equation:
\begin{equation}
m^4 + M m^3 + W = 0
\end{equation}
Since $W < 0$, this equation has a single positive real solution for the planetary mass $m$. Subsequently, we insert the obtained mass $m$ into Equation \ref{eq:pl2} to obtain a lower limit on the orbital semi-major axis $a$.

\section{Results: New discoveries \label{sec:results1}}

\subsection{Substellar companions HD 148284 B, HD 214823 B, and HD 217850 B \label{sec:K11}}

\subsubsection{Stellar properties}

SIMBAD classifies HD 148284 as a K0-dwarf. However, this is inconsistent with a surface temperature of 5572 K reported by \citet{Brewer2016}. Furthermore, based on equation \ref{eq:sp1}, the star is 1.26 mag above the main sequence. Therefore, HD 128284 is more likely a G-type main sequence star or a subgiant. Its derived age from Yale-Yonsei isochrones is $8.7 \pm 1.2$ Gyr and its distance from the Sun is slightly above 94 pc. Using the process outlined in \citet{IsaacsonFischer2010} and described in Section \ref{sec:stellarpars}, we calculate the stellar velocity jitter as 2.15 \ms. We also derive a stellar rotation period of 42 days based on the chromospheric activity. HD 148284 was observed for a total of 30 times over more than 11 years by the N2K consortium.

HD 214823 is a G0-type star. Its position of $\Delta M_V = 1.56$ mag above the main sequence suggests that HD 214823 could be a subgiant. This claim is further supported by the stellar radius estimate of $R = 2.04 \pm 0.15$ \rsun\ reported by \citet{Brewer2016}. The star has a mass of $M = 1.31 \pm 0.24$ \msun\ and it is at a distance of 98 pc from the Sun. HD 214823 is Sun-like in age: we estimate the star to be $4.3 \pm 0.5$ Gyr old. Based on the S-value $S_{\rm HK} = 0.15$, we derive a radial velocity jitter estimate of 2.33 \ms. We also derive a stellar rotation period of 25 days. We report a total of 28 observational visits of HD 214823 spanning almost exactly 11 years.

HD 217850 is a G8-type main sequence star according to the Hipparcos catalog and SIMBAD. Its estimated mass is $M = 1.03 \pm 0.16$ \msun\ and its age is $7.6 \pm 1.3$ Gyr. It has a high metallicity of \feh\ = 0.26 \citep{Brewer2016} and its distance from the Sun is roughly 61 pc. Using the activity index $S_{\rm HK} = 0.16$ we obtain a jitter estimate of 2.13 \ms. HD 217850 was observed at Keck over 11 years on 27 different nights.

Table \ref{t_starswithplanets} summarizes the parameters for all N2K stars.

\subsubsection{Keplerian models}

\begin{figure}
	\plotone{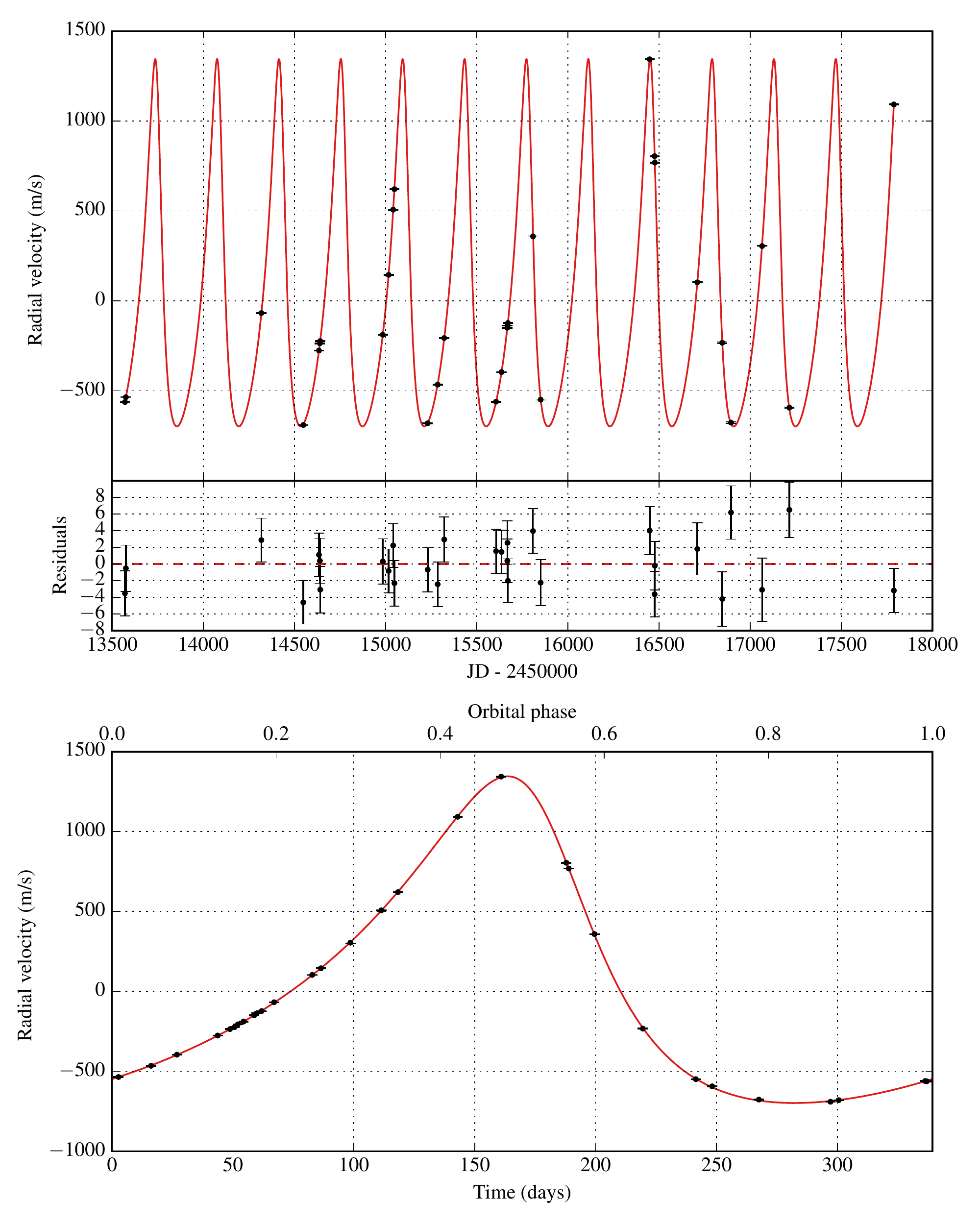}
	\caption{Keplerian model for HD 148284 B. The first plot shows the best Keplerian fit (in red) and the observed data points with an inset display of the residual velocities after fitting. The second plot shows a phase-folded version of the first plot.\label{figK11}}
\end{figure}

\begin{figure}
	\plotone{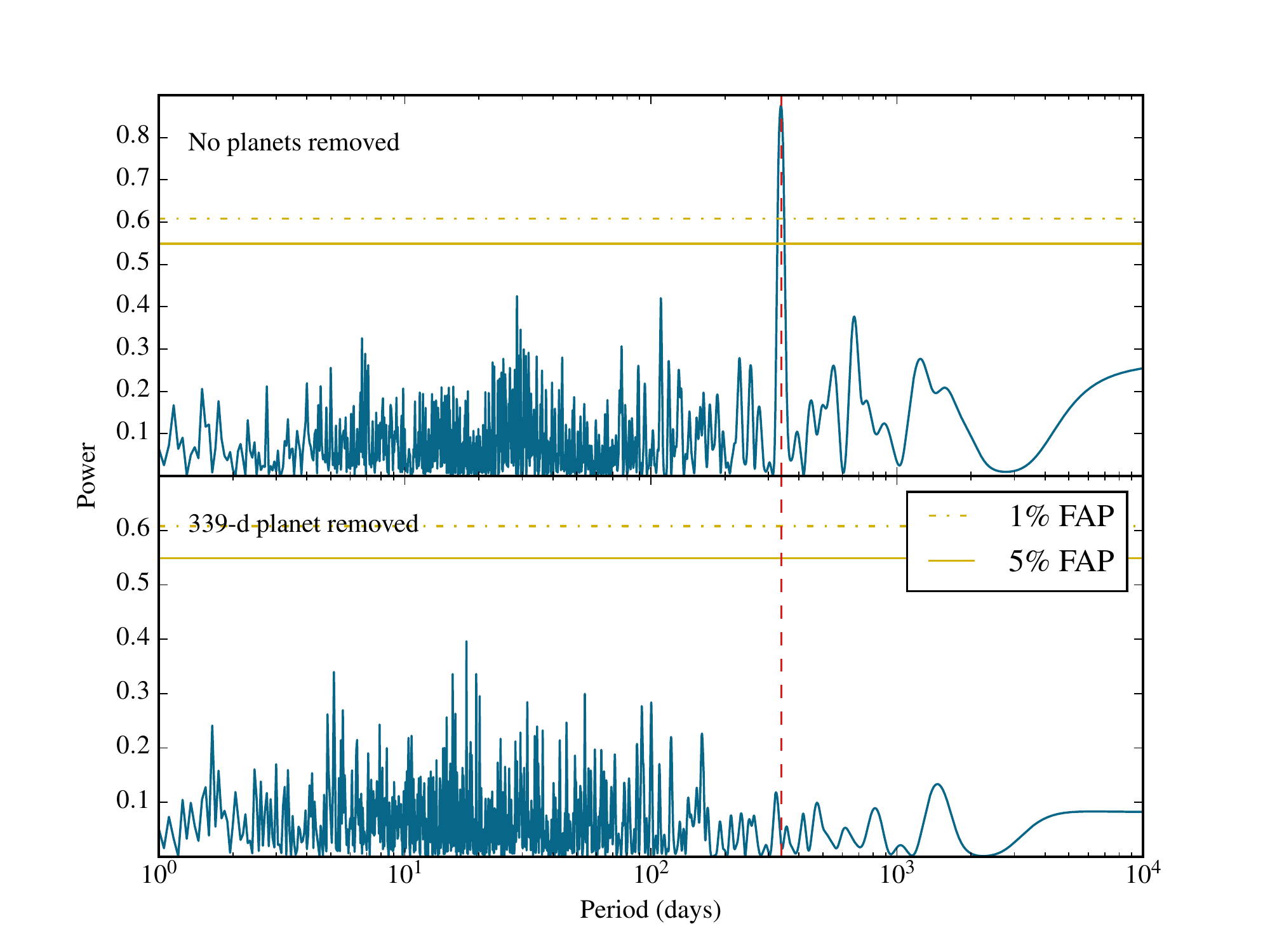}
	\caption{Periodograms for HD 148284. There is a strong signal at 339 days in the raw data. After fitting for a single Keplerian companion, the remaining peaks all have a high false alarm probability (FAP).\label{figK11P}}
\end{figure}

The generalized Lomb-Scargle periodogram finds a highly significant peak around 339 days in the radial velocity data of HD 148284. Indeed, after running the data through our Keplerian fitter, we recover a plausible companion HD 148284 B with a lower mass limit of $33.7 \pm 5.5$ \mjup\ and an orbital period of $P = 339.33 \pm 0.02$ days. Thanks to a well-sampled time baseline of over 11 years, we can constrain the eccentricity of this companion to $e = 0.3893 \pm 0.0009$. We infer an orbital semi-major axis of $a = 0.97 \pm 0.08$ AU. After adding 2.15 \ms of jitter in quadrature, the binary model has a \chisq\ of 1.31. The Keplerian fit can be seen in Figure \ref{figK11}. After subtracting the planetary signal from the data, no significant periodogram peaks remain - refer to Figure \ref{figK11P} for the periodograms. Finally, including a linear trend of $0.4 \pm 0.3$ m/s/day brings the $\chi^2$ down only marginally from 31.4 to 29.6. Plugging $\Delta\chi^2 = 1.8$ into a 7-dimensional $\chi^2$-distribution, we recover a p-value of 0.97 for the null hypothesis that this improvement is merely statistical. Thus, we stick with the zero-trend model.

\begin{figure}
	\plotone{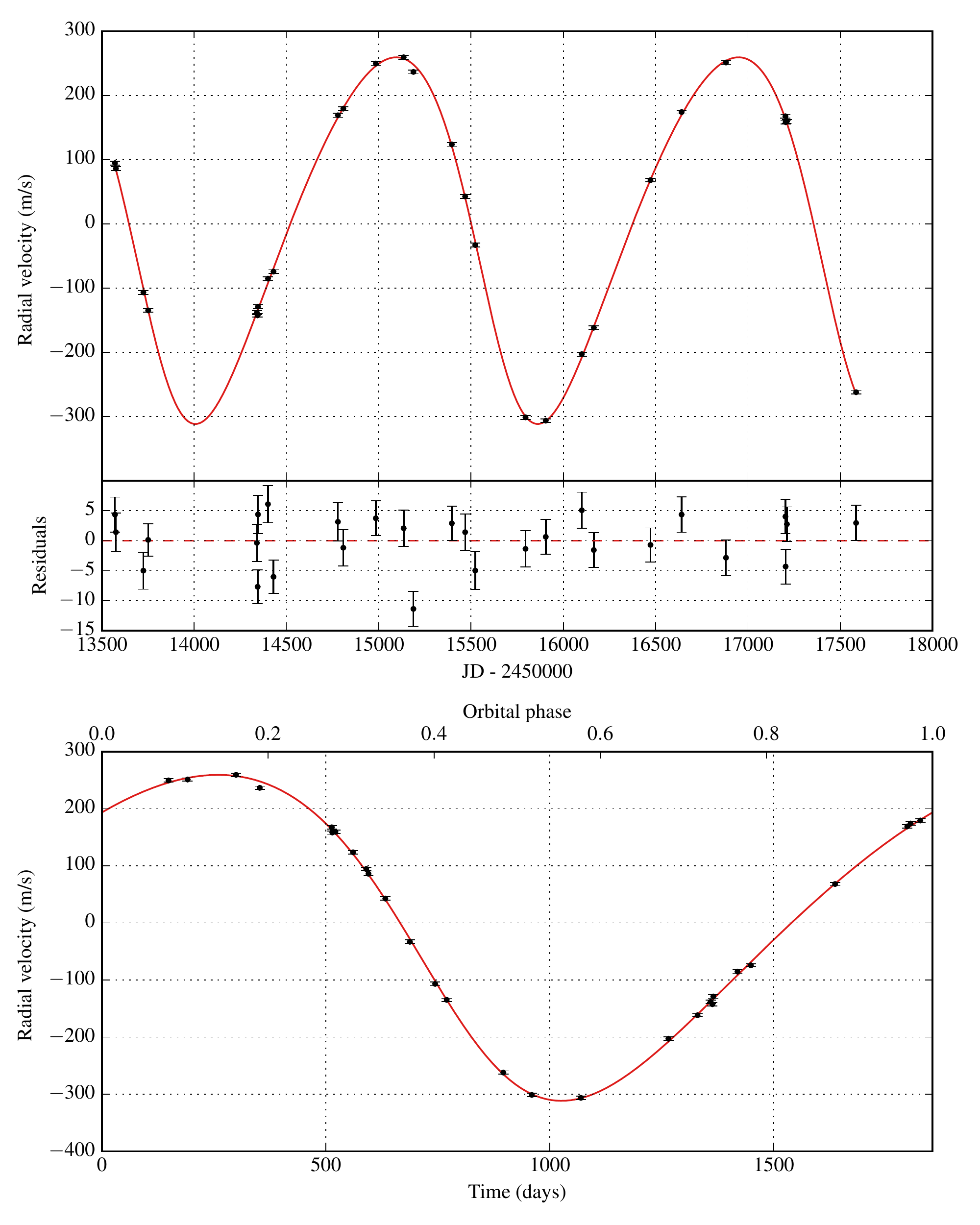}
	\caption{Keplerian model for HD 214823 B. The first plot shows the best Keplerian fit (in red) and the observed data points with an inset display of the residual velocities after fitting. The second plot shows a phase-folded version of the first plot.\label{figK22}}
\end{figure}

\begin{figure}
	\plotone{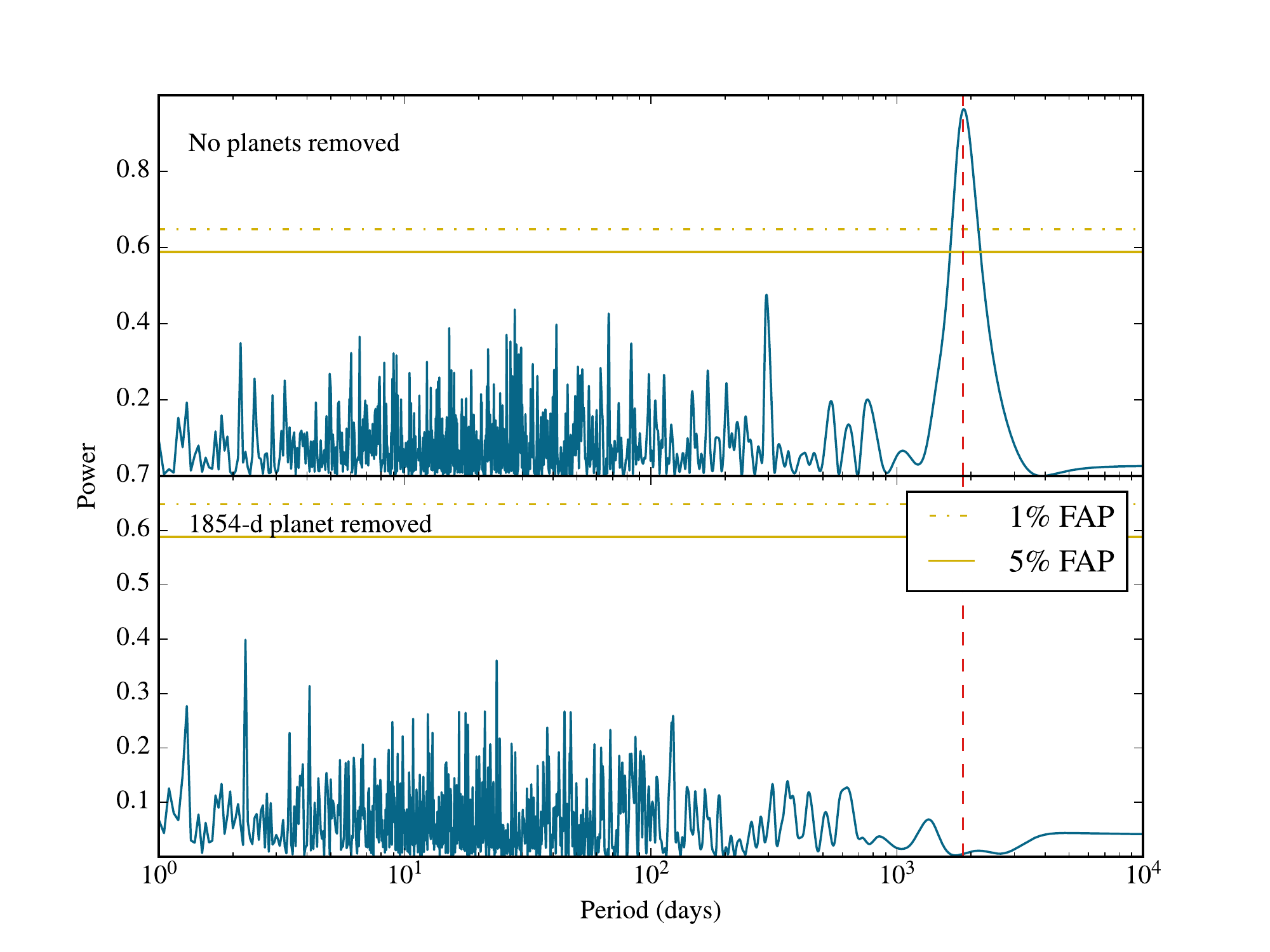}
	\caption{Periodograms for HD 214823. There is a strong signal at 1854 days in the raw data. After fitting for a single Keplerian companion, the remaining peaks all have a high false alarm probability (FAP).\label{figK22P}}
\end{figure}

Similarly, we find a 1854-day companion with a lower mass limit of $M \sin i = 20.3 \pm 0.1$ \mjup\ around HD 214823. It is on a slightly eccentric orbit ($e = 0.164 \pm 0.003$) at a distance of $a = 3.2 \pm 0.2$ from the star. After adding 2.33 \ms of jitter to the reported radial velocity errors, our fit has a \chisq\ of 2.61 with a residual RMS of 4.23 \ms. However, no additional significant peaks can be detected in the Lomb-Scargle periodogram after removing the 1854-day signal (see Figure \ref{figK22P}). Neither is there any compelling evidence for a linear trend. Thus, the relatively high value for \chisq\ suggests that we probably underestimated the radial velocity jitter caused by chromospheric activity. The Keplerian fit is displayed in Figure \ref{figK22}.

\begin{figure}
	\plotone{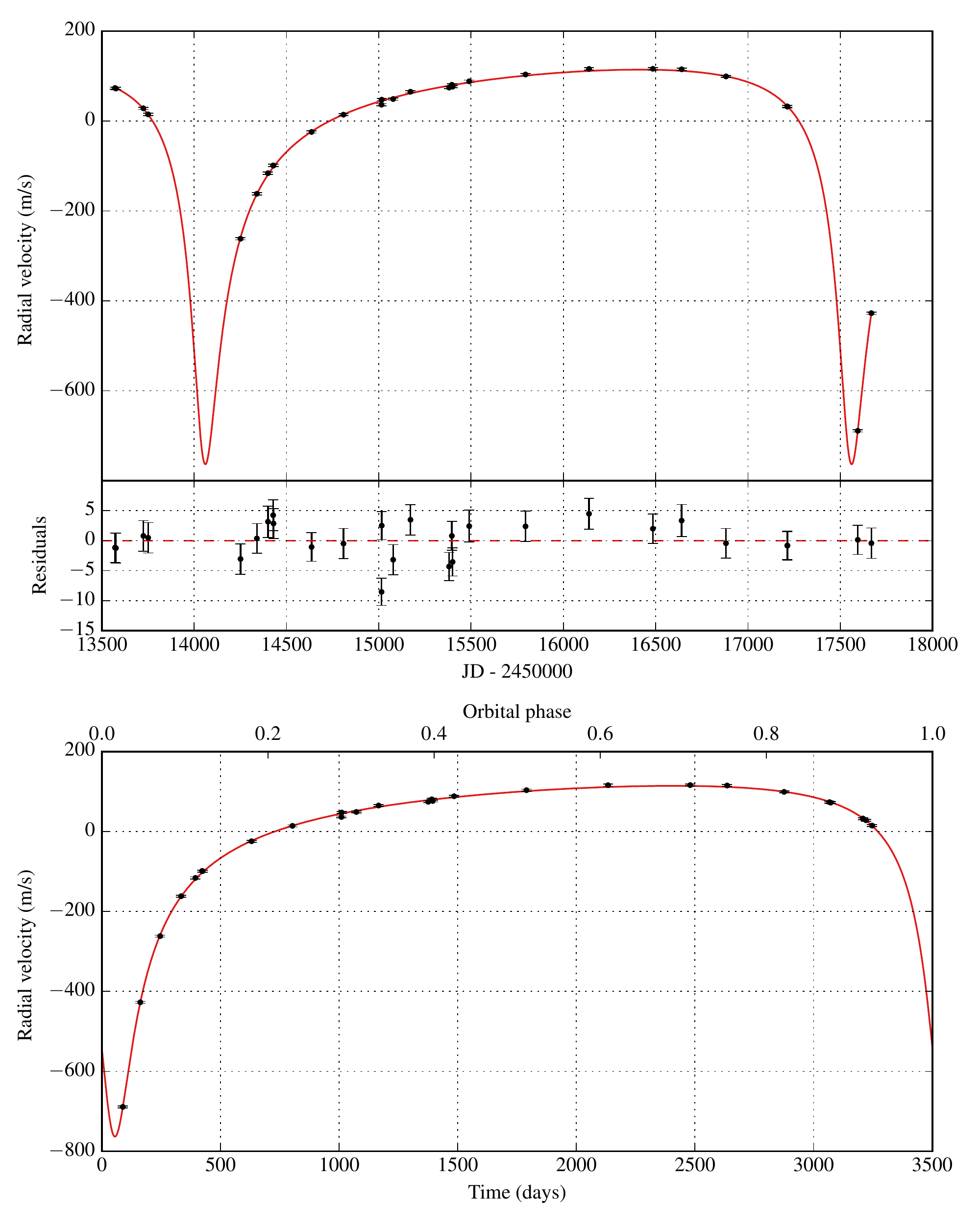}
	\caption{Keplerian model for HD 217850 B. The first plot shows the best Keplerian fit (in red) and the observed data points with an inset display of the residual velocities after fitting. The second plot shows a phase-folded version of the first plot.\label{figK26}}
\end{figure}

We also report the discovery of a companion around HD 217850 with a mass of $M \sin i = 21.6 \pm 1.4$ \mjup\ and an orbital period of $3501.3 \pm 2.1$ days. The orbit is highly eccentric at $e = 0.762 \pm 0.002$. Our best fit had a \chisq\ of 1.81 after including the chromospheric jitter, and the RMS of the radial velocity residuals after subtracting the Keplerian signal was 2.9 \ms. We demonstrate the Keplerian fit in Figure \ref{figK26}.

\begin{figure}
	\plotone{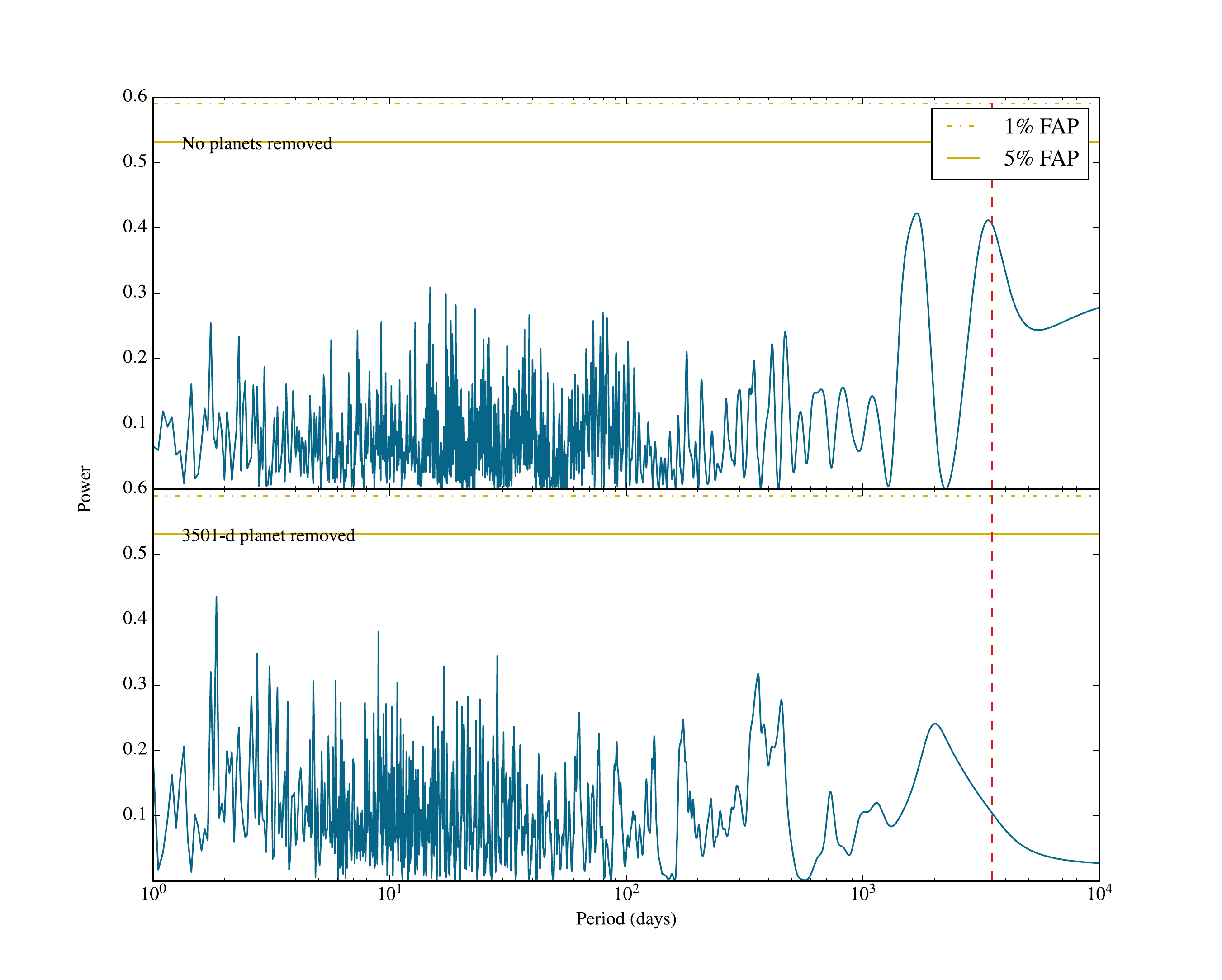}
	\caption{Periodograms for HD 217850. There is a significant signal at 3501 days in the raw data. After fitting for a single Keplerian companion, the remaining peaks all have a high false alarm probability (FAP).\label{figK26P}}
\end{figure}

\begin{figure}
	\plotone{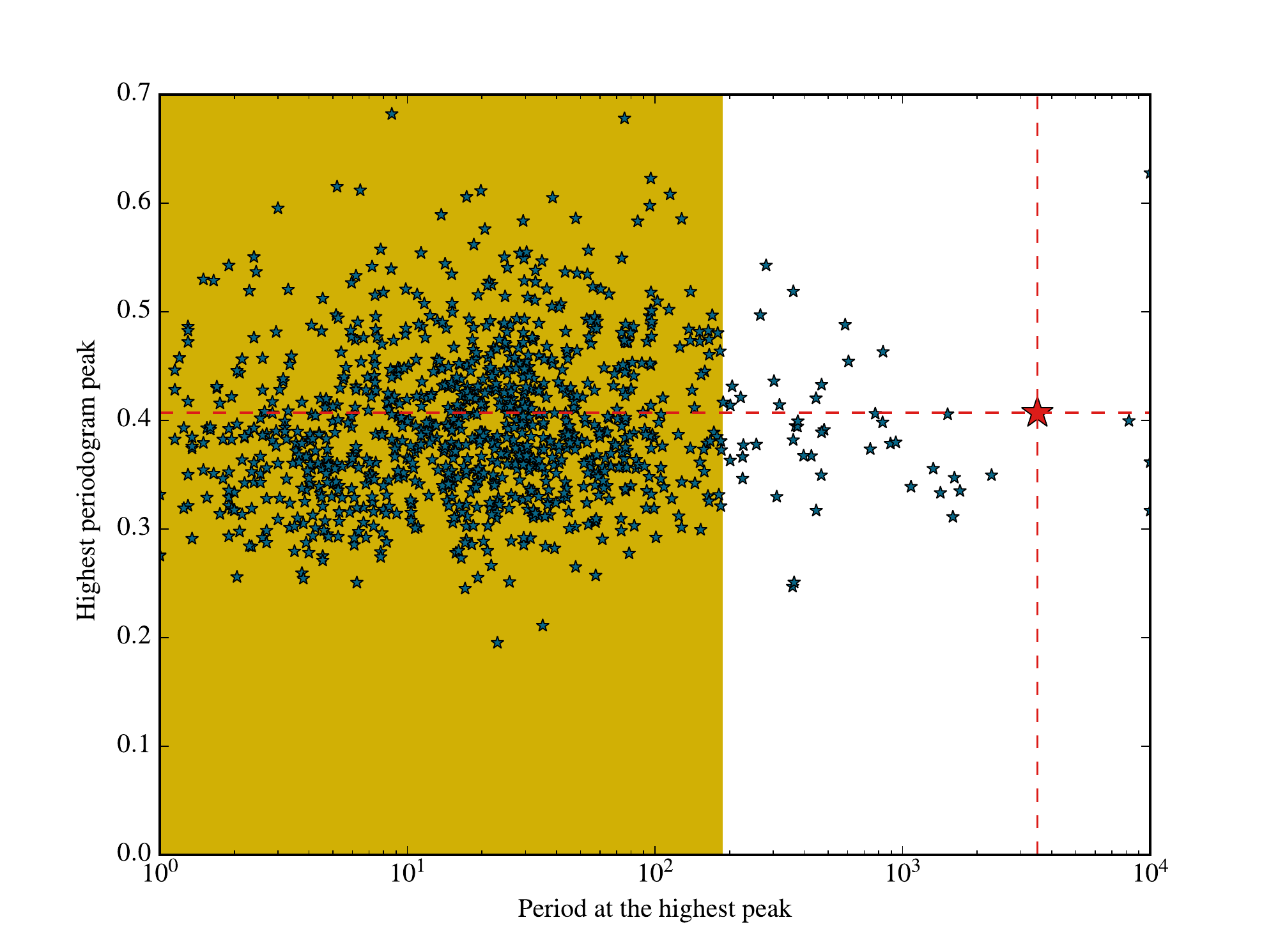}
	\caption{The distribution of the highest periodogram peaks and their corresponding periods for 1000 bootstrap samples of the radial velocities of HD 217850. The red star (and the dashed lines) indicates the location of the best Keplerian fit for HD 217850 B with an orbital period of 3501 days. The area shaded in gold covers 95\% of the samples which all have periods below 185 days.\label{figK26a}}
\end{figure}

Interestingly, when using the Lomb-Scargle periodogram to look for companions around HD 217850, we recover two peaks at 1688 and 3384 days (thus, differing by a factor of two), given in Figure \ref{figK26P}. Based on the bootstrap method described in Section \ref{sec:fitting}, both peaks have a false alarm probability (FAP) much greater than 5\%. However, closer investigation lends us confidence that the signal is not caused by a window function in the data. For each of the 1000 bootstrap samples, we have plotted the highest periodogram peak and the corresponding period in Figure \ref{figK26a}. We find that over 95\% of the highest peaks have a period below 200 days. Based on our time baseline of over 3000 days and Figure \ref{figK26}, the observations would have to include a streak of over 20 measurements over more than 10 orbital periods that exhibit monotonously increasing radial velocities. Given that the cadence of N2K follow-up of HD 217850 varied from days to years over the observation period, this coincidence would be extremely unlikely, thus disfavoring the window function interpretation. To this end, the periodogram fails to properly take into account the continuous increase in orbital phase between consecutive measurements. Finally, we are also encouraged by the fact that the model looks extremely visually convincing, reducing the RMS of the residuals from 178 \ms\ to 2.9 \ms. We also tried fitting for the 1688-day peak which would induce an additional unobserved radial velocity minimum into the data. We discovered a local probability maximum near $P = 1719$ days with a reduced chi-squared of 564 compared to the 1.81 of our 3501-period model. In this particular case, the Lomb-Scargle periodogram is clearly not a very successful goodness-of-fit proxy.

In order to eliminate the possibility that any of these signals is caused by magnetic activity that can occasionally mimic planetary signals, we also produced Lomb-Scargle periodograms for the S-values. For all three stars, the highest periodogram peaks had a false alarm probability (FAP) well above 5\%. Subsequently, explaining the above radial velocity signals in terms of magnetic activity is significantly less convincing than using Keplerian models. For HD 217850, we also calculated the Pearson correlation coefficient between the observed radial velocities and the S-values; we obtained $\rho = -0.29$ with a p-value of 0.11, derived by bootstrapping the velocities with replacement and calculating the fraction of times we obtained a higher correlation coefficient $\left|\rho\right|$. Thus, there is no significant correlation.

Table \ref{tblK11} summarizes the parameters of each of the three companions in greater detail.

\begin{deluxetable}{lccc}
	\tablewidth{0pt}
	\tablecaption{Orbital parameters for HD 148284 B, HD 214823 B, HD 217850 B\label{tblK11}}
	\tablehead{
		\colhead{Parameter} & \colhead{HD 148284 B} & \colhead{HD 214823 B} & \colhead{HD 217850 B}
	}
	\startdata
	P (days) & 339.331 $\pm$ 0.018 & 1854.4 $\pm$ 1.1 & 3501.3 $\pm$ 2.1\\
	K (\ms) & 1022.0 $\pm$ 1.2 & 285.47 $\pm$ 0.96 & 439.0 $\pm$ 5.8\\
	e & 0.38926 $\pm$ 0.00089 & 0.1641 $\pm$ 0.0026 & 0.7621 $\pm$ 0.0019\\
	$\omega$ (deg) & 35.56 $\pm$ 0.14 & 124.0 $\pm$ 1.2 & 165.95 $\pm$ 0.22\\
	T$_P$ (JD) & 13750.96 $\pm$ 0.21 & 13793.1 $\pm$ 5.9 & 14048.4 $\pm$ 3.9\\
	\msini (\mjup) & 33.7 $\pm$ 5.5 & 20.3 $\pm$ 2.6 & 21.6 $\pm$ 2.6\\
	a (AU) & 0.974 $\pm$ 0.079 & 3.23 $\pm$ 0.2 & 4.56 $\pm$ 0.24\\
	\chisq & 1.31 & 2.61 & 1.81\\
	RMS (\ms) & 2.97 & 4.23 & 2.88\\
	N$_{\rm obs}$ & 30 & 28 & 28\\
	\enddata
\end{deluxetable}

\subsection{A double-planet model for HD 148164}

HD 148164 is a young F8 star at a distance of 72 pc from Earth. Its estimated age is $2.4 \pm 0.8$ Gyr and mass $M = 1.21 \pm 0.24$ \msun. It has a high metallicity of $\rm [Fe/H] = 0.24$ and an activity index of $S_{\rm HK} = 0.15$, yielding a relatively low chromospheric jitter estimate of 2.37 \ms\ which we added to the radial velocity errors in quadrature. We also estimated the rotation period to be 18 days. Refer to Table \ref{t_starswithplanets} for a complete overview of the stellar characteristics.

HD 148164 was observed 43 times, with the observations distributed over a time baseline of more than 12 years. The radial velocity measurements have an RMS of 34 \ms, suggesting a strong likelihood for the presence of companions. We present a two-planet model for HD 148164 that reduces the RMS from 34 \ms\ to 5.6 \ms.

The outer planet HD 148164 b has a period comparable to our time baseline: $P = 5062 \pm 114$ days. Its estimated mass is $M \sin i = 5.16 \pm 0.82$ \mjup, its orbit has an eccentricity of $e = 0.125 \pm 0.017$ and a semi-major axis of $6.15 \pm 0.50$ AU. The inner planet HD 148164 c has an orbital period of $328.55 \pm 0.41$ days and a mass of $M \sin i = 1.23 \pm 0.25$ \mjup. It is traveling on an eccentric orbit with $e = 0.587 \pm 0.026$ and semi-major axis $a = 0.99 \pm 0.07$ AU.

\begin{figure}
	\epsscale{0.7}
	\plotone{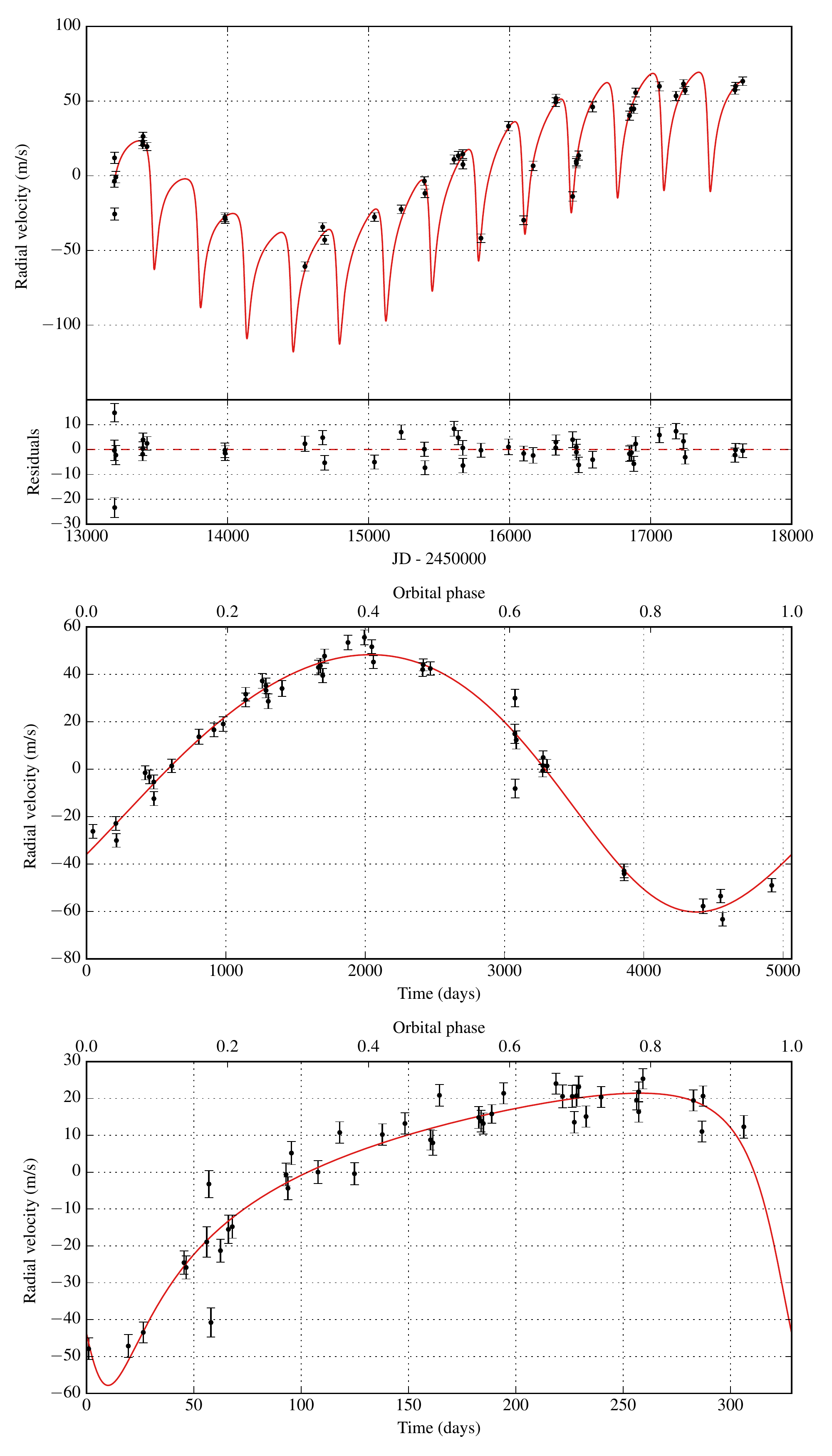}
	\caption{Keplerian model for HD 148164 b and c. The first plot shows the best Keplerian fit (in red) and the observed data points with an inset display of the residual velocities after fitting. The remaining plots show HD 148164 b and c, respectively, phase-folded by their orbital periods after removing the other planets.\label{figK54}}
\end{figure}

\begin{figure}
	\epsscale{1.0}
	\plotone{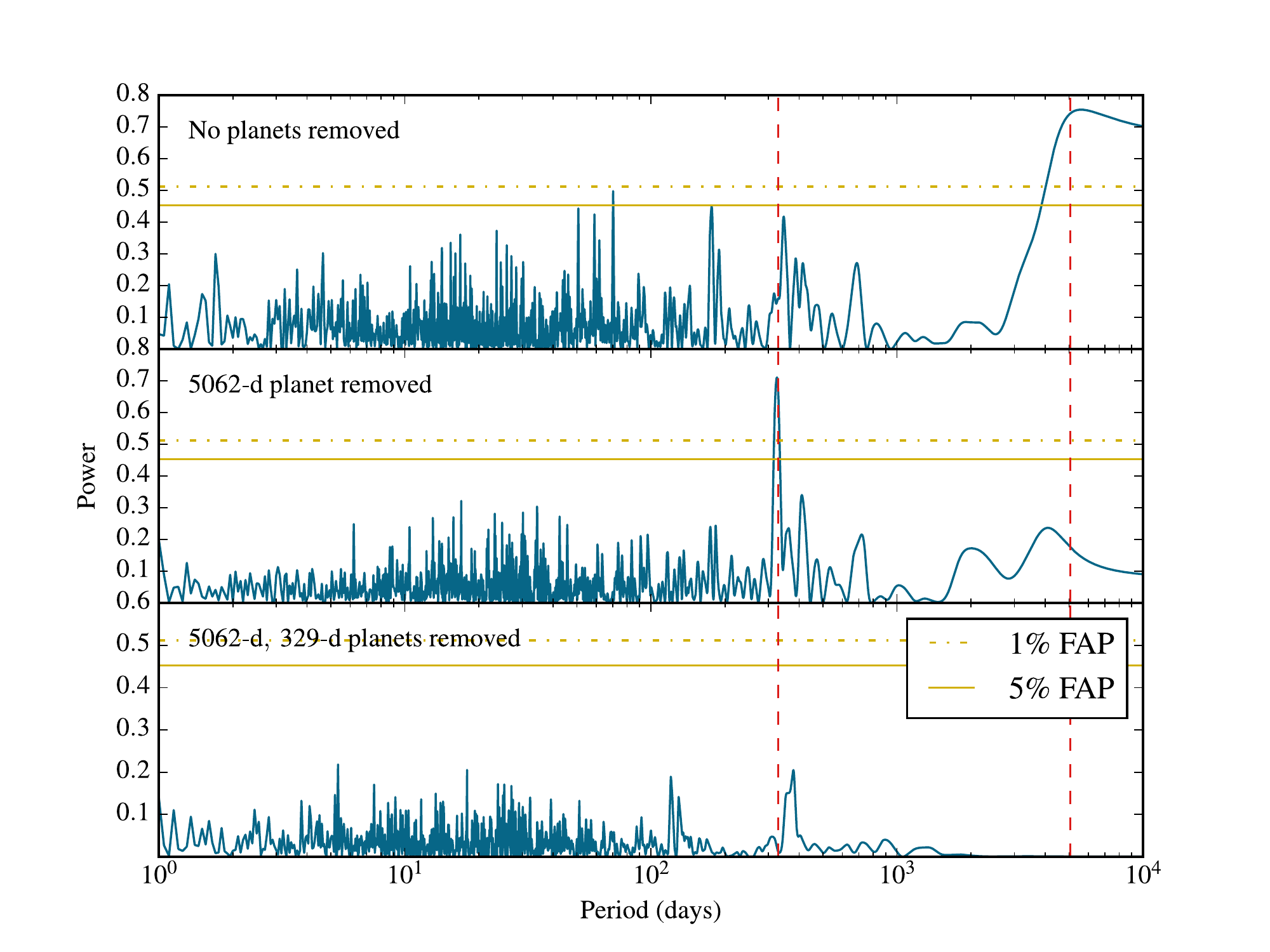}
	\caption{Periodograms for HD 148164. There are two significant peaks in the data. After fitting for the two Keplerian signals near these periods, the remaining peaks all have a high false alarm probability (FAP).\label{figK54P}}
\end{figure}

A complete set of orbital parameters is given in Table \ref{tblK54} while the Keplerian fit is displayed in Figure \ref{figK54}. We also kept track of the Lomb-Scargle periodograms while iteratively removing planets - both companions had false alarm probabilities (FAPs) less than 1\%, see Figure \ref{figK54P}. After removing both planets, no significant peaks remain. However, our current model still has a \chisq\ of 3.73 which suggests that the chromospheric jitter for HD 148164 might be underestimated.

\begin{figure}
	\plotone{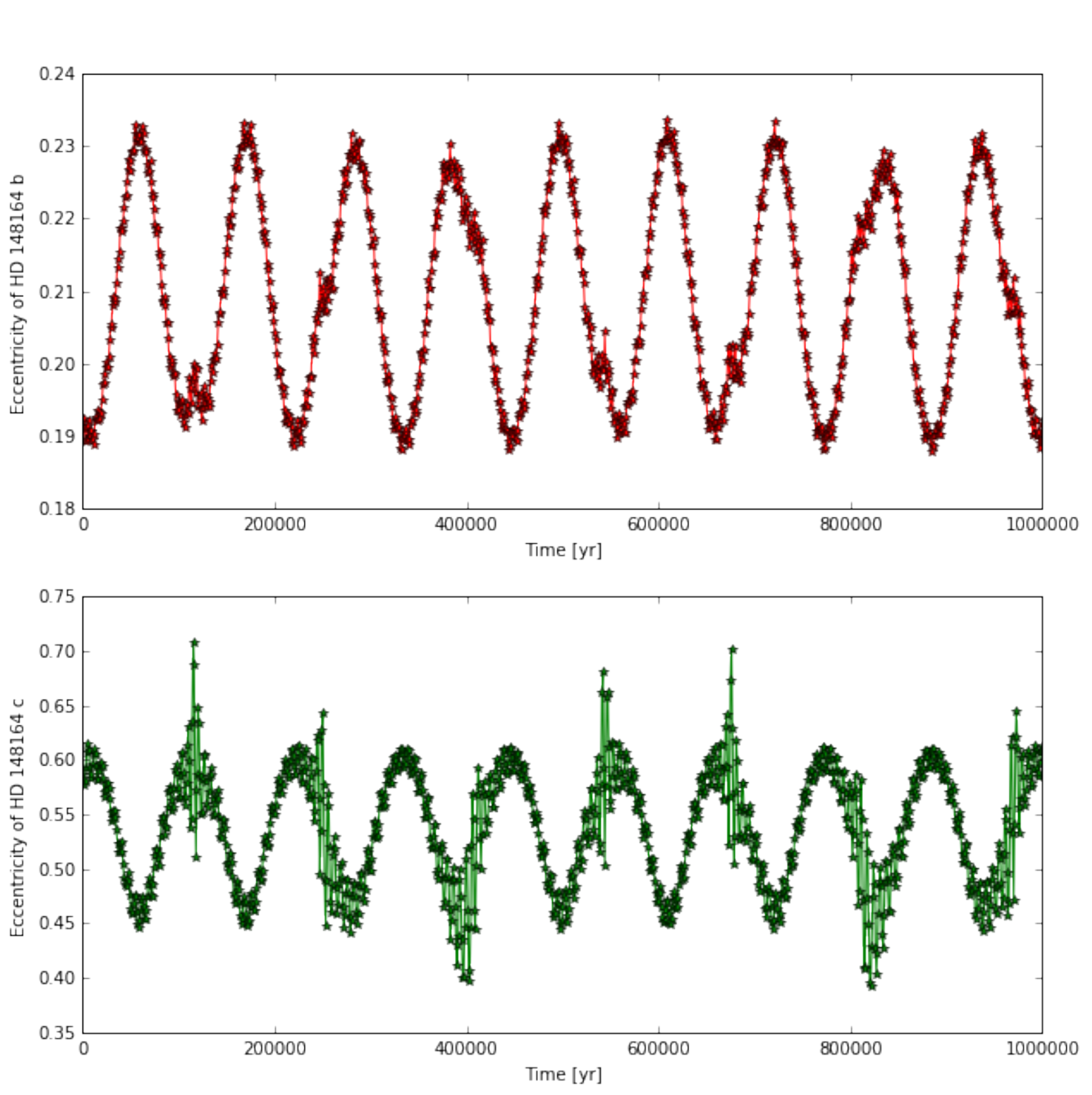}
	\caption{The eccentricities of the two planets HD 148164 a and b over 1,000,000 years of the dynamical simulation exhibiting synchronized libration. The simulation was carried out using the REBOUND integrator.\label{figK54sim}}
\end{figure}

Finally, we tested the dynamical stability of the system using the WHFast integrator in the REBOUND package (\citet{Rein2012}, \citet{Rein2015}). We concluded that the orbits are stable over at least a million years, although the orbital eccentricities exhibit a degree of synchronized libration as can be seen in Figure \ref{figK54sim}.

\begin{deluxetable}{lcc}
	\tablewidth{0pt}
	\tablecaption{Orbital parameters for HD 148164 b, HD 148164 c\label{tblK54}}
	\tablehead{
		\colhead{Parameter} & \colhead{HD 148164 b} & \colhead{HD 148164 c}
	}
	\startdata
	P (days) & 328.55 $\pm$ 0.41 & 5062 $\pm$ 114\\
	K (\ms) & 39.6 $\pm$ 1.7 & 54.28 $\pm$ 0.89\\
	e & 0.587 $\pm$ 0.026 & 0.125 $\pm$ 0.017\\
	$\omega$ (deg) & 141.5 $\pm$ 2.7 & 152 $\pm$ 11\\
	T$_P$ (JD) & 13472.7 $\pm$ 4.7 & 14193 $\pm$ 155\\
	\msini (\mjup) & 1.23 $\pm$ 0.25 & 5.16 $\pm$ 0.82\\
	a (AU) & 0.993 $\pm$ 0.066 & 6.15 $\pm$ 0.5\\
	\chisq & 3.73 & \\
	RMS (\ms) & 5.62 & \\
	N$_{\rm obs}$ & 43 & \\
	\enddata
\end{deluxetable}

\subsection{A cold Jupiter around HD 211810}

HD 211810 is classified as a G5 star. The estimate of its distance from the Sun is 61 pc, and it has a derived mass of $M = 1.03 \pm 0.02$ \msun. We use the chromospheric jitter estimate of 2.19 \ms\ given by \citet{IsaacsonFischer2010}, and we calculate a stellar rotation period estimate of 37 days. Please refer to Table \ref{t_starswithplanets} for additional information about the star.

We report the discovery of a new companion HD 211810 b in an eccentric orbit with $e = 0.68 \pm 0.14$. The planet completes its orbit once every $1558 \pm 22$ days. We estimate that the companion has a mass of $M \sin i = 0.66 \pm 0.42$ \mjup\ and its orbital semi-major axis is $a = 2.63 \pm 0.03$ AU.

\begin{figure}
	\plotone{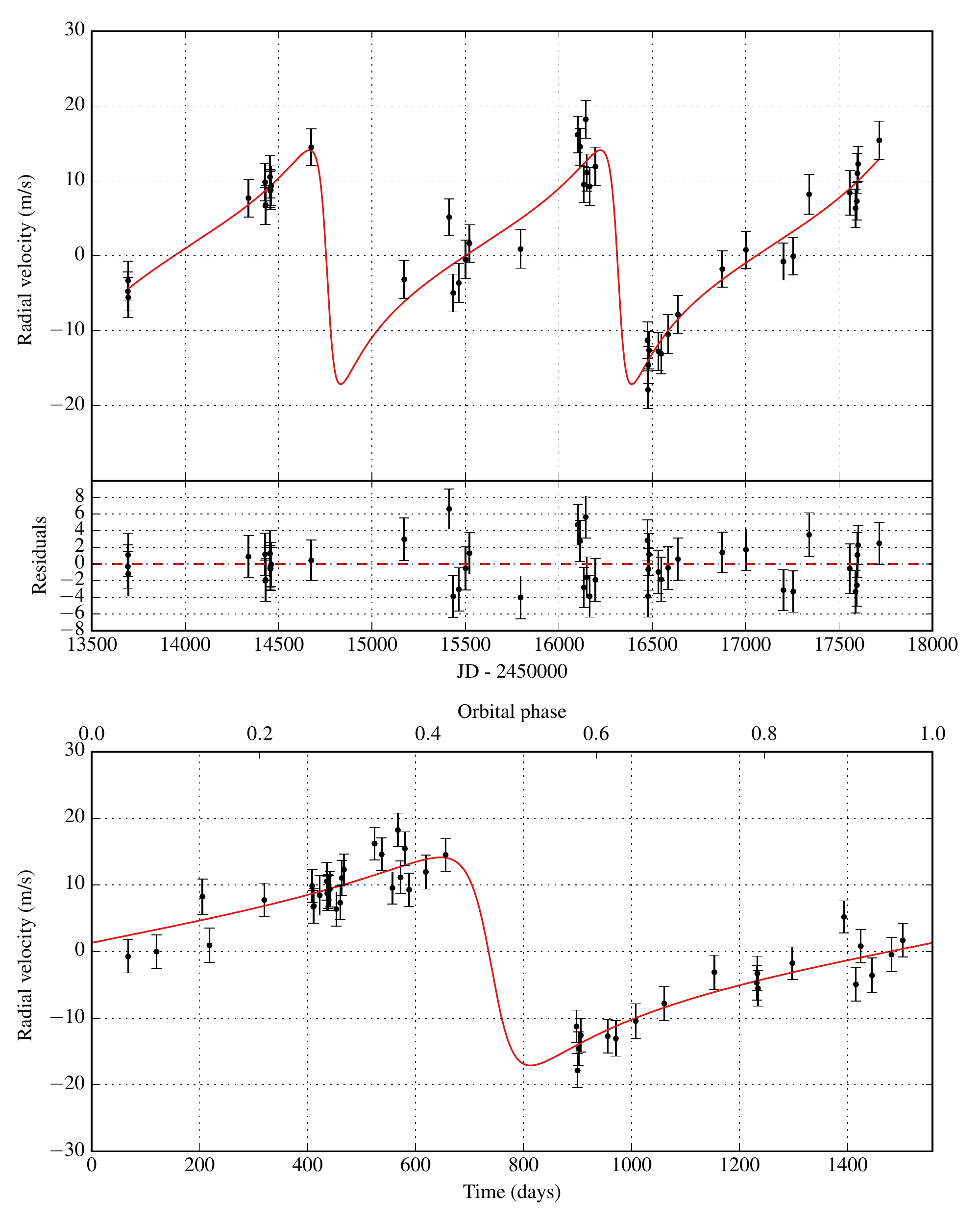}
	\caption{Keplerian model for HD 211810 b. The first plot shows the best Keplerian fit (in red) and the observed data points with an inset display of the residual velocities after fitting. The second plot shows a phase-folded version of the first plot.\label{figK94}}
\end{figure}

\begin{figure}
	\plotone{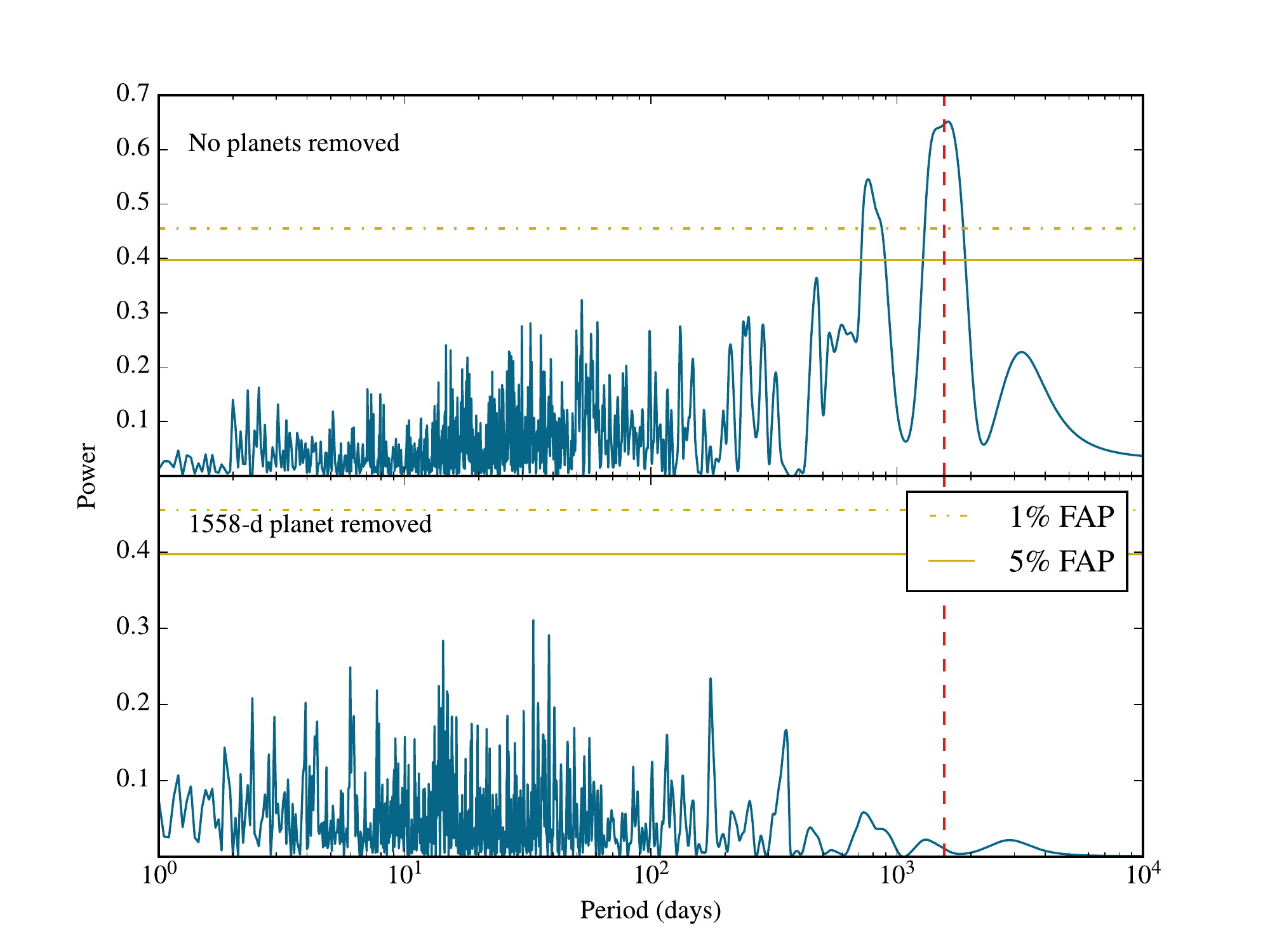}
	\caption{Periodograms for HD 211810. The red dashed line displays the location of the fitted planet. There is a significant signal near 1558 days in the raw data. After fitting for a single Keplerian companion, the remaining peaks all have a high false alarm probability (FAP).\label{figK94P}}
\end{figure}

A comprehensive list of parameters for HD 211810 is given in Table \ref{tblK94}. Refer to Figure \ref{figK94} for the Keplerian fit. We detected no significant linear trend in our data, and the \chisq-parameter of our best fit is 1.20, based on 46 observations by Keck/HIRES. We also tested the hypothesis that our data might be caused by a window function by generating a Lomb-Scargle periodogram of the data, given in Figure \ref{figK94P}. There is a significant peak close to the fitted period with a FAP much less than 1\%, and no significant peaks remain after subtracting the single-planet model, suggesting no additional companions.

\begin{figure}
	\plotone{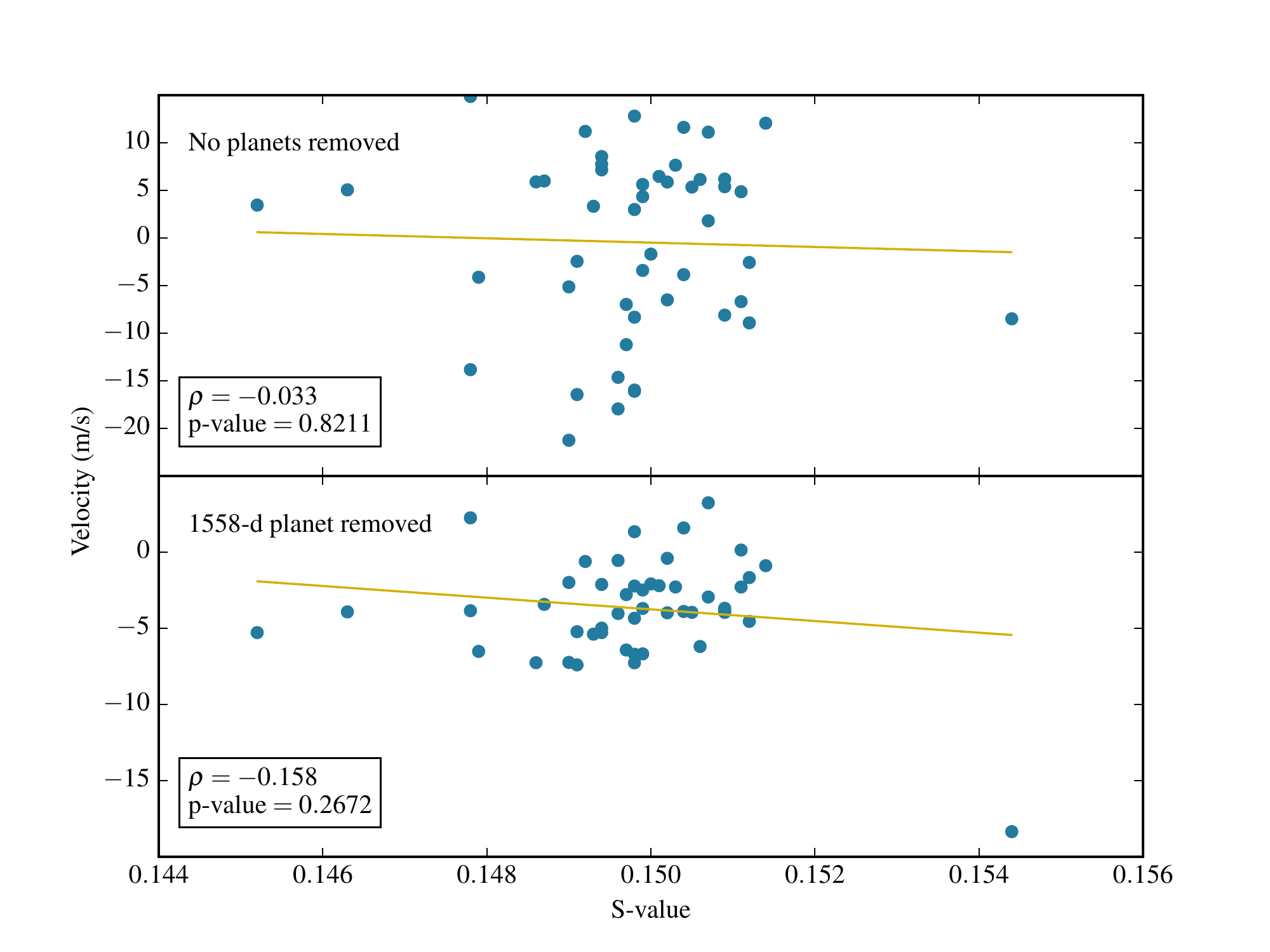}
	\caption{There is no significant correlation between the radial velocities and the S-values of HD 211810. The correlation increases somewhat after removing the 1558-day planet as the effects of stellar activity become more pronounced. Furthermore, removing the outlier at $S_{\rm HK} = 0.1544$ from the bottom plot further increases the significance of the correlation by giving a p-value of 0.107.\label{figK94SC}}
\end{figure}

Finally, we investigated if the RV signal could have been caused by stellar activity masquerading as a Keplerian companion by looking at the correlation between the radial velocity measurements and their corresponding S-values. As can be seen in Figure \ref{figK94SC}, there is no significant correlation in the raw data. However, the correlation becomes much more significant for the residual velocities (p-value: 0.14) once we remove the Keplerian signal, suggesting that the RV residuals might be at least partly generated by stellar activity.

\begin{deluxetable}{lcccc}
	\tablewidth{0pt}
	\tablecaption{Orbital parameters for HD 55696 b, HD 98736 b, HD 203473 b, HD 211810 b\label{tblK94}}
	\tablehead{
		\colhead{Parameter} & \colhead{HD 55696 b} & \colhead{HD 98736 b} & \colhead{HD 203473 b} & \colhead{HD 211810 b}
	}
	\startdata
	P (days) & 1827 $\pm$ 10 & 968.8 $\pm$ 2.2 & 1552.9 $\pm$ 3.4 & 1558 $\pm$ 22\\
	K (\ms) & 76.7 $\pm$ 3.9 & 52 $\pm$ 12 & 133.6 $\pm$ 2.4 & 15.6 $\pm$ 7.2\\
	e & 0.705 $\pm$ 0.022 & 0.226 $\pm$ 0.064 & 0.289 $\pm$ 0.01 & 0.68 $\pm$ 0.14\\
	$\omega$ (deg) & 137.0 $\pm$ 2.4 & 162 $\pm$ 22 & 18.0 $\pm$ 1.1 & 98 $\pm$ 14\\
	T$_P$ (JD) & 13648 $\pm$ 26 & 13541 $\pm$ 67 & 13333.6 $\pm$ 9.1 & 14763 $\pm$ 86\\
	\msini (\mjup) & 3.87 $\pm$ 0.72 & 2.33 $\pm$ 0.78 & 7.8 $\pm$ 1.1 & 0.67 $\pm$ 0.44\\
	a (AU) & 3.18 $\pm$ 0.18 & 1.864 $\pm$ 0.091 & 2.73 $\pm$ 0.17 & 2.656 $\pm$ 0.043\\
	Trend (m/s/yr) & 1.34 $\pm$ 0.34 & -3.08 $\pm$ 0.17 & -24.9 $\pm$ 2.0 & \\
	Curvature (m/s/yr$^2$) &  &  & 3.91 $\pm$ 0.3 & \\
	\chisq & 4.4 & 2.05 & 2.0 & 1.2\\
	RMS (\ms) & 7.18 & 3.08 & 3.34 & 2.55\\
	N$_{\rm obs}$ & 28 & 20 & 36 & 46\\
	\enddata
\end{deluxetable}

\subsection{A cold eccentric companion orbiting HD 55696}

HD 55696 is classified as a G0V star by the Hipparcos Catalogue. It is located at a distance of 72 pc from the Sun and it has a very high metallicity of $\rm [Fe/H] = 0.37$. The star has a derived mass of $M = 1.29 \pm 0.20$ \msun\ and a radius of $R = 1.52 \pm 0.07$ \rsun. It is relatively young compared to the rest of our sample with an age of $2.6 \pm 0.5$ Gyr. Based on the chromospheric activity indices \lrphk $= -4.97$ and \shk $= 0.16$, we estimate the star's rotation period as 19 days and the chromospheric RV jitter as 2.53 \ms. HD 55696 was observed on 28 short visits between 2004 and 2016 by the N2K Consortium.

\begin{figure}
	\plotone{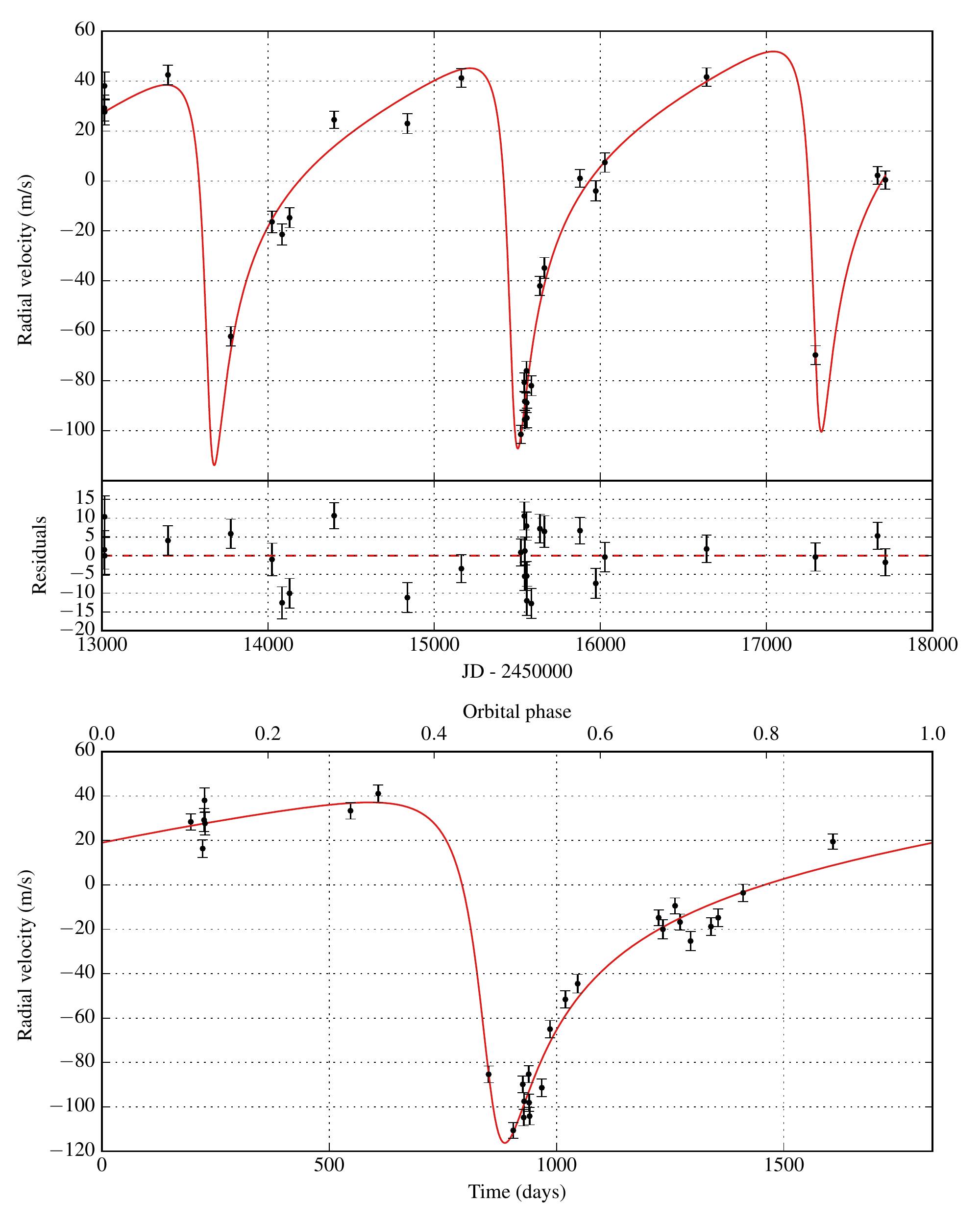}
	\caption{Keplerian model for HD 55696 b. The first plot shows the best Keplerian fit (in red) and the observed data points with an inset display of the residual velocities after fitting. The second plot shows a phase-folded version of the first plot. The model includes a linear trend.\label{figK343}}
\end{figure}

\begin{figure}
	\plotone{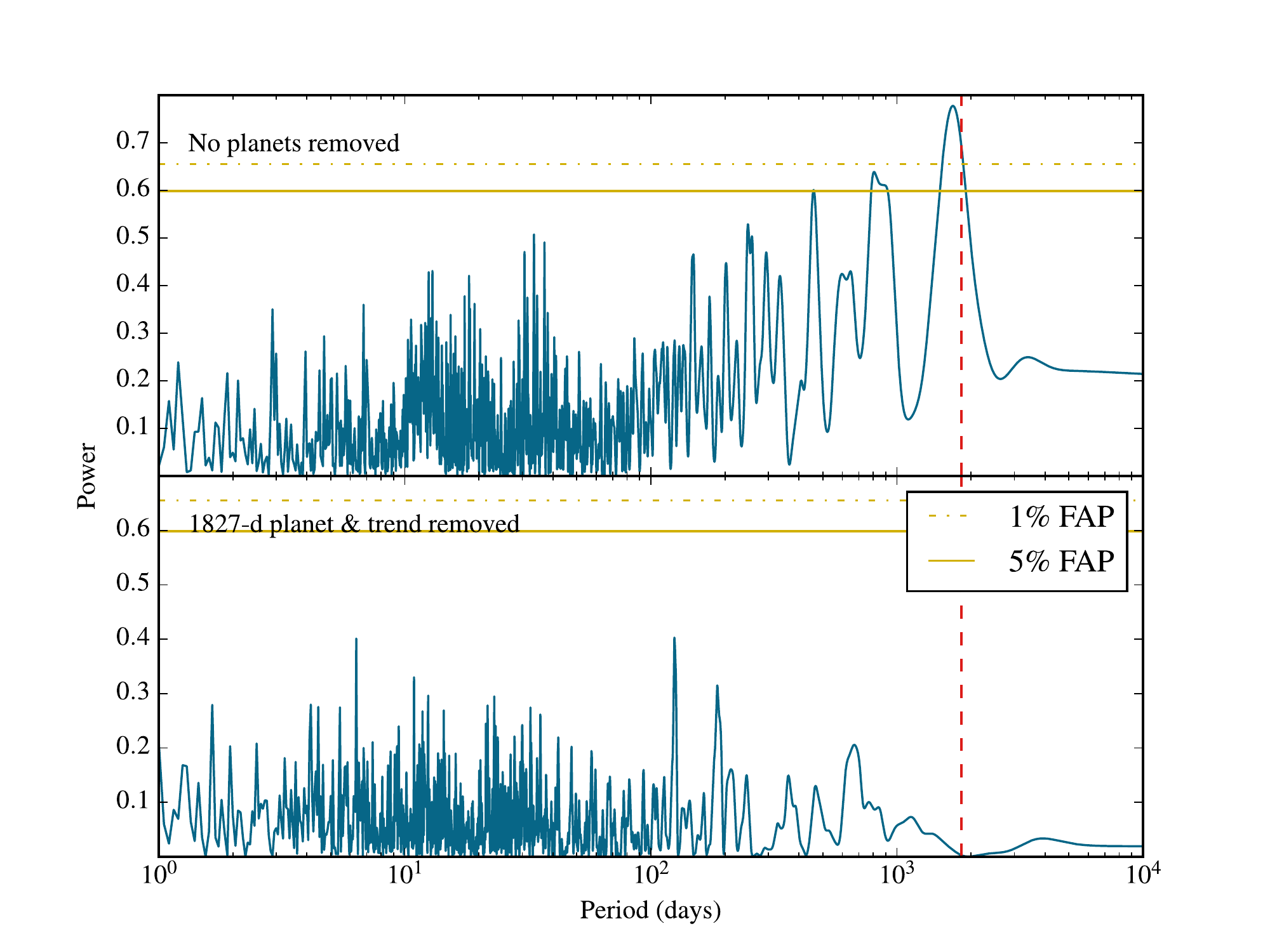}
	\caption{Periodograms for HD 55696. The red dashed line displays the location of the fitted planet. There is a significant signal near 1827 days in the raw data. After fitting for a single Keplerian companion, the remaining peaks all have a high false alarm probability (FAP).\label{figK343P}}
\end{figure}

Our analysis finds evidence for a plausible companion HD 55696 b with an orbital period of $P = 1827 \pm 10$ days. The mass of the companion is at least $M \sin i = 3.87 \pm 0.72$ \mjup, the orbital eccentricity is $e = 0.705 \pm 0.022$, and the orbital semi-major axis is $a = 3.18 \pm 0.18$ AU. We also recover a somewhat significant linear trend of 1.3 m/s/yr. Our Keplerian model, plotted in Figure \ref{figK343}, has a \chisq\ of 4.40. The high value of \chisq\ might be due to additional stellar activity. A comprehensive overview of the fitted parameters is given in \ref{tblK94}. We also generated a Lomb-Scargle periodogram for HD 55696, which can be seen in Figure \ref{figK343P}. After removing the 1827-day signal, the rest of the periodogram peaks all have a high false alarm probability, thereby prompting us to abandon the search for any additional companions.

\subsection{A cold planet in a binary system HD 98736}

SIMBAD classifies HD 98736 as a binary star with a bright, G-type component A and a much fainter, M-dwarf component B. According to Yale-Yonsei isochrones, the main component has a mass of $M = 0.92 \pm 0.13$ \msun\ and a radius of $R = 0.93 \pm 0.04$ \rsun. The age of the system is estimated as $7.4 \pm 3.0$ Gyr and its distance from the Sun is 31 pc. \citet{Vogt2015} measured a separation of 5.1'' between the binary components. Combining this with the distance estimate, we find that the components are currently at least 158 AU apart. This system was observed on 20 different nights by the N2K Consortium. From this point forward, we will associate the radial velocity measurements of HD 98736 with the G-type component of the system and model the influence of the M-dwarf on the main component's radial velocities as a linear trend. We introduced 2.2 \ms\ of chromospheric jitter into our models based on Equation \ref{eq:jit4b} and the measured \shk $= 0.19$; this jitter was added to all radial velocities in quadrature.

\begin{figure}
	\plotone{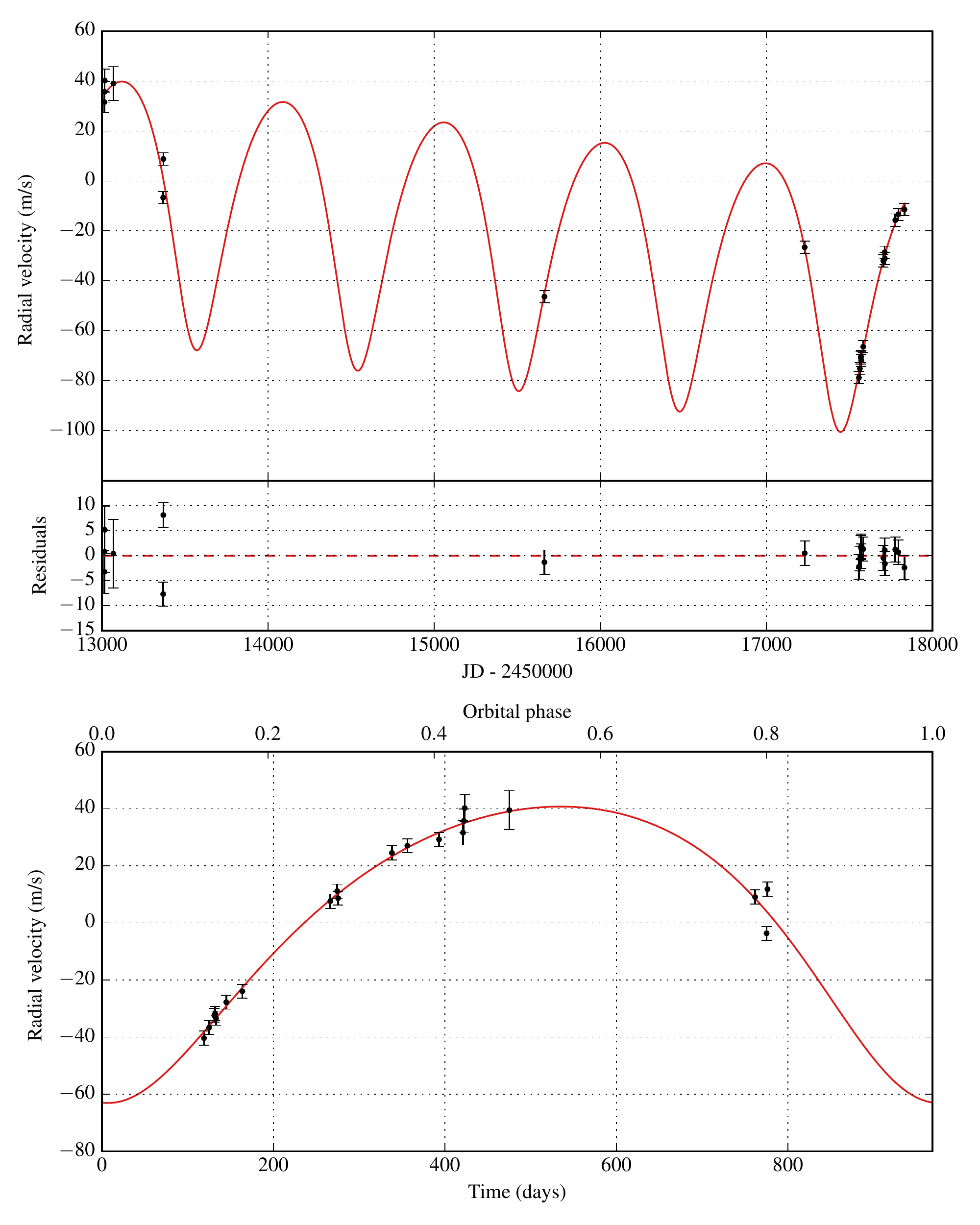}
	\caption{Keplerian model for HD 98736 b. The first plot shows the best Keplerian fit (in red) and the observed data points with an inset display of the residual velocities after fitting. The second plot shows a phase-folded version of the first plot. The model includes a linear trend.\label{figK399}}
\end{figure}

\begin{figure}
	\plotone{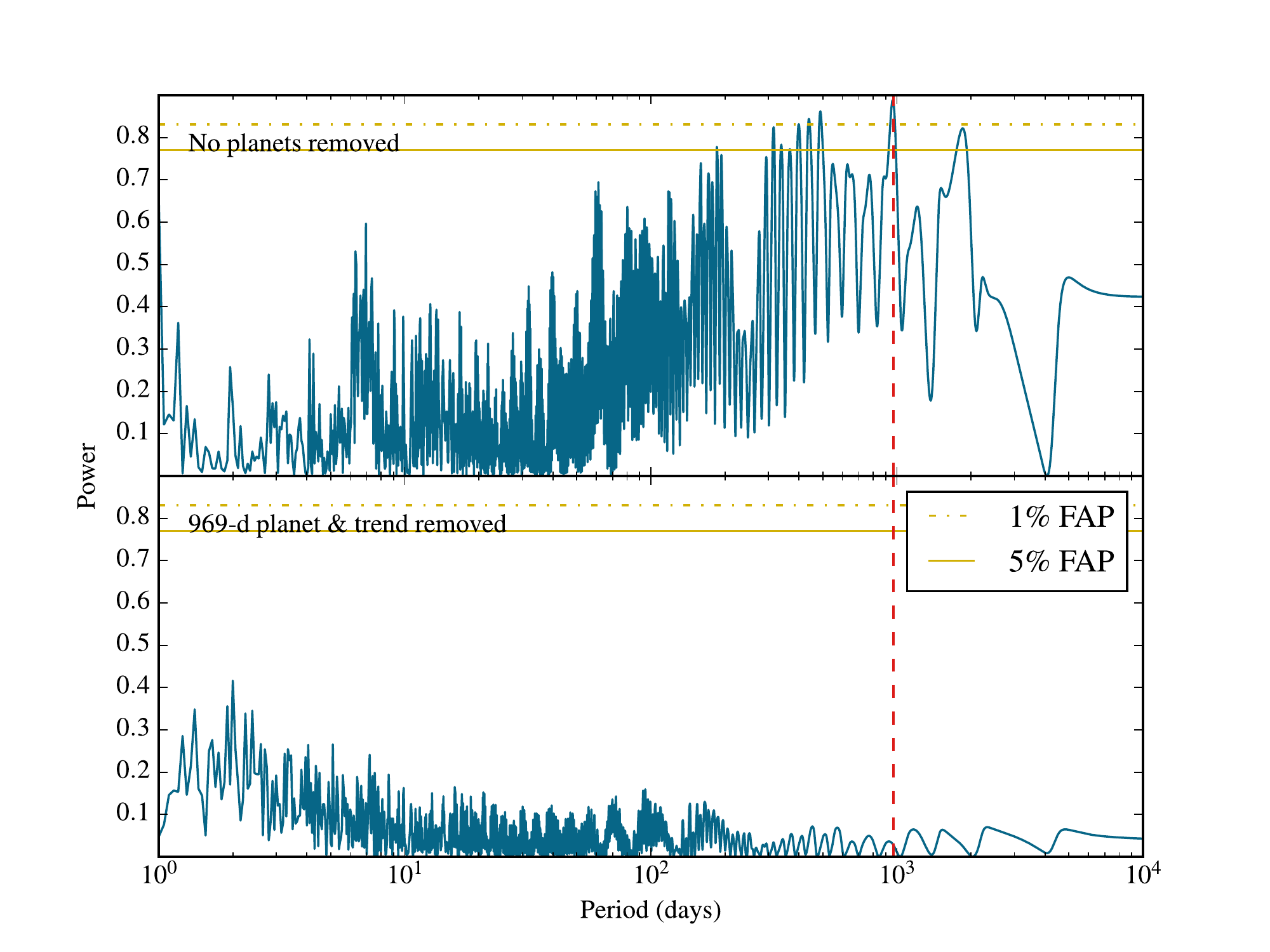}
	\caption{Periodograms for HD 98736. The red dashed line displays the location of the fitted planet. There is a significant signal near 969 days in the raw data. After fitting for a single Keplerian companion, the remaining peaks all have a high false alarm probability (FAP).\label{figK399P}}
\end{figure}

Our analysis yields convincing evidence for an additional, planetary companion with a mass of $M \sin i = 2.33 \pm 0.76$ \mjup. We estimate the orbital period of the planet as $P = 968.8 \pm 2.2$ days, the eccentricity as $e = 0.226 \pm 0.061$, and the semi-major axis as $a = 1.86 \pm 0.09$ AU, well below the separation between the two stellar components. We also recover a significant linear trend of $-3.1 \pm 0.2$ m/s/yr which might be due to the M-dwarf companion. In order to examine the possible influence of a window function in our radial velocity data, we generated a Lomb-Scargle periodogram of the velocities, seen in Figure \ref{figK399P}. We find that the putative orbital period of 969 days is well below the 1\% false alarm probability (FAP) threshold, and there are no additional significant signals after removing the 969-day orbit. We also looked for long-term magnetic activity via Lomb-Scargle analysis of the \shk values, but found no significant periodogram peaks. Our final model has a \chisq\ of 2.05 and can be seen in Figure \ref{figK399}. A complete overview of the modeled parameters is given in Table \ref{tblK94}.

\subsection{A massive cold planet orbiting HD 203473 with a residual RV acceleration}

HD 203473 is a G5-type star with a mass of $1.12 \pm 0.21$ \msun. It has a derived age $5.2 \pm 1.0 \,Gyr$ and its location is approximately 64 pc from the Sun. It is relatively inactive: the Ca II core line emission indices $\log R_{\rm HK}^{`} = -5.05$ and $S_{\rm HK} = 0.15$ yield a radial velocity jitter estimate of 2.3 \ms\ as well as a rotation period of 28 days. HD 203473 was observed 36 times during the N2K project over a total of 12.5 years, one of the longest time baselines in our sample.

\begin{figure}
	\plotone{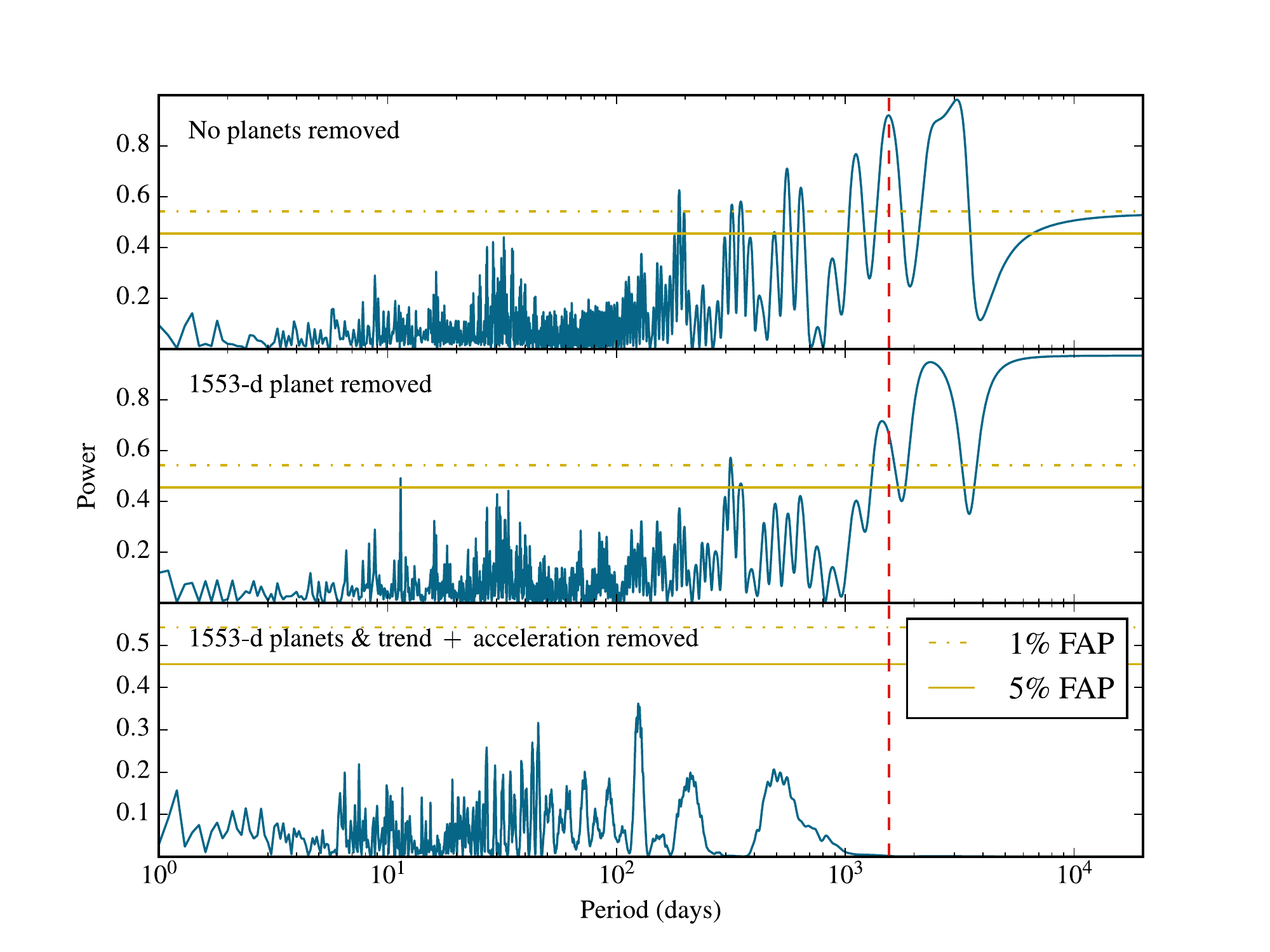}
	\caption{Periodograms for HD 203473. The red dashed line displays the location of the fitted planet. After fitting for a single Keplerian companion on a 1553-day orbit, a significant long-period trend emerges. Consequently, we included a linear trend as well as an acceleration term in our RV models. After subtracting the final model from the raw data, the periodogram peaks of the RV residuals all have a high false alarm probability (FAP).\label{figK37P}}
\end{figure}

The Lomb-Scargle periodogram of HD 203473 (given in Figure \ref{figK37P}) shows significant peaks for a period of $\sim$3000 days as well as for its higher-frequency harmonics (1500 days, 1000 days, 750 days, etc). However, we find that the 1500-day peak produces the best Keplerian fit. Thus, we present a model for a substellar companion HD 203473 b with an orbital period of $P = 1553 \pm 3$ days, an eccentricity of $e = 0.289 \pm 0.010$, a semi-major axis of $a = 2.73 \pm 0.17$ AU, and a companion mass of $m \sin i = 7.8 \pm 1.2$ \mjup.

\begin{figure}
	\plotone{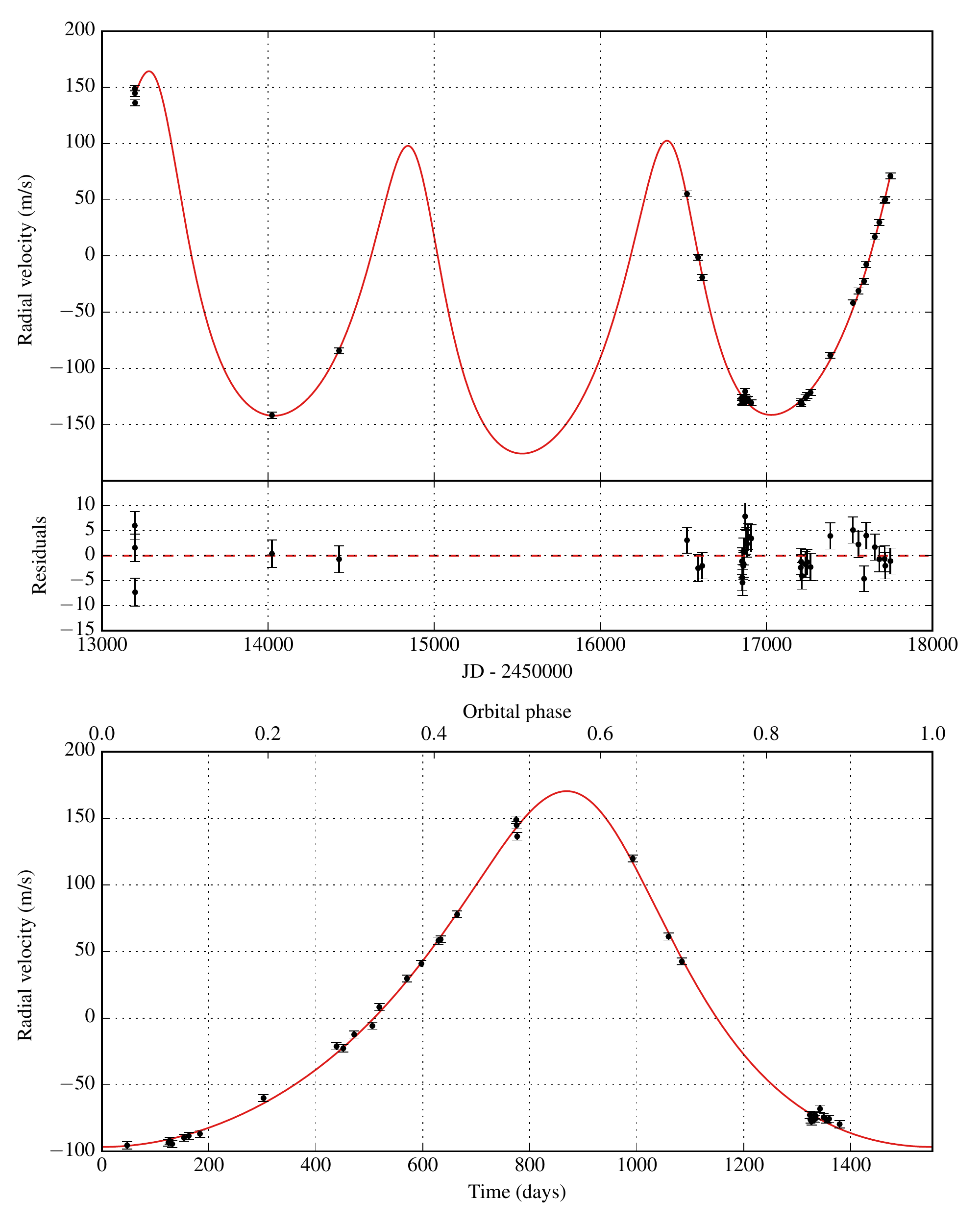}
	\caption{Keplerian model for HD 203473 b. The first plot shows the best Keplerian fit (in red) and the observed data points with an inset display of the residual velocities after fitting. The second plot shows a phase-folded version of the first plot. The model includes a linear trend as well as a curvature term.\label{figK37}}
\end{figure}

In addition, we recover a significant linear trend of $-25 \pm 2$ m/s/yr as well as an acceleration of $3.9 \pm 0.3$ m/s/yr$^2$ in the radial velocity residuals. These are likely due to an additional previously unseen long-period companion whose orbit can not be resolved at the current time baseline (fitting for an additional long-period planet failed to converge unless the eccentricity of the planet was held constant at $e = 0$). Subsequently, we decided to adopt a single-planet model for the time being. A comprehensive overview of the modelled parameters is given in Table \ref{tblK94} and the model can be seen in Figure \ref{figK37}. The \chisq-value of the final model was 2.00. We also looked for signs of magnetic activity by generating a Lomb-Scargle periodogram of the measured \shk values, but failed to find any significant peaks in the periodogram.

\section{Results: Interesting systems \label{sec:results2}}

\subsection{Binary star systems HD 3404, HD 24505, HD 98630 \& HD 103459}

As explained in Section \ref{sec:pl}, we can only give lower limits on the companion mass $m$ and the orbital semi-major axis $a$ for the following companions heavier than 83.8 \mjup\ (equivalent to 0.08 \msun).

\subsubsection{HD 3404}

HD 3404 is a G2 subgiant at a distance of 71 pc from the Sun. It has an estimated mass of $M = 1.17 \pm 0.22$ \msun, a radius of $R = 2.05 \pm 0.15$ \rsun, and an age of $6.7 \pm 1.3$ Gyr. Based on the chromospheric activity indices, we derive a stellar rotation period of 44 days and a radial velocity jitter estimate of 4.27 \ms.

\begin{figure}
	\plotone{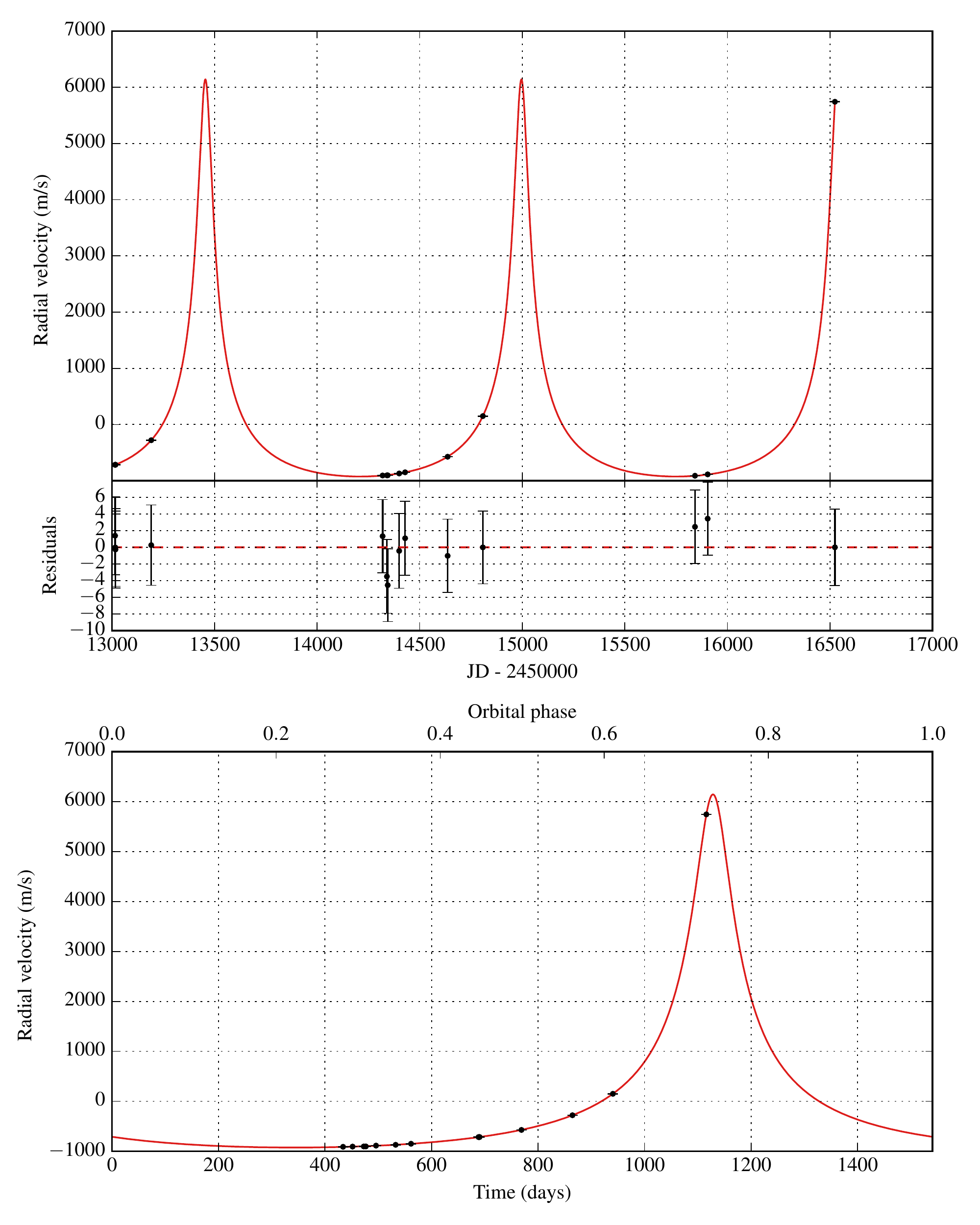}
	\caption{Keplerian model for HD 3404 B. The first plot shows the best Keplerian fit (in red) and the observed data points with an inset display of the residual velocities after fitting. The second plot shows a phase-folded version of the first plot.\label{figK322}}
\end{figure}

HD 3404 has been observed during 14 visits by the N2K Consortium. We find compelling evidence for an eccentric binary companion HD 3404 B with a mass of $M \geq 145 \pm 28$ \mjup\ (or $0.14 \pm 0.03$ \msun). The orbit has a period of $P = 1540.8 \pm 1.9$ days, a semi-major axis of $a \geq 2.86 \pm 0.16$ AU, and an eccentricity of $e = 0.738 \pm 0.004$. A more detailed overview of the model can be seen in Table \ref{tblK242}. The model has a \chisq\ of 0.37. An illustration of the radial velocity signal is given in Figure \ref{figK322}.

\subsubsection{HD 24505}

HD 24505 A is classified as a G5 giant in the Hipparcos catalog. However, its height of $\Delta M_V = 1.50$ magnitudes above the main sequence implies that it might be a subgiant instead. It has a derived mass $M = 1.11 \pm 0.20$ \msun, radius $R = 1.64 \pm 0.11$ \rsun, and age $7.3 \pm 0.8$ Gyr. The star has a rotation period of 40 days, a radial velocity jitter estimate of 4.23 \ms, and it is located 70 pc from the Solar System.

\begin{figure}
	\plotone{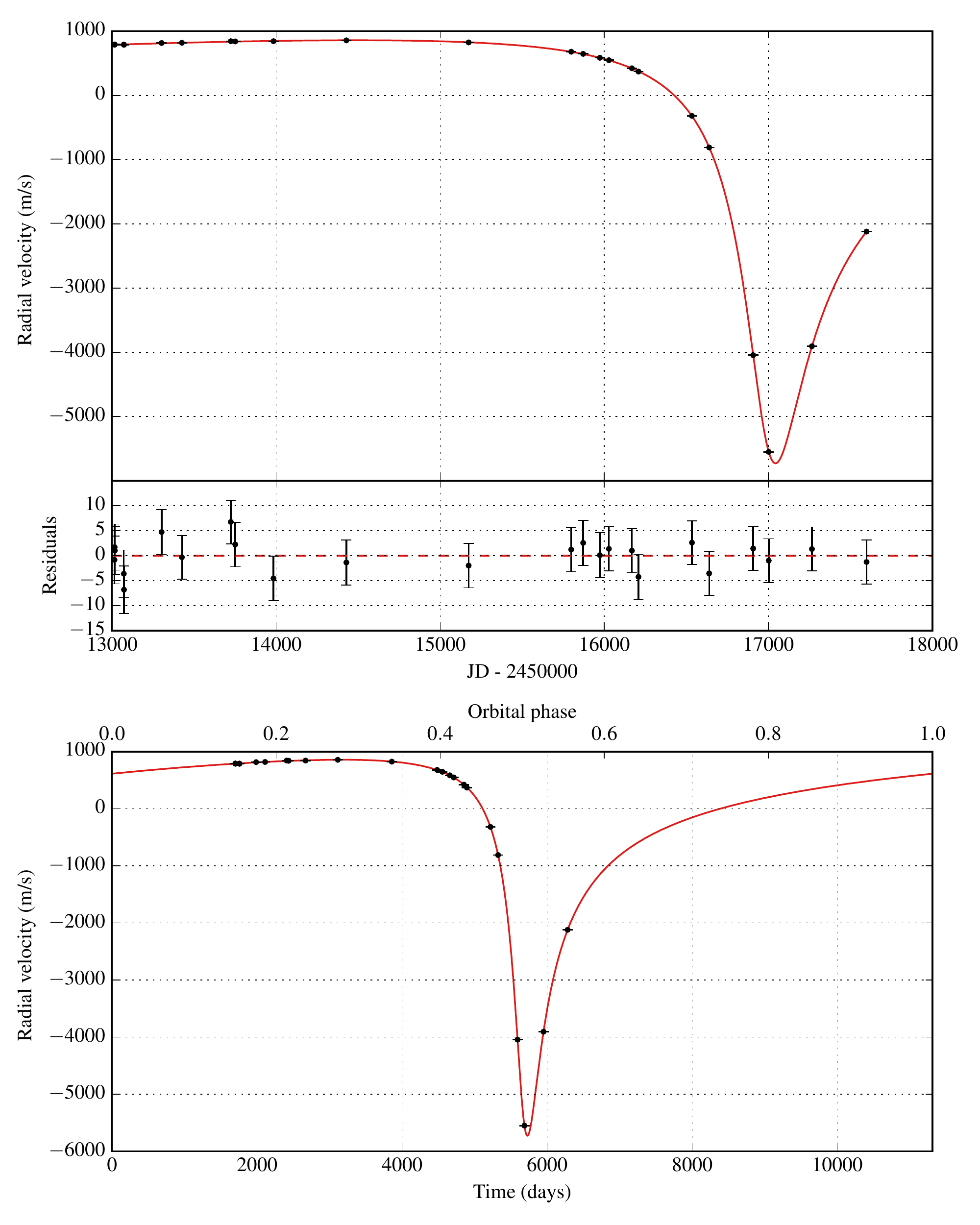}
	\caption{Keplerian model for HD 24505 B. The first plot shows the best Keplerian fit (in red) and the observed data points with an inset display of the residual velocities after fitting. The second plot shows a phase-folded version of the first plot.\label{figK305}}
\end{figure}

Based on 24 Keck/HIRES radial velocity observations, we recover a stellar companion HD 24505 B with a mass of $m \geq 222 \pm 30$ \mjup\ (equivalent to $0.21 \pm 0.03$ \msun). Its orbit is eccentric with $e = 0.798 \pm 0.001$ and has a semi-major axis of at least $a \geq 10.8 \pm 0.6$ AU. The orbital period works out to $P = 11315 \pm 92$ days, making this one of the longest-period companions discovered by the N2K project. The orbital model has a \chisq\ of 0.59 and can be seen in Figure \ref{figK305}. A more detailed overview of orbital parameters is given in Table \ref{tblK242}.

\subsubsection{HD 98630}

HD 98630 is a G0 star at a distance of 102 pc from Earth. It has a derived mass of $M = 1.44 \pm 0.07$ \msun\ and a radius of $1.94 \pm 0.19$ \rsun. Thus, HD 98360 might be a subgiant (we estimate its height above the main sequence $\Delta M_V = 1.42$ mag). It has an estimated age of 9.9 Gyr, a rotation period of 12 days, and 2.50 \ms\ of RV jitter. The star has been observed by Keck/HIRES on 21 separate nights.

\begin{figure}
	\plotone{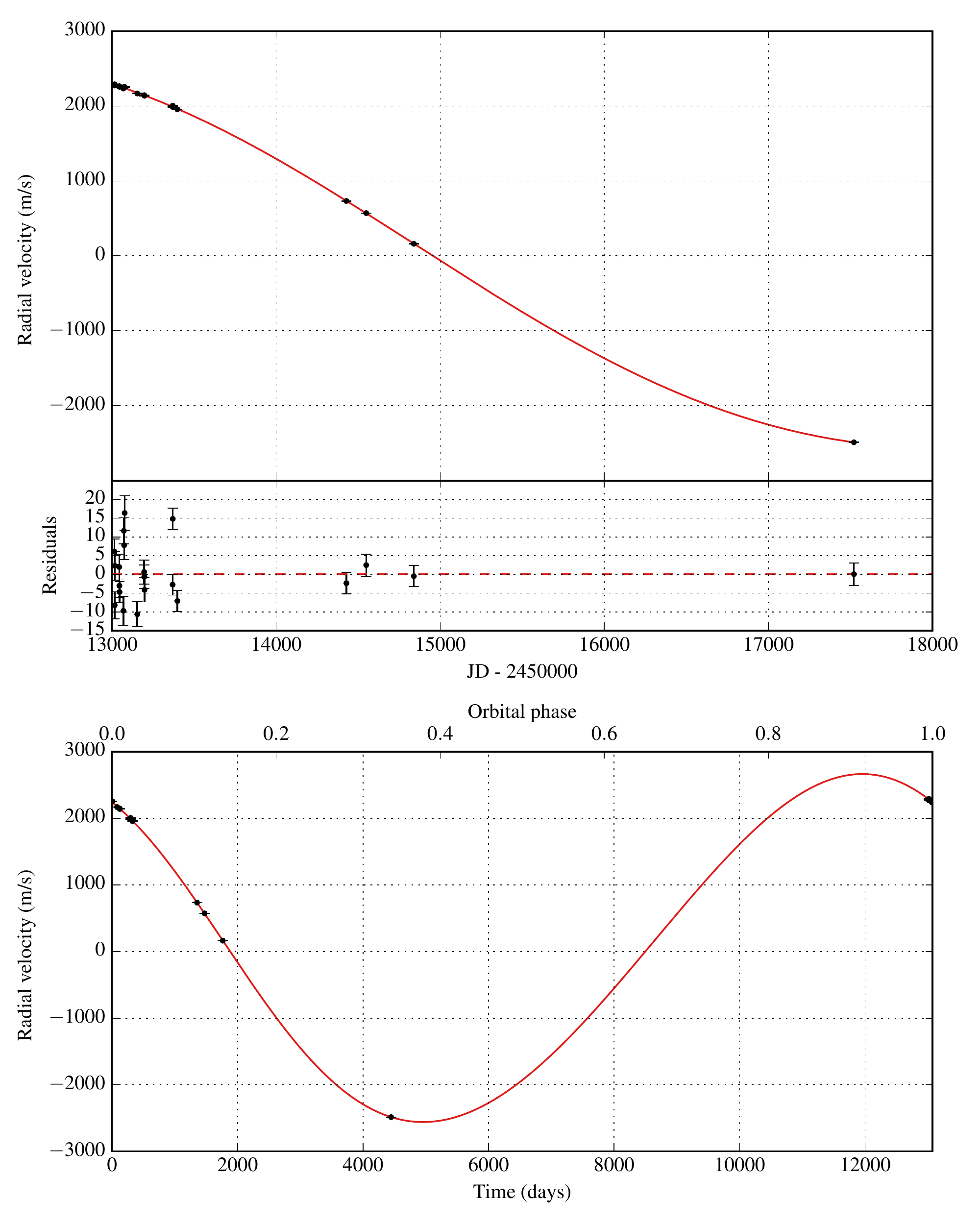}
	\caption{Keplerian model for HD 98630 B. The first plot shows the best Keplerian fit (in red) and the observed data points with an inset display of the residual velocities after fitting. The second plot shows a phase-folded version of the first plot.\label{figK398}}
\end{figure}

We recover two models for a binary companion HD 98630 B. The preferred one gives an orbital period of $P = 13074 \pm 982$ days (or 35.8 years) and a mass of $m \geq 280 \pm 11$ \mjup\ for the companion (equivalent to $0.27 \pm 0.01$ \msun). The orbit has a semi-major axis of $a \geq 11.8 \pm 0.6$ AU and an eccentricity of $e = 0.059 \pm 0.032$. The model has a \chisq\ of 6.25 and is displayed in Figure \ref{figK398}. More information about the fitted parameters is given in Table \ref{tblK242}.

\begin{figure}
	\plotone{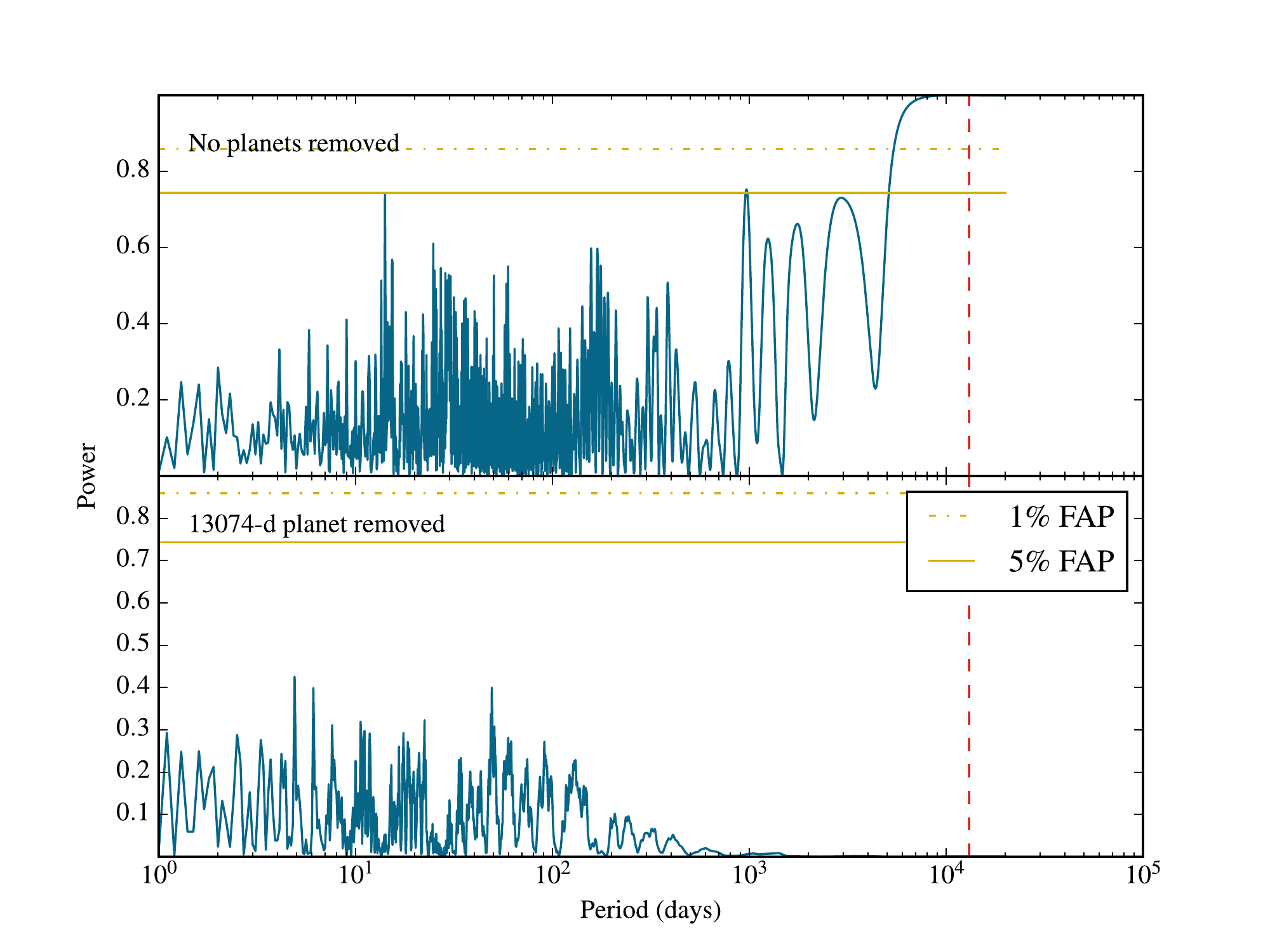}
	\caption{Periodograms for HD 98630. The red dashed line displays the location of the fitted planet in the preferred model. A shorter, 4000-day orbital period would have a significantly higher false alarm probability (FAP). After removing the Keplerian signal, all of the remaining peaks have a high FAP.\label{figK398P}}
\end{figure}

We also recover a competing model with \chisq\ of 6.66 where the companion has an orbital period near 4138 days and a mass of $M \sin i = 25 \pm 8$ \mjup, and the model also includes a linear trend of -394 m/s/yr. However, we discard this model based on the higher \chisq\ value as well as the fact that the Lomb-Scargle periodogram of radial velocities (displayed in Figure \ref{figK398P}) also strongly favors the longer-period model.

\subsubsection{HD 103459}

HD 103459 A is a G5-type metal-rich star located 60 pc from the Sun. It has a mass of $M = 1.18 \pm 0.20$ \msun\ and a radius of $R = 1.69 \pm 0.10$ \rsun, which suggests that HD 103459 A is likely a subgiant. However, since it is only $\Delta M_V = 1.33$ magnitudes above the main sequence, we utilize Equation \ref{eq:jit4a} to obtain a chromospheric jitter estimate of 2.31 \ms\ for HD 103459 A. The star has a derived age of $6.0 \pm 0.6$ Gyr and a rotation period is 32 days. We observed HD 103459 at Keck during 29 short visits.

\begin{figure}
	\plotone{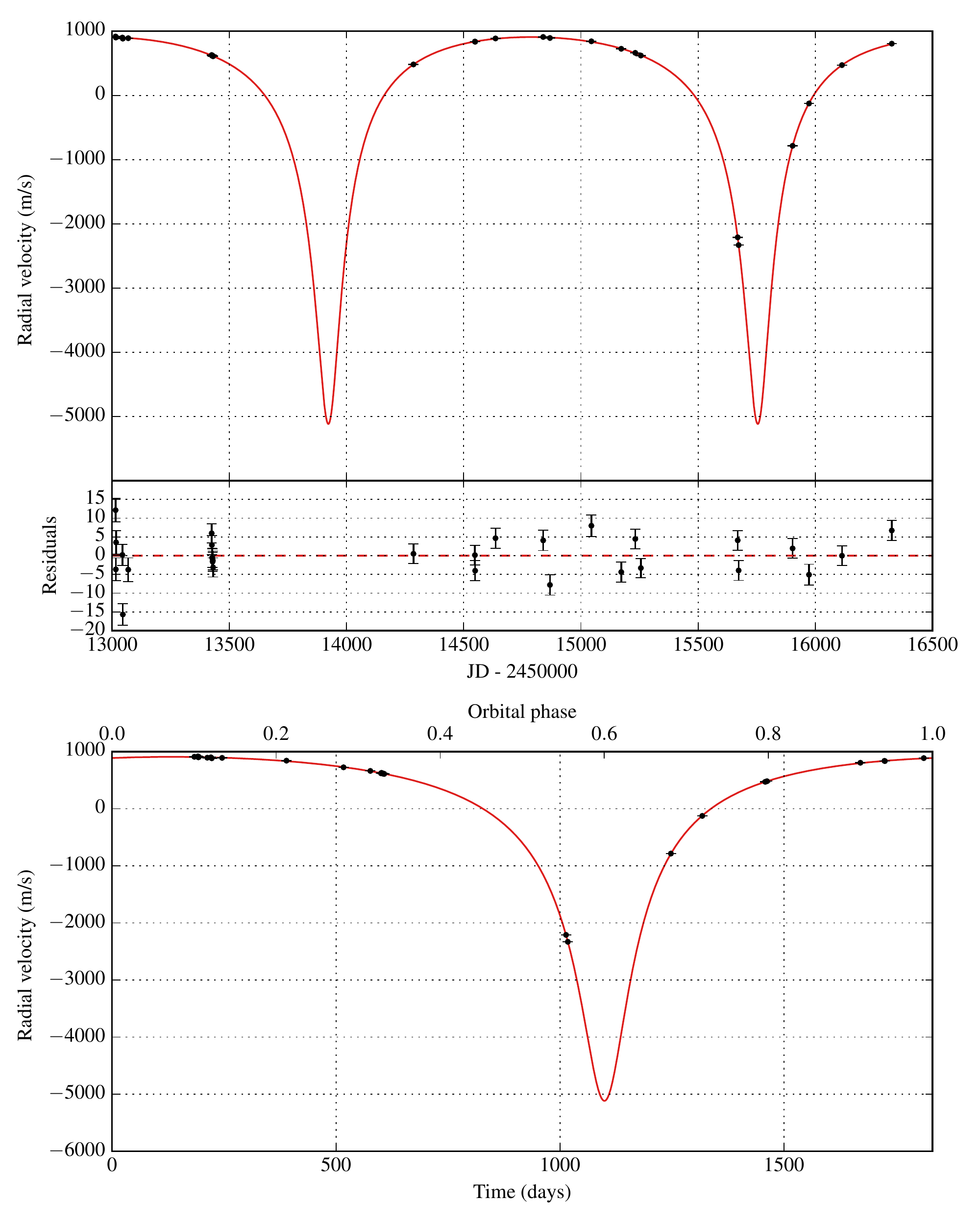}
	\caption{Keplerian model for HD 103459 B. The first plot shows the best Keplerian fit (in red) and the observed data points with an inset display of the residual velocities after fitting. The second plot shows a phase-folded version of the first plot.\label{figK242}}
\end{figure}

We find persuasive evidence for a stellar binary companion HD 103459 B with a mass of $m \geq 140 \pm 20$ \mjup\ (or $0.13 \pm 0.02$ \msun) on an eccentric orbit with an eccentricity of $e = 0.699 \pm 0.005$ and a semi-major axis of $a \geq 3.21 \pm 0.16$ AU. The orbital period is $P = 1831.9 \pm 0.9$ days. The Keplerian fit is given in Figure \ref{figK242} and an overview of the orbital parameters is given in Table \ref{tblK242}. Our model has a \chisq\ of 4.51 which is partly due to using Equation \ref{eq:jit4a} instead of Equation \ref{eq:jit5} for chromospheric jitter.

\begin{deluxetable}{lcccc}
	\tablewidth{0pt}
	\tablecaption{Orbital parameters for HD 3404 B, HD 24505 B, HD 98360 B, HD 103459 B\label{tblK242}}
	\tablehead{
		\colhead{Parameter} & \colhead{HD 3404 B} & \colhead{HD 24505 B} & \colhead{HD 98360 B} & \colhead{HD 103459 B}
	}
	\startdata
	P (days) & 1540.8 $\pm$ 1.9 & 11315 $\pm$ 92 & 13074 $\pm$ 982 & 1831.91 $\pm$ 0.87\\
	K (\ms) & 3535 $\pm$ 188 & 3294.0 $\pm$ 2.8 & 2613 $\pm$ 24 & 3013 $\pm$ 60\\
	e & 0.7381 $\pm$ 0.0044 & 0.798 $\pm$ 0.0012 & 0.059 $\pm$ 0.032 & 0.6993 $\pm$ 0.0046\\
	$\omega$ (deg) & 0.86 $\pm$ 0.9 & 157.895 $\pm$ 0.072 & 71 $\pm$ 14 & 182.585 $\pm$ 0.068\\
	T$_P$ (JD) & 13455.3 $\pm$ 9.0 & 16995 $\pm$ 65 & 14297 $\pm$ 542 & 13924.4 $\pm$ 1.9\\
	\msini (\mjup) & 145 $\pm$ 28 & 222 $\pm$ 30 & 280 $\pm$ 11 & 140 $\pm$ 20\\
	a (AU) & 2.86 $\pm$ 0.16 & 10.82 $\pm$ 0.61 & 11.75 $\pm$ 0.62 & 3.21 $\pm$ 0.16\\
	\chisq & 0.37 & 0.59 & 6.25 & 4.51\\
	RMS (\ms) & 2.02 & 3.02 & 7.30 & 5.35\\
	N$_{\rm obs}$ & 14 & 24 & 21 & 29\\
	\enddata
\end{deluxetable}

\subsection{Possible second companion around HD 38801}

\subsubsection{Stellar characteristics}

HD 38801 is a K0-type subgiant at a distance of 91 pc from the Sun. It is slightly more massive than the Sun with $M = 1.3 \pm 0.3$ \msun\ and it has a radius of $R = 2.4 \pm 0.2$ \rsun. \citet{Brewer2016} estimated its age to be within $4.9 \pm 1.0$ Gyr. HD 38801 has a very high metallicity at $\mathrm{[Fe/H]} = 0.32$. The star being a subgiant, we use Equation \ref{eq:jit5} to estimate its chromospheric jitter at 4.32 \ms. We also calculate its rotation period, obtaining $P_{rot} = 46$ days. Please refer to Table \ref{t_starswithplanets} for a comprehensive overview of the stellar parameters.

\subsubsection{Single-companion model}

\begin{figure}
	\plotone{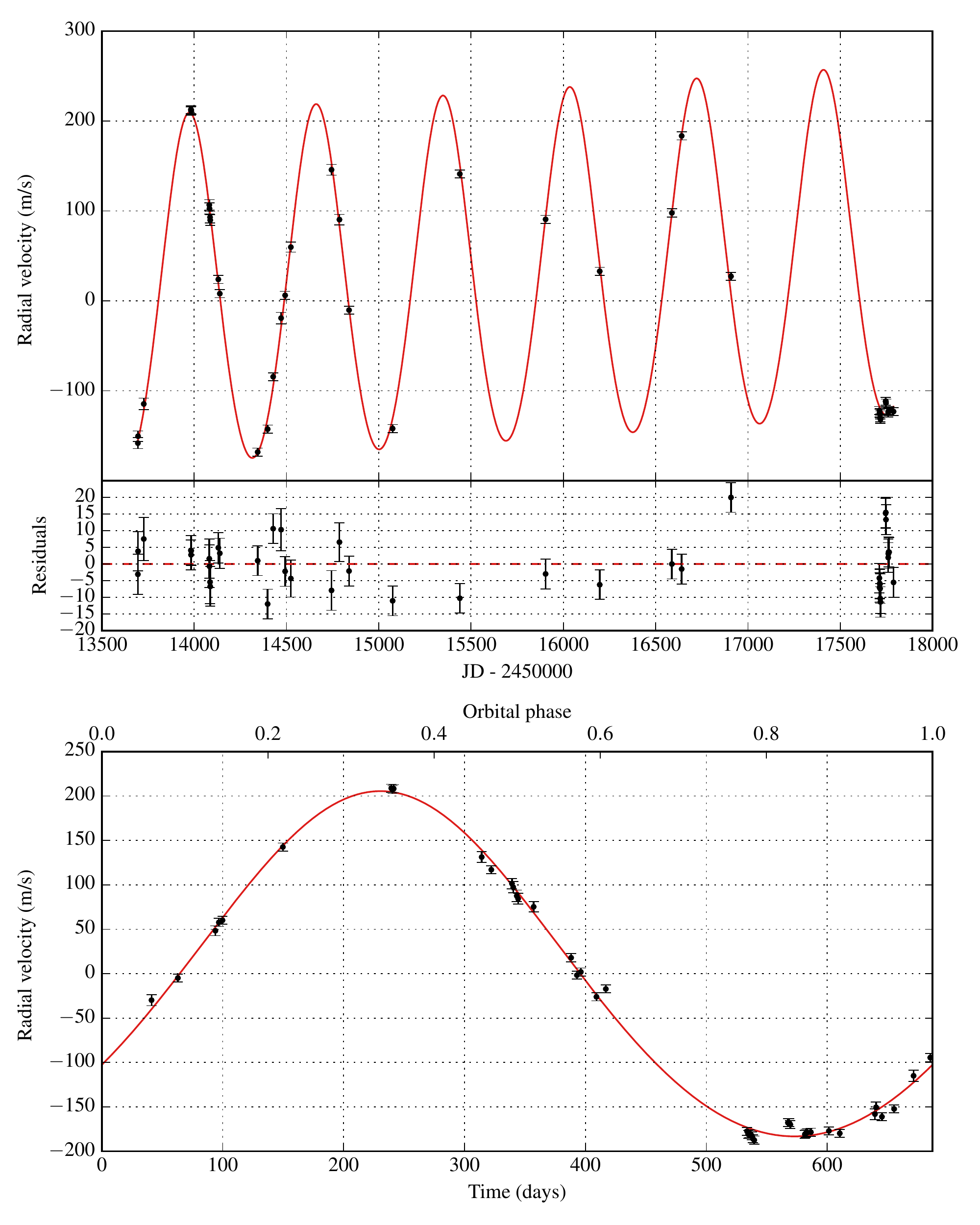}
	\caption{Keplerian model for HD 38801 b. The first plot shows the best Keplerian fit (in red) and the observed data points with an inset display of the residual velocities after fitting. The second plot shows a phase-folded version of the first plot. The model includes a linear trend.\label{figK17}}
\end{figure}

The discovery of HD 38801 b was first announced by \citet{Harakawa2010} using 10 observations at Keck as well another 11 at the Subaru Telescope. We provide a refined Keplerian fit by adding 22 additional measurements with the Keck/HIRES spectrograph, bringing the length of the time baseline up to 11 years. As given in greater detail in Table \ref{tblK17}, the star has a putative companion with a mass of $M \sin i = 10.0 \pm 1.4$ \mjup\ traveling at an orbit with semi-major axis $a = 1.66 \pm 0.11$ AU, period $P = 687.1 \pm 0.5$ days, and eccentricity $e = 0.057 \pm 0.006$. We also find a significant linear radial velocity trend of $5.1 \pm 0.3$ \ms\ per year. This single-companion solution can be seen in Figure \ref{figK17}. After adding the stellar jitter in quadrature, the model has a \chisq\ value of 3.41 with a residuals' RMS of 7.7 \ms. This is the model we report in Table \ref{t_planets} at the end of this paper. The orbital parameters of this model are also given in Table \ref{tblK17}.

\subsubsection{Double-companion model}

\begin{figure}
	\plotone{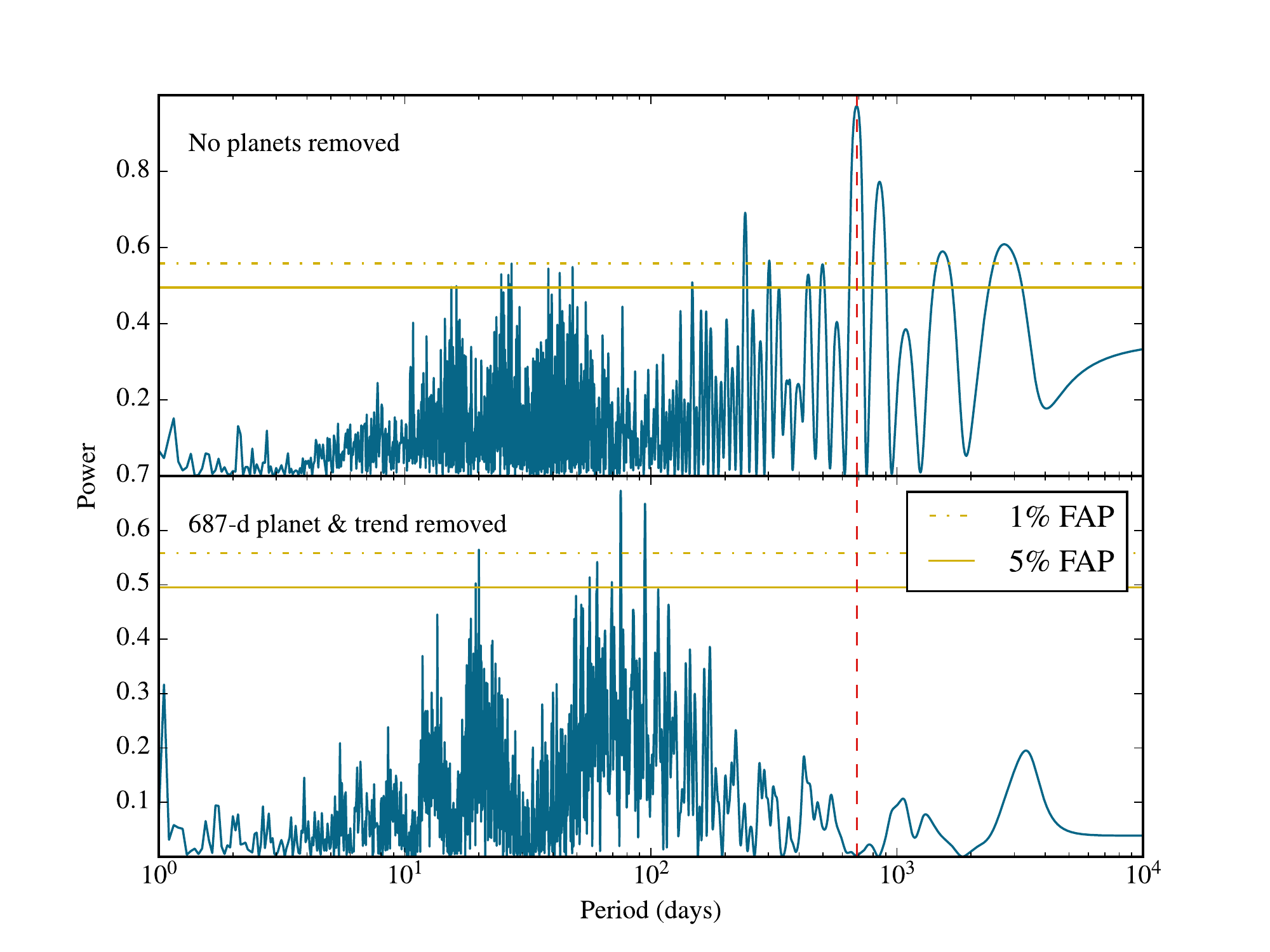}
	\caption{Periodograms for HD 38801. The red dashed line displays the location of the fitted planet. After fitting for a single Keplerian companion and de-trending, the periodogram still displays several peaks at periods of 20-100 days with a false alarm probability (FAP) less than 1\%. This is suggestive of a presence of additional signals.\label{figK17P}}
\end{figure}

The high values for \chisq\ as well as the RMS of the residuals suggests that we might have severely underestimated the chromospheric jitter (4.32 \ms). However, it might also indicate the presence of additional companions. Indeed, the Lomb-Scargle periodogram given in Figure \ref{figK17P} portrays several significant peaks after subtracting out the single companion and the linear trend. Consequently, we tried adding a second substellar companion to our model.

\begin{figure}
	\epsscale{0.7}
	\plotone{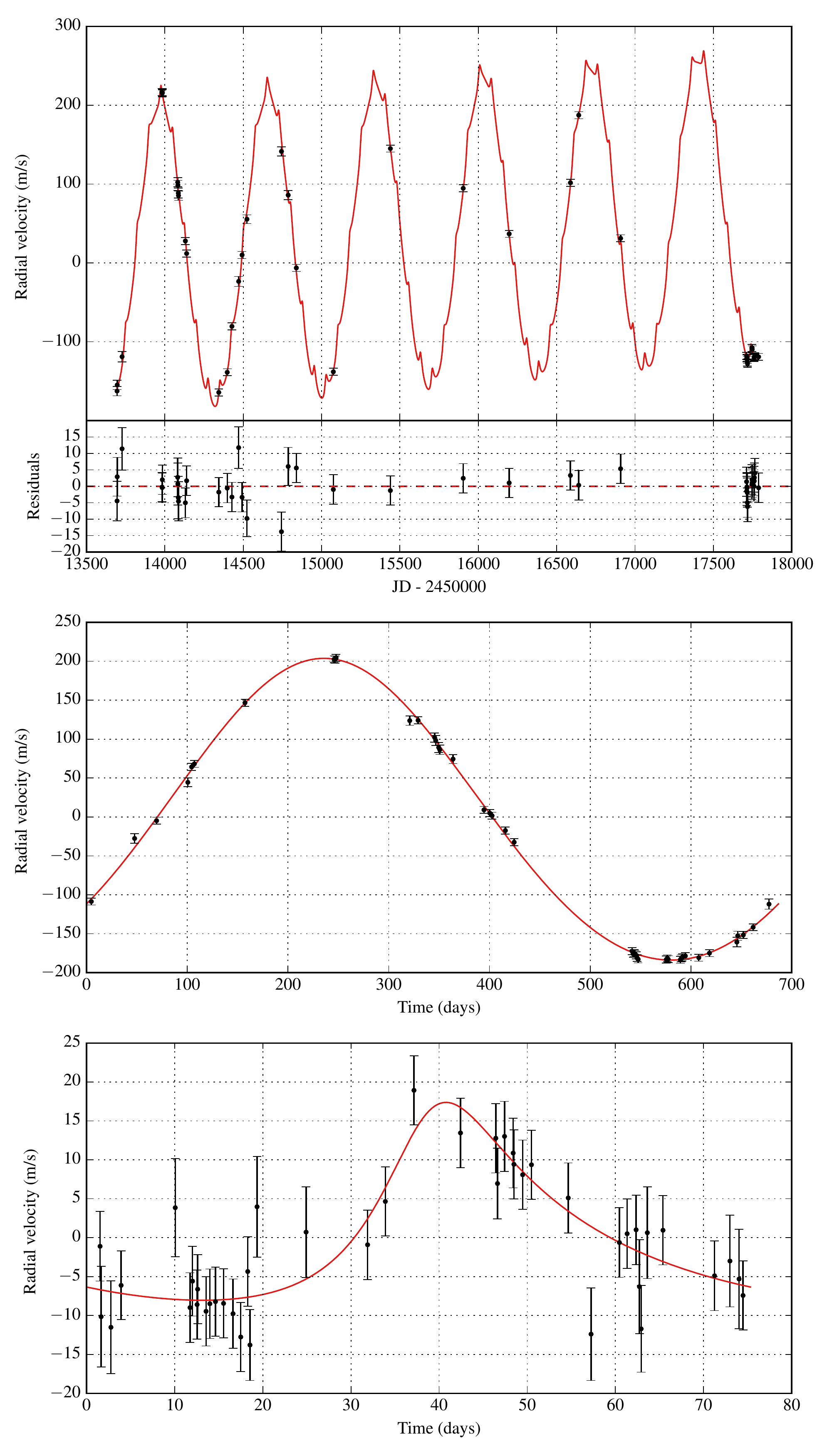}
	\caption{An alternative, double-planet Keplerian model for HD 38801. The first plot shows the best Keplerian fit (in red) and the observed data points with an inset display of the residual velocities after fitting. The two remaining plots are versions of the first folded by the orbital periods of each proposed companion. The model includes a linear trend.\label{figK17a}}
\end{figure}

Our double-planet model returns similar parameters for HD 38801 b but also includes an inner planet HD 38801 c with a mass of $M \sin i = 0.28 \pm 0.15$ \mjup\ and an orbital semi-major axis of $a = 0.38 \pm 0.03$ AU. The orbit has a period of $P = 12.7 \pm 3.4$ days and it is noticeably eccentric with $e = 0.42 \pm 0.27$. The full set of parameter values for both companions is given in Table \ref{tblK17} and the model is displayed in Figure \ref{figK17a}. This time, the \chisq\ parameter has a value of 1.00 and the RV residuals have a root mean square of 4.6 \ms.

By multiplying \chisq\ with the number of degrees of freedom, we find that the original $\chi^2$ dropped from 119.2 to 30.0 by introducing a second companion. At the same time, the number of degrees of freedom dropped by 5. Using a chi-square difference test, we calculated that this change corresponds to a p-value of $\sim 10^{-17}$, suggesting a very high significance. However, this assumes that the stellar jitter value of 4.32 \ms\ is accurate. Investigating the phase-folded RV curve of the proposed companion HD 38801 c in Figure \ref{figK17a}, one can notice that while the observations have good phase coverage overall, the precise shape of the narrower peak due to $e > 0$ cannot been conclusively established. Thus, our verdict is that more observations need to be accumulated before the nature of this secondary signal can be decisively resolved.

In addition, we tested the dynamical stability of the three-body system by using the symplectic WHFast integrator in the REBOUND package \citep{Rein2015}. We established that the system is stable over one million years. Furthermore, we also tried to attribute the signal to magnetic activity in the star by generating a Lomb-Scargle periodogram of the $S_{\rm HK}$ values. While some of the periodogram peaks were near the fitted shorter period, none of them proved to be significant in a bootstrap test. This lends credibility to the double-companion model.

\begin{table*}[t]
	\begin{center}
		\caption{Two competing models for HD 38801: a single-companion solution versus a double-companion one\label{tblK17}}
		\begin{tabular}{lccc}
			\tableline\tableline
			& 1-planet model\tablenotemark{1} & \multicolumn{2}{c}{2-planet model}\\
			Parameter & HD 38801 b & \shortstack{\\HD 38801 b\\(alternative)} & \shortstack{\\HD 38801 c\\(alternative)}\\
			\tableline
			P (days) & 687.14 $\pm$ 0.46 & 686.82 $\pm$ 0.62 & 75.32 $\pm$ 0.15\\
			K (\ms) & 194.4 $\pm$ 1.8 & 193.8 $\pm$ 1.9 & 12.7 $\pm$ 3.4\\
			e & 0.0572 $\pm$ 0.0063 & 0.052 $\pm$ 0.012 & 0.42 $\pm$ 0.27\\
			$\omega$ (deg) & 2 $\pm$ 11 & 356 $\pm$ 13 & 332 $\pm$ 55\\
			T$_P$ (JD) & 13976 $\pm$ 20 & 13966 $\pm$ 24 & 13746.2 $\pm$ 9.9\\
			\msini (\mjup) & 9.966 $\pm$ 0.096 & 9.94 $\pm$ 0.11 & 0.28 $\pm$ 0.12\\
			a (AU) & 1.66 $\pm$ 0.11 & 1.66 $\pm$ 0.11 & 0.38 $\pm$ 0.026\\
			Trend (m/s/yr) & 5.07 $\pm$ 0.32 & \multicolumn{2}{c}{5.69 $\pm$ 0.45}\\
			\chisq & 3.41 & \multicolumn{2}{c}{1.00}\\
			RMS (\ms) & 7.71 & \multicolumn{2}{c}{4.58}\\
			N$_{\rm obs}$ & 43 & \multicolumn{2}{c}{43}\\
			\tableline
		\end{tabular}
	\tablenotetext{1}{Currently, we opt for the single-planet model in our final analysis.}
	\end{center}
\end{table*}

\subsection{Possible third, long-period companion around HD 163607 \label{sec:K40}}

The metal-rich type G5 subgiant HD 163607 can be found at a distance of 69 pc from the Sun. The star is slightly older than the Sun with an age of $7.8 \pm 0.7$ Gyr, and it has a mass of $1.12 \pm 0.16$ \msun. Using the magnetic activity indices $\log R_{\rm HK}^{`} = -5.01$ and $S_{\rm HK} = 0.16$, we derived a stellar rotation period of 39 days and a chromospheric jitter estimate of 4.25 \ms. Our data set consists of 68 RV observations over 11 years.

\citet{Giguere2012} announced the discovery of a double-planet system around HD 163607 by the N2K consortium. This system is unique because the inner planet, completing a full orbit around the star in $75.195 \pm 0.034$ days in our best model, has an unusually high eccentricity $e = 0.744 \pm 0.012$. The planet (HD 163607 b) has a mass of $M \sin i = 0.79 \pm 0.11$ \mjup\ and an orbital semi-major axis of $a = 0.36 \pm 0.02$ AU. The outer planet is more massive with $M \sin i = 2.16 \pm 0.27$ \mjup\ and has a lower eccentricity $e = 0.076 \pm 0.023$. HD 163607 c has an orbital period of $P = 1267 \pm 7$ days and a semi-major axis of $a = 2.38 \pm 0.12$ AU.

\begin{figure}
	\epsscale{0.7}
	\plotone{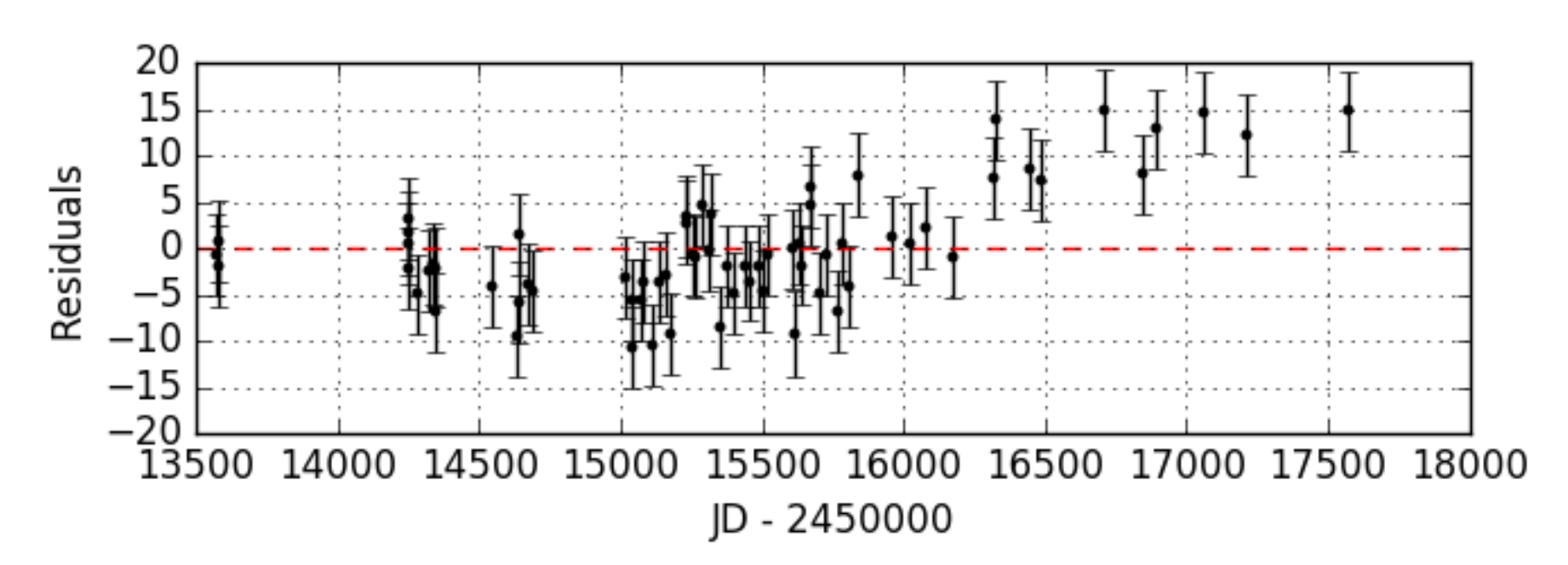}
	\caption{Radial velocity residuals (\ms) in a two-planet model for HD 163607 after removing the two planets.\label{figK40resid}}
\end{figure}

However, after removing the two planets, the Lomb-Scargle periodogram still displays a broad but significant long-period signal which suggests a possible third companion whose period cannot be resolved to a sufficient accuracy at this point. Furthermore, the radial velocity residuals in the double-planet model display a clear upward trend with some possible curvature, as can be seen in Figure \ref{figK40resid}. Therefore, we fitted for a linear trend term as well as for an acceleration, and obtained $-2.7 \pm 1.0$ m/s/yr for the former and $0.96 \pm 0.18$ m/s/yr$^2$ for the latter, both included in our best model. This brought the \chisq\ down from 2.39 to 0.54.

\begin{figure}
	\epsscale{0.7}
	\plotone{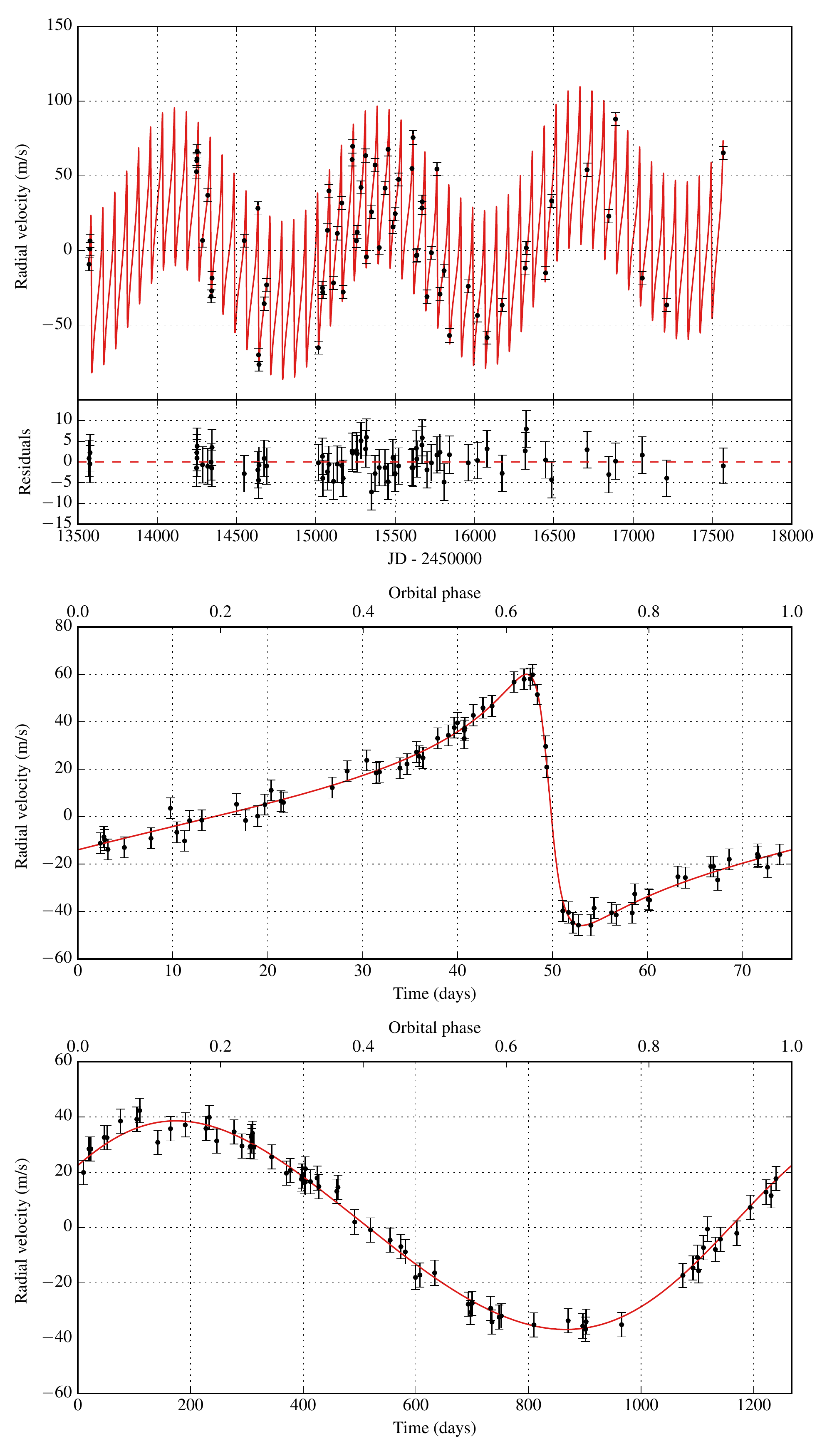}
	\caption{Keplerian model for HD 163607 b and c. The first plot shows the best Keplerian fit (in red) and the observed data points with an inset display of the residual velocities after fitting. The model includes a linear trend as well as a curvature term. The two remaining plots show HD 163607 b and c, respectively, phase-folded by the orbital period after removing the other planet as well as the linear trend and the curvature.\label{figK40}}
\end{figure}

\begin{figure}
	\epsscale{1.0}
	\plotone{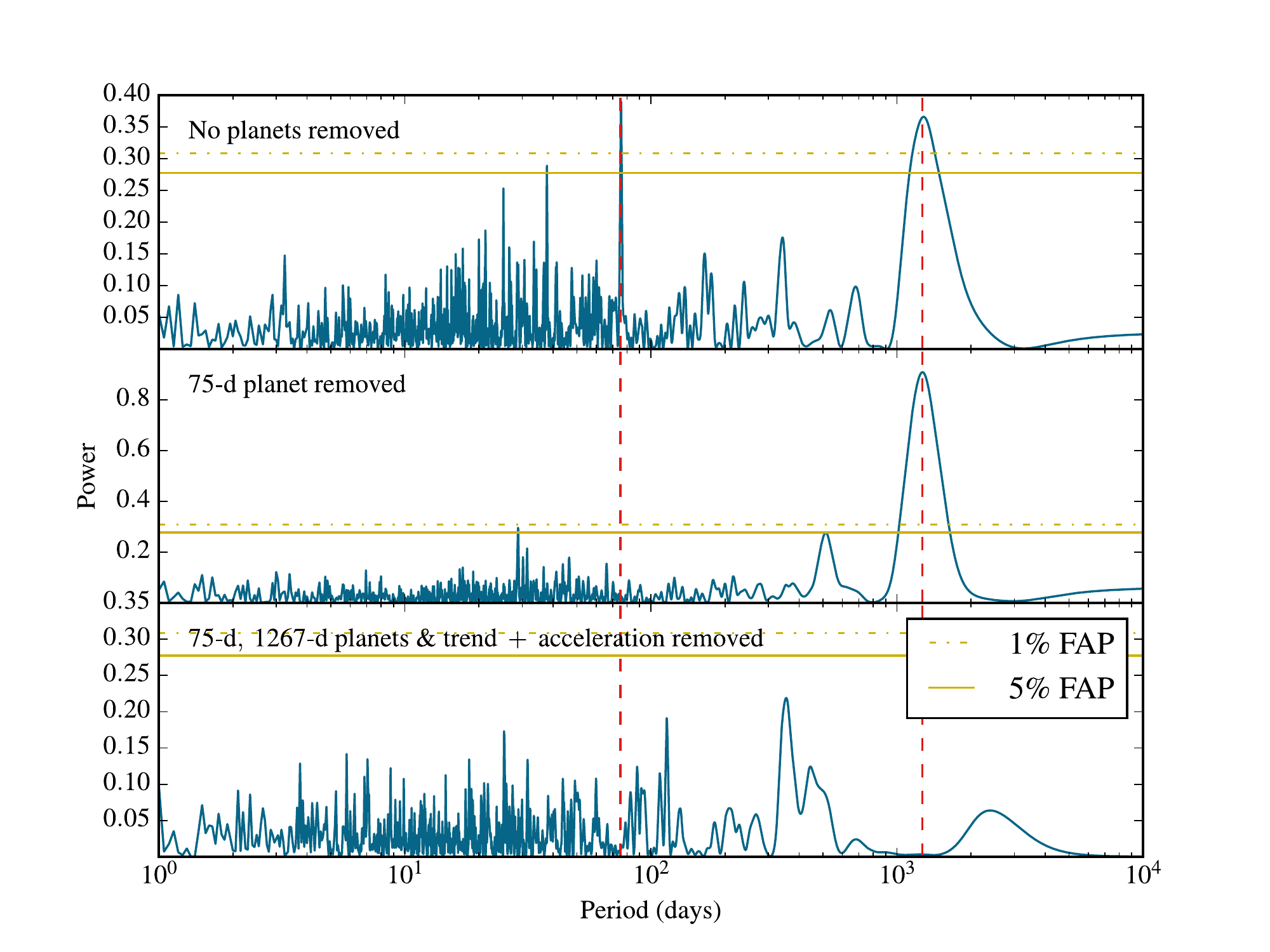}
	\caption{Periodograms for HD 163607. The dashed lines depict the periods of the fitted planets. After removing the Keplerian companions, the linear trend and the curvature, the remaining peaks all have a high false alarm probability (FAP).\label{figK40P}}
\end{figure}

The Keplerian model can be seen in Figure \ref{figK40} and its parameters are fully described in Table \ref{t_planets}. The Lomb-Scargle periodogram no longer shows significant signals, as can be seen in Figure \ref{figK40P}. We also tried fitting for a third companion, but with the current time baseline ($\sim$11 years) it is not possible to put sufficient constraints on the orbital period. However, we can conclude that the period must be at least 6000 days which translates into a semi-major axis of at least 7 AU, or an angular separation of 100 mas or more as viewed from Earth, making this a potential target for direct imaging.

\subsection{A planetary system around the young F8 star HD 147506 aka HAT-P-2}

HD 147506, also known as HAT-P-2, is an F8-type young star at a distance of 114 pc from the Sun. Its fitted mass from Yale-Yonsei isochrones is $1.33 \pm 0.03$ days and its age is $1.4 \pm 0.5$ Gyr. The star is a relatively fast rotator, with an estimated rotation period of 5 days. Based on the Ca II core line emission index $S_{\rm HK} = 0.19$, we calculate a chromospheric jitter estimate of 3.03 \ms.

HD 147506 has a known transiting planet first discovered by \citet{Bakos2007a}. An update to the system parameters was provided by \citet{Pal2010} using additional radial velocity and photometric data. The radial velocity data of \citet{Pal2010} included a combined total of 35 observations with Keck/HIRES, Lick/Hamilton, and OHP/SOPHIE, all of which we included in our fitting as well. In addition, we add 38 Keck/HIRES measurements, bringing the time baseline up from 1.7 years to 10.4 years. Since the transiting planet has an orbital period of 5 days which is many orders of magnitude smaller than the time baseline, we can achieve great precision while fitting for the orbital period even without including the photometric data.

In particular, our best single-planet model yields $P = 5.6335158 \pm 0.0000036$ days while \citet{Pal2010} reported $P = 5.6334729 \pm 0.0000061$ days. This is more than a 7-sigma difference. To test the significance of our model, we also generated a model where we fixed the period to the value of \citet{Pal2010}, but left the rest of the parameter set identical. This model yielded a $\chi^2$ of 1498 compared to 1369 for our best model. Using a $\chi^2$-difference distribution for 1 additional degree of freedom, we find that the fixed-period model has a p-value of $< 10^{-29}$ of being the correct one, thus we overwhelmingly prefer our new model, supported by our long time baseline spanning almost 700 orbital periods.

\begin{figure}
	\plotone{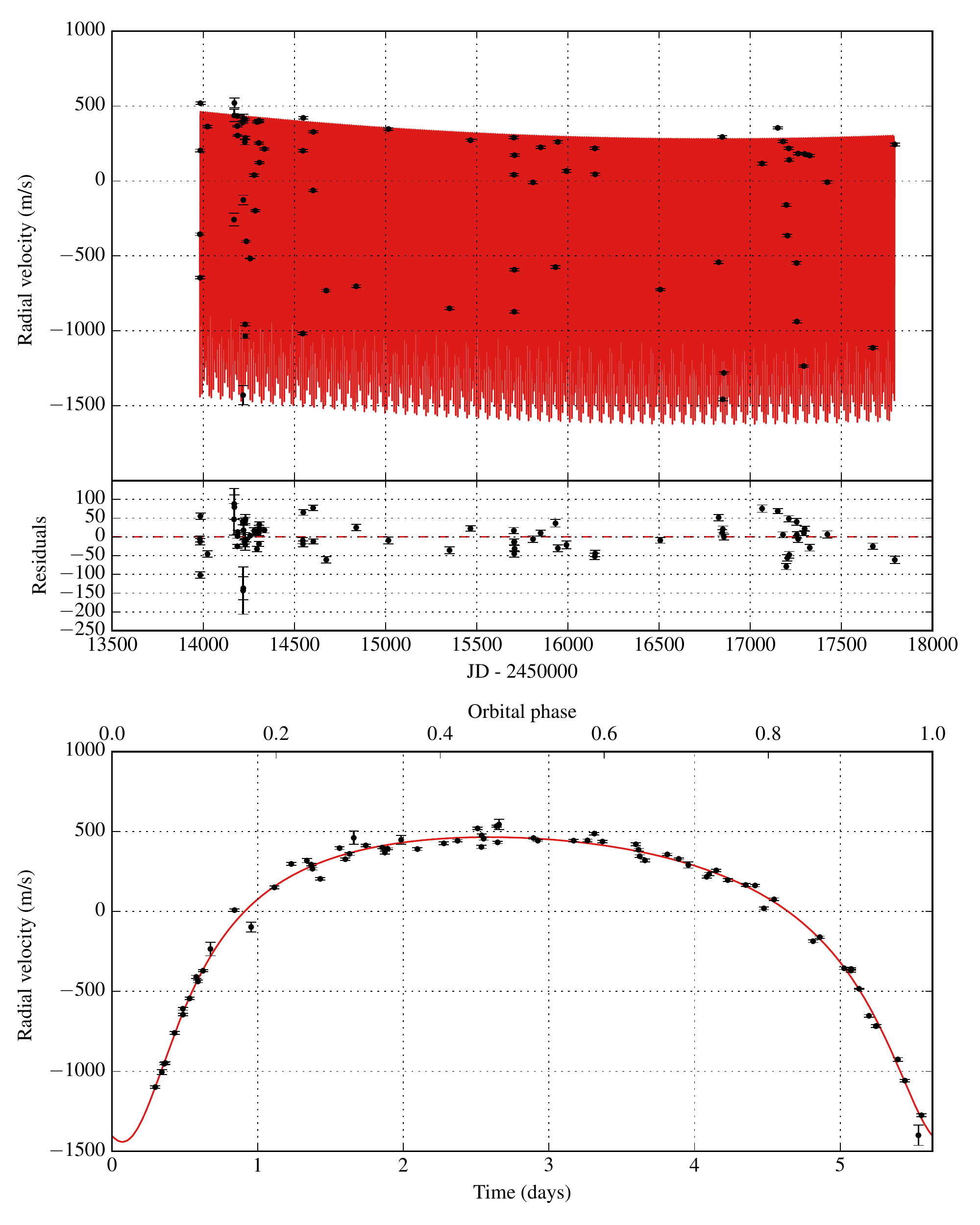}
	\caption{Keplerian model for HD 147506 b. The first plot shows the best Keplerian fit (in red) and the observed data points with an inset display of the residual velocities after fitting. The second plot shows a phase-folded version of the first plot. The model includes a linear trend as well as a curvature term.\label{figK30}}
\end{figure}

In addition to the orbital period $P = 5.6335158 \pm 0.0000036$, our model yielded an eccentricity of $e = 0.5172 \pm 0.0019$. The derived mass of the planet is $M \sin i = 8.62 \pm 0.17$ \mjup\ and the orbital semi-major axis is $a = 0.0681 \pm 0.0005$ AU. A complete overview of the fitted parameters is given in Table \ref{t_planets}. The Keplerian model is plotted in Figure \ref{figK30}.

\begin{figure}
	\plotone{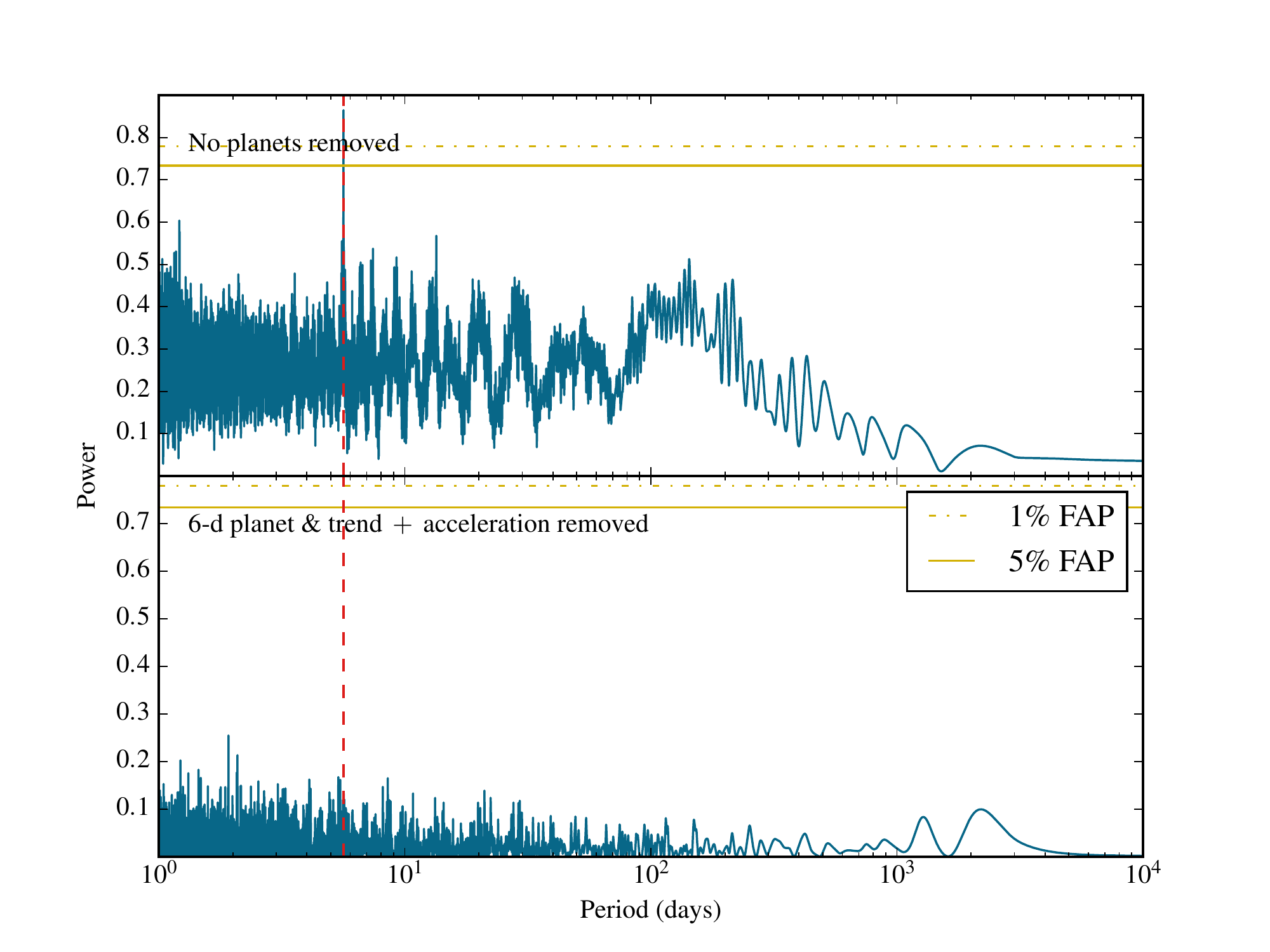}
	\caption{Periodograms for HD 147506. The dashed lines depict the periods of the fitted planets. After fitting for three Keplerian companions, the remaining peaks all have a high false alarm probability (FAP).\label{figK30P}}
\end{figure}

We also found a very significant linear radial velocity trend of $-47 \pm 1$ m/s/yr as well as a curvature of $6.1 \pm 0.3$ m/s/yr$^2$. Therefore, the system likely contains additional long-period companions, but using the current data we were not able to constrain their period sufficiently. In particular, fixing $e = 0$ for a second long-period planet led to a model with a higher \chisq\ whereas allowing all parameters to vary freely resulted in a rapid increase in eccentricity for a marginal decrease in \chisq. Thus, we currently recommend a single-planet model with a linear trend and a RV curvature. After removing these, the periodogram of the residuals shows no additional peaks, as can be seen in Figure \ref{figK30P}.

Interestingly, HAT-P-2 b has been proposed as a candidate for the detection of general relativistic precession of the orbital periastron due to its large orbital eccentricity and small semi-major axis. In particular, \citet{JordanBakos2008} calculate a longitude of periastron $\omega = 184.6$ deg and predict a GR precession rate of $\dot \omega_{\rm GR} = 1.8$ deg/century. We obtained $\omega = 188.0 \pm 0.2$ deg, and thus the predicted $\dot \omega_{\rm GR}$ is not large enough to explain the discrepancy between the two modelled values of $\omega$. Furthermore, GR precession would not be able to account for the large curvature detected in the RV data set. Therefore, the difference between the two models is better explained by an additional long-period companion or long-term stellar activity.

\subsection{One planet for HD 207832 instead of two}

Using Keck/HIRES data, \citet{Haghighipour2012} reported the discovery of a two-planet system around the G5 dwarf HD 207832. However, our analysis concludes that only one detectable planet exists in that system whereas the second signal is likely due to stellar activity.

HD 207832 is a young G5 dwarf (age $1.4 \pm 1.3$ Gyr) at a distance of 54 pc from the Sun. \cite{Brewer2016} derived the star's mass $M = 1.05 \pm 0.17$ \msun, radius $R = 0.89 \pm 0.05$ \rsun, and luminosity $L = 0.78 \pm 0.09$ \lsun. Using the stellar activity proxies $\log R_{\rm HK}^{`} = -4.67$ and $S_{\rm HK} = 0.25$, we calculate a rotation period of 19 days and a chromospheric jitter estimate of 4.03 \ms. We included a total of 64 Keck/HIRES observations of HD 207832.

\begin{figure}
	\plotone{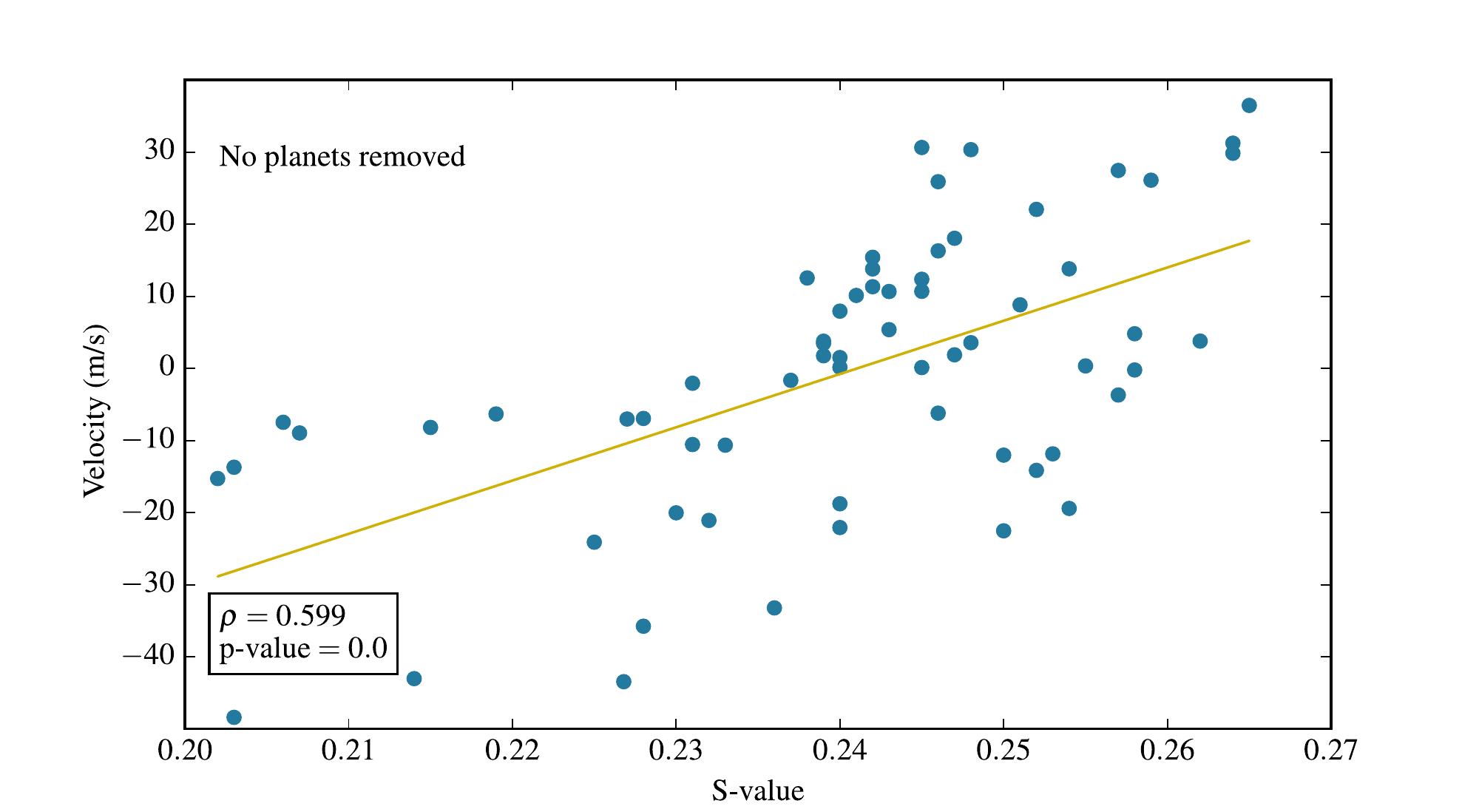}
	\caption{The radial velocities of HD 207832 are highly correlated with the respective S-values related to stellar activity (the correlation has a p-value less than 0.0001). Therefore, we decorrelate the velocities with the linear trend given above.\label{figK73SC}}
\end{figure}

\begin{figure}
	\plotone{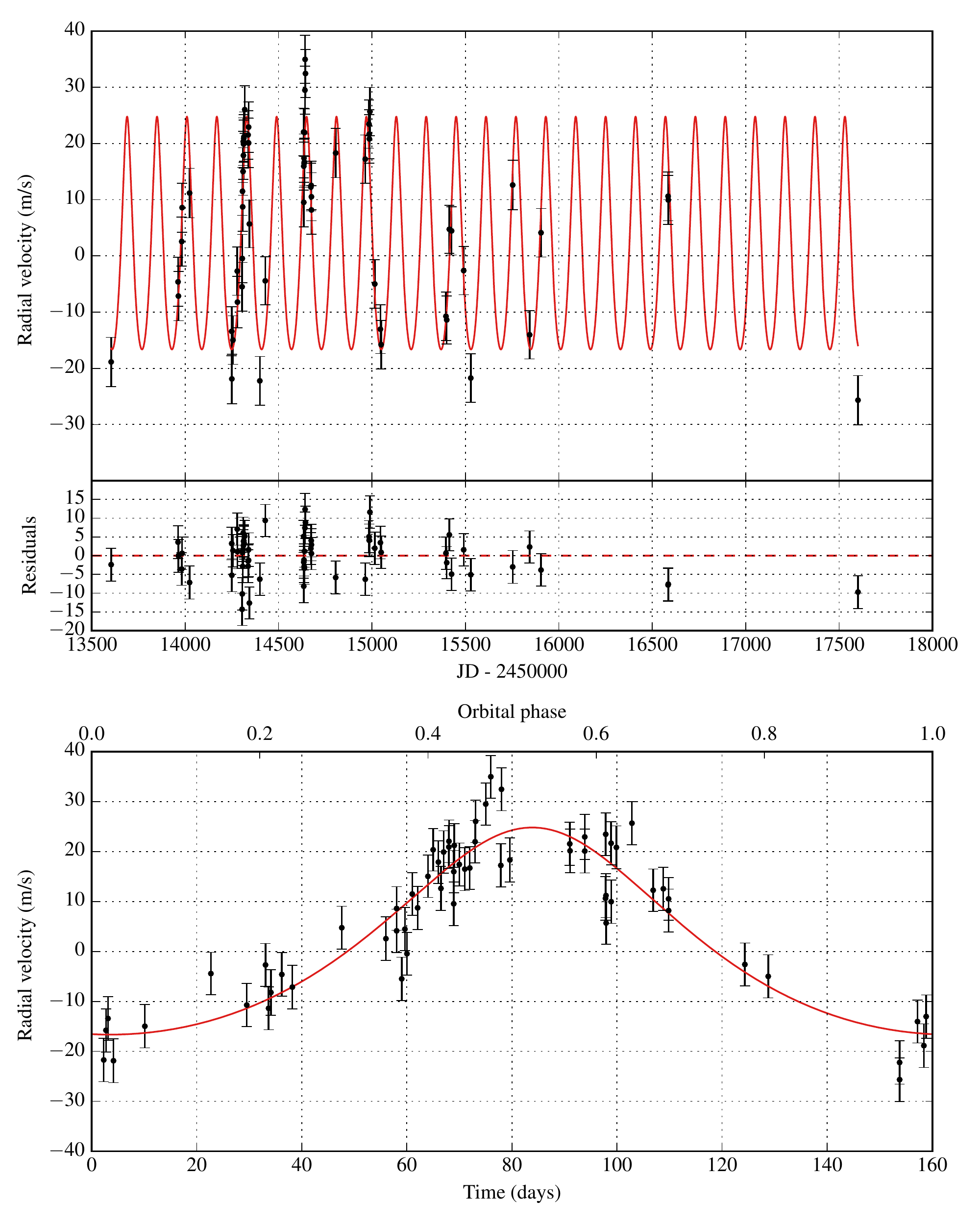}
	\caption{Keplerian model for HD 207832 b. The first plot shows the best Keplerian fit (in red) and the observed data points with an inset display of the residual velocities after fitting. The second plot shows a phase-folded version of the first plot.\label{figK73}}
\end{figure}

The measured radial velocities of HD 207832 are highly correlated with the respective stellar \shk values ($p < 0.0001$, see Figure \ref{figK73SC}). This suggests that the velocities are significantly influenced by stellar activity. Therefore, we decorrelated the velocities by the fitted linear correlation given in Figure \ref{figK73SC} before proceeding with our analysis. Subsequently, we detected a planet identified as HD 207832 b with a period of $P = 160.07 \pm 0.23$ days, eccentricity $e = 0.196 \pm 0.053$, mass $M \sin i = 0.56 \pm 0.09$ \mjup, and orbital semi-major axis $a = 0.58 \pm 0.03$ AU. The fitted planet can be seen in Figure \ref{figK73} and its orbital parameters are given in detail in Table \ref{t_planets}. Our final model has a \chisq\ of 1.82.

\begin{figure}
	\plotone{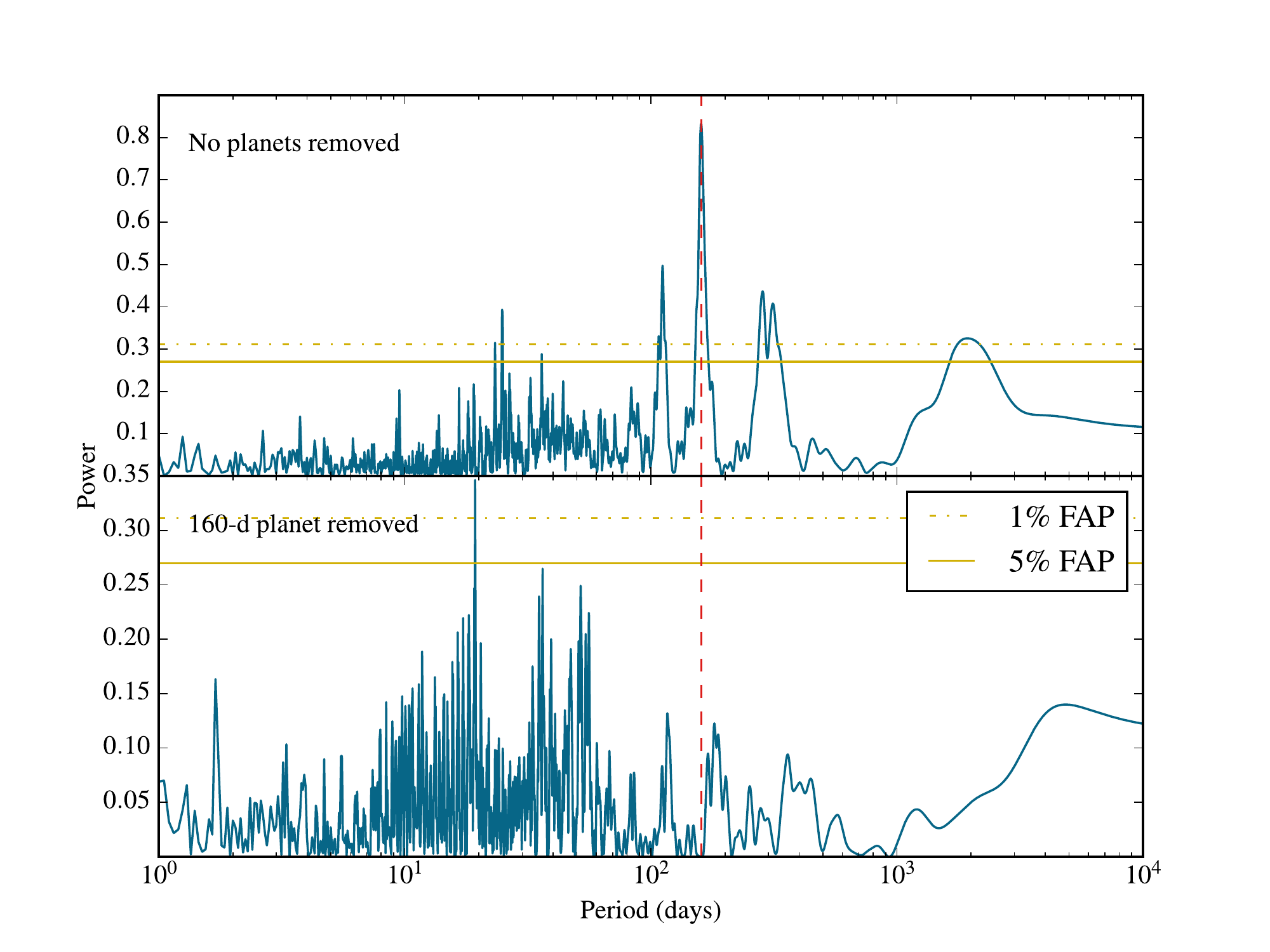}
	\caption{Periodograms for HD 207832. The dashed line depicts the period of the fitted planet. After fitting for a single Keplerian companions, the remaining 19-day peak is likely due to stellar activity.\label{figK73P}}
\end{figure}

After removing the one-planet Keplerian signal from the decorrelated velocities, the Lomb-Scargle periodogram (plotted in Figure \ref{figK73P}) shows one more significant peak with a period close to the stellar rotation period of 19 days. Therefore, we predict that this additional peak is most likely related to stellar activity and it is not a real companion. We also found some evidence for a second, 1156-day companion reported by \citet{Haghighipour2012} in the Lomb-Scargle periodogram in the original data, however after decorrelating the velocities the signal disappeared completely.

\subsection{Failure to recover the Keplerian signal of HD 73256 b}

HD 73256 is a main-sequence star between the spectral classes of G8 and K0 located 38 pc from the Sun. It has a derived mass of $M = 1.00 \pm 0.14$ \msun, a radius of $R = 0.94 \pm 0.03$ \rsun, and an age of $3.2 \pm 2.0$ Gyr. The star is chromospherically active with \lrphk $= -4.44$ and \shk $= 0.41$, yielding a rotation period of 11 days and an RV jitter estimate of 3.31 \ms. HD 73256 was observed for a total of 17 times between 2004 and 2016 by the N2K Consortium.

\begin{figure}
	\plotone{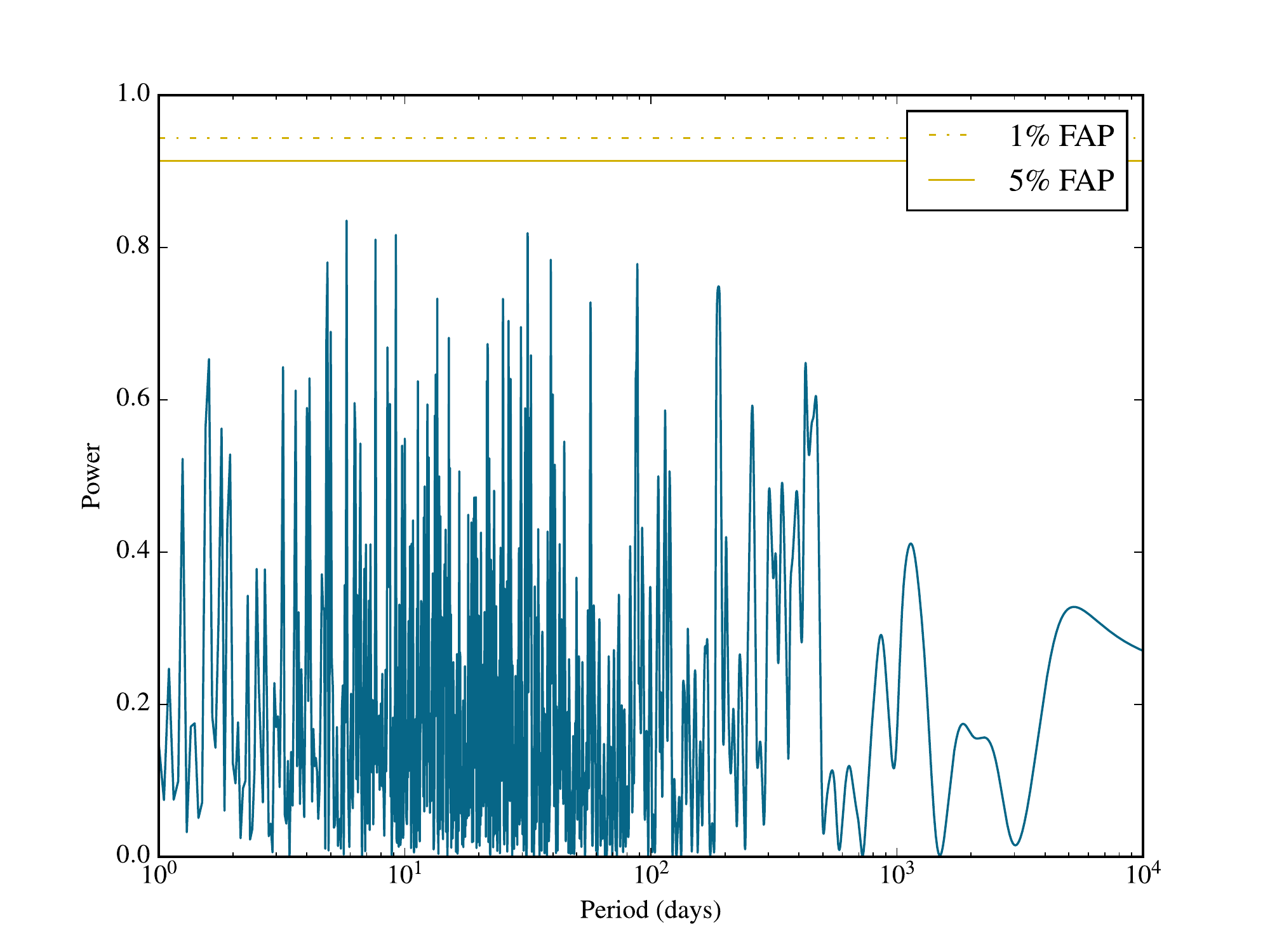}
	\caption{Periodograms for HD 73256. All tested periods exhibit a high false alarm probability (FAP), including the putative 2.5-day orbital period of HD 73256 b.\label{figK360P}}
\end{figure}

More than a decade ago, \citet{Udry2003} presented evidence for a hot Jupiter companion HD 73256 b on a 2.5-day orbit based on two months of observations with the CORALIE spectrograph on the 1.2-m Euler Swiss Telescope at La Silla. Unfortunately, we are unable to reproduce a similar Keplerian model using Keck/HIRES data. Searching for such a signal while holding the orbital period constant at $P = 2.54858$ days, as given by \citet{Udry2003}, as well as treating it as a free parameter yield extraordinarily poor fits with \chisq\ $> 200$. Furthermore, generating a Lomb-Scargle periodogram of the radial velocities (displayed in Figure \ref{figK360P}) fails to reveal any significant signals at all, suggesting that any periodicity in the data is due to a window function. We also did not succeed in combining our data with the CORALIE data used by \citet{Udry2003} since we were unable to find a publicly available version of the latter.

We can identify a somewhat stable local minimum near 2.5 days if we inflate the RV jitter value to 15 m/s. However, this still yields a \chisq\ of $\sim$20, and leads to the emergence of at least one competing model with an orbital period close to 6 days. Adding to the confusion, \citet{Wright2012} lists HD 73256 b with an orbital period of 5.2 days. Finally, we reproduced a potential Keplerian signal using the orbital parameters from \citet{Udry2003} with N2K observation times and uncertainties, and easily recovered the planet using our fitting algorithm. Thus, it is likely that we would have been able to detect the signal of HD 73256 b, particularly given the very large RV semi-amplitude of $K = 269 \pm 8$ m/s claimed by \citet{Udry2003}. Consequently, our data does not support the existence of a massive planet on a tight orbit around HD 73256.

\subsection{Long-term magnetic activity of HD 75898 masquerading as a Keplerian signal}

Even though HD 75898 is labelled a G0 star in the Hipparcos catalog, \citet{Robinson2007} remark that the luminosity, temperature, and surface gravity are more consistent with a metal-rich F8V star. The star has a derived mass of $M = 1.26 \pm 0.23$ \msun, a radius of $R = 1.58 \pm 0.11$ \rsun, and an age of $3.6 \pm 0.6$ Gyr. It is at a distance of 76 pc from the Solar System. We estimate the rotation period as 24 days, and the chromospheric RV jitter as 2.33 \ms. However, \citet{Robinson2007} caution that these might be underestimated since the star seems to be on the verge of moving off the main sequence. The first 20 N2K observations of HD 75898 were published by \citet{Robinson2007} whereas we bring the total number of observations up to 55.

Due to an extended time baseline, we are able to refine the orbital parameters of HD 75898 b, previously reported by \citet{Robinson2007}. In particular, our model yields an orbital period of $P = 422.90 \pm 0.29$ days, an eccentricity of $e = 0.110 \pm 0.010$, a planetary mass of $M \sin i = 2.71 \pm 0.36$ \mjup, and an orbital semi-major axis of $a = 1.19 \pm 0.07$ AU. These results are in good agreement with the values given by \citet{Robinson2007}.

\begin{figure}
	\epsscale{0.7}
	\plotone{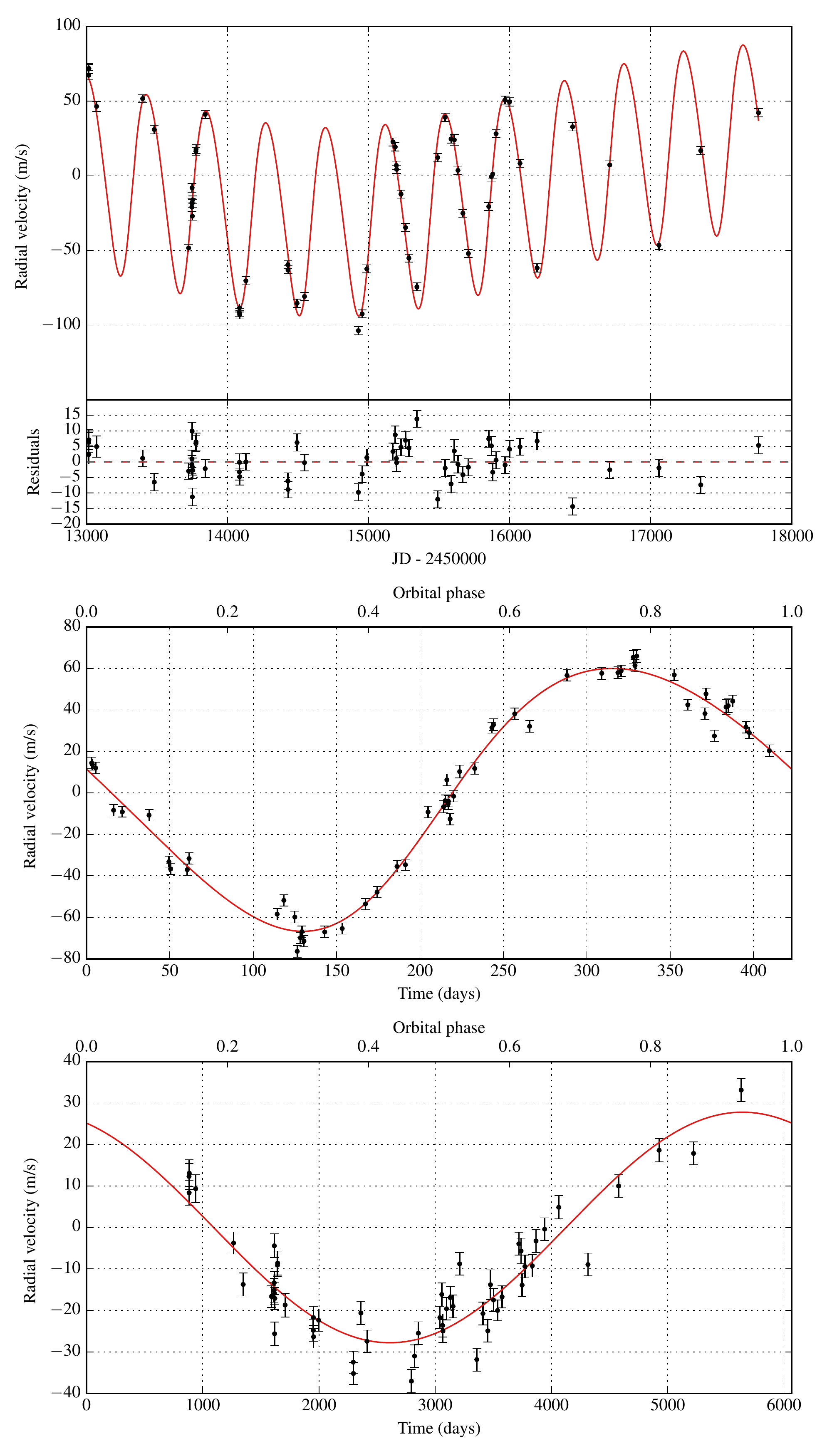}
	\caption{Keplerian model for HD 75898 b and the long-term activity of the star. The first plot shows the best Keplerian fit (in red) and the observed data points with an inset display of the residual velocities after fitting. The second plot is a phase-folded RV curve of the planet whereas the final plot illustrates the modeled magnetic activity of the star.\label{figK364}}
\end{figure}

\begin{figure}
	\epsscale{1.0}
	\plotone{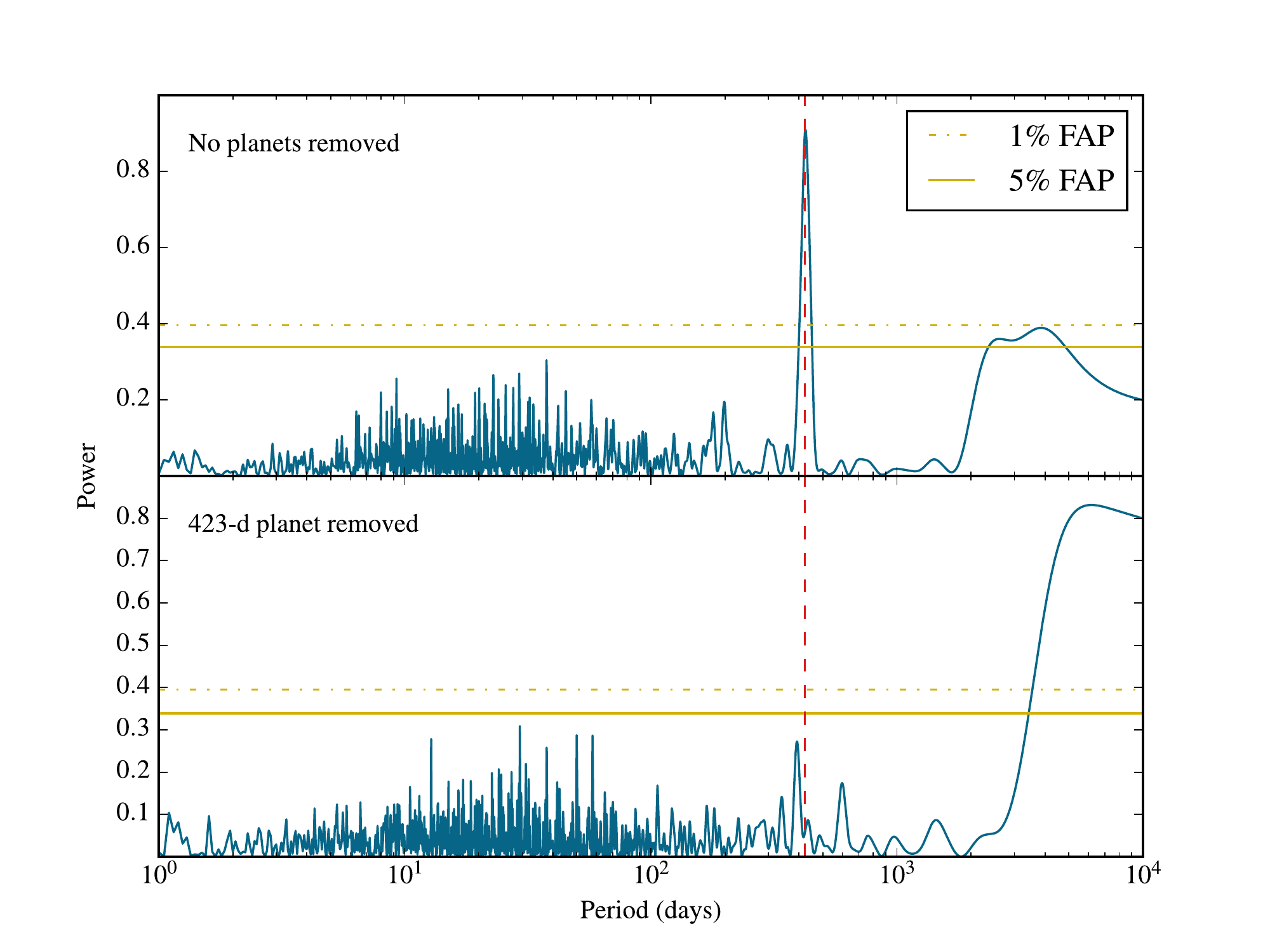}
	\caption{Periodograms for HD 75898. The red dashed line displays the location of the fitted planet. There is a significant signal near 433 days in the raw data. After fitting for a single Keplerian companion, a significant long-period signal emerges.\label{figK364P}}
\end{figure}

\begin{figure}
	\plotone{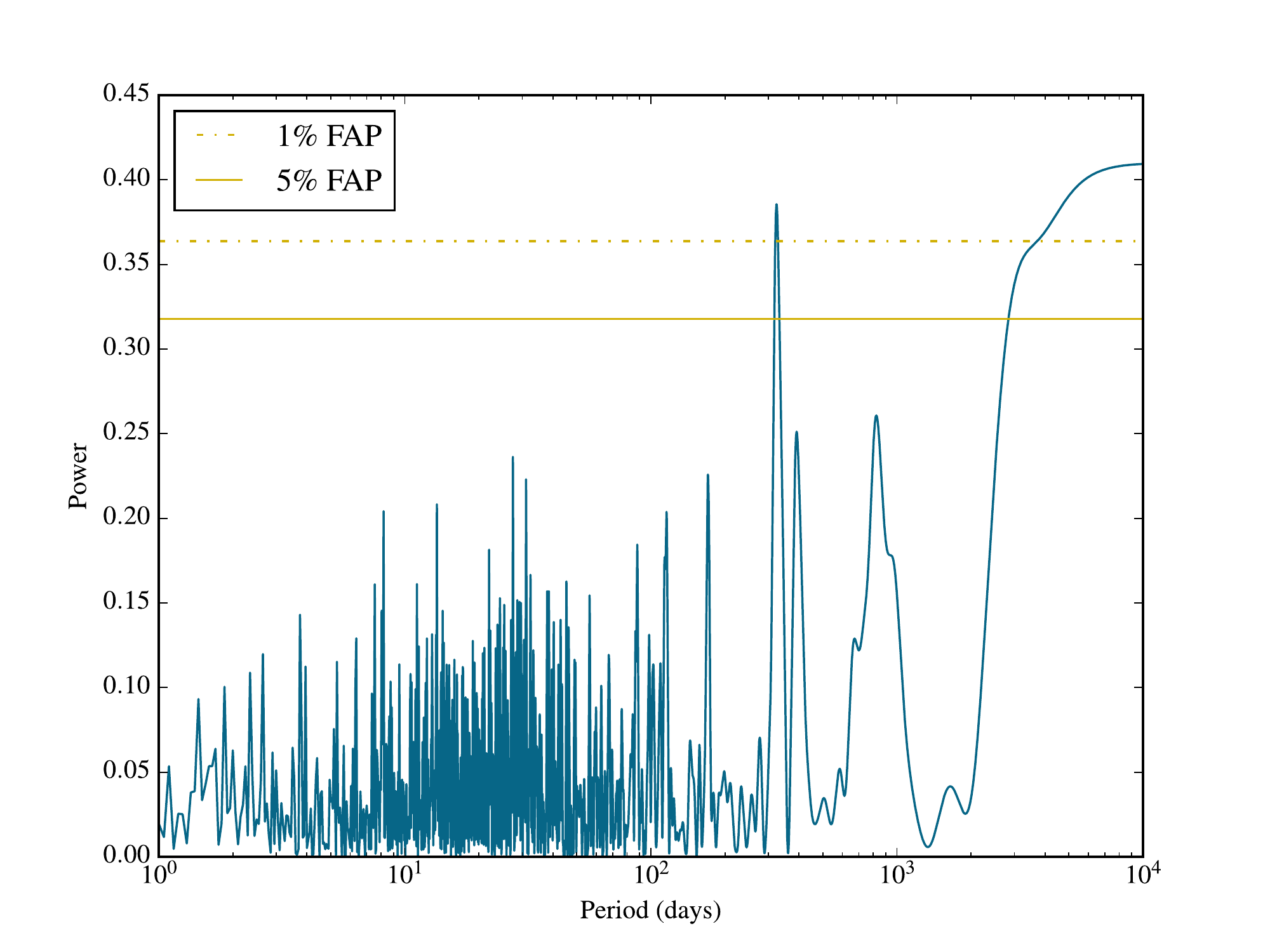}
	\caption{Periodogram for the \shk values of HD 75898, suggesting that any signal with a period above 4000 days is significantly correlated with the star's magnetic activity.\label{figK364S}}
\end{figure}

After removing the 423-day orbit from the radial velocities, a significant long-period signal emerges on the Lomb-Scargle periodogram of the radial velocities, seen in Figure \ref{figK364P}. This signal was also described by \citet{Robinson2007} as evidence for a plausible additional companion. However, after performing a Lomb-Scargle analysis of the \shk values, we conclude that this signal is likely due to long-term magnetic activity of HD 75898. The periodogram of \shk values, given in Figure \ref{figK364S}, suggests that any signal with a period above 4000 days is significantly correlated with the \shk values. Subsequently, we modeled the stellar activity as a sinusoidal Keplerian signal. Our best fit, given in Figure \ref{figK364}, included a period of $P = 6066 \pm 337$ days and a semi-amplitude of $K = 44 \pm 12$ for the long-term activity. The model has a \chisq\ of 5.25 and is described in greater detail in Table \ref{tblK364}.

\begin{deluxetable}{lcc}
	\tablewidth{0pt}
	\tablecaption{Orbital parameters for the HD 75898 system, including stellar activity\label{tblK364}}
	\tablehead{
		\colhead{Parameter} & \colhead{HD 75898 b} & \colhead{Stellar activity}
	}
	\startdata
	P (days) & 422.9 $\pm$ 0.29 & 6066 $\pm$ 337\\
	K (\ms) & 63.39 $\pm$ 0.71 & 27.8 $\pm$ 1.5\\
	e & 0.11 $\pm$ 0.01 & 0.0\\
	$\omega$ (deg) & 241.1 $\pm$ 5.2 & 77.5 $\pm$ 5.7\\
	T$_P$ (JD) & 13299.0 $\pm$ 5.9 & 13014.946\\
	\msini (\mjup) & 2.71 $\pm$ 0.36 & 2.9 $\pm$ 0.57\\
	a (AU) & 1.191 $\pm$ 0.073 & 7.03 $\pm$ 0.69\\
	\chisq & 5.49 & \\
	RMS (\ms) & 5.82 & \\
	N$_{\rm obs}$ & 55 & \\
	\enddata
\end{deluxetable}

\section{Results: Updated Orbital Parameters \label{sec:results3}}

\begin{figure}
	\epsscale{0.5}
	\plotone{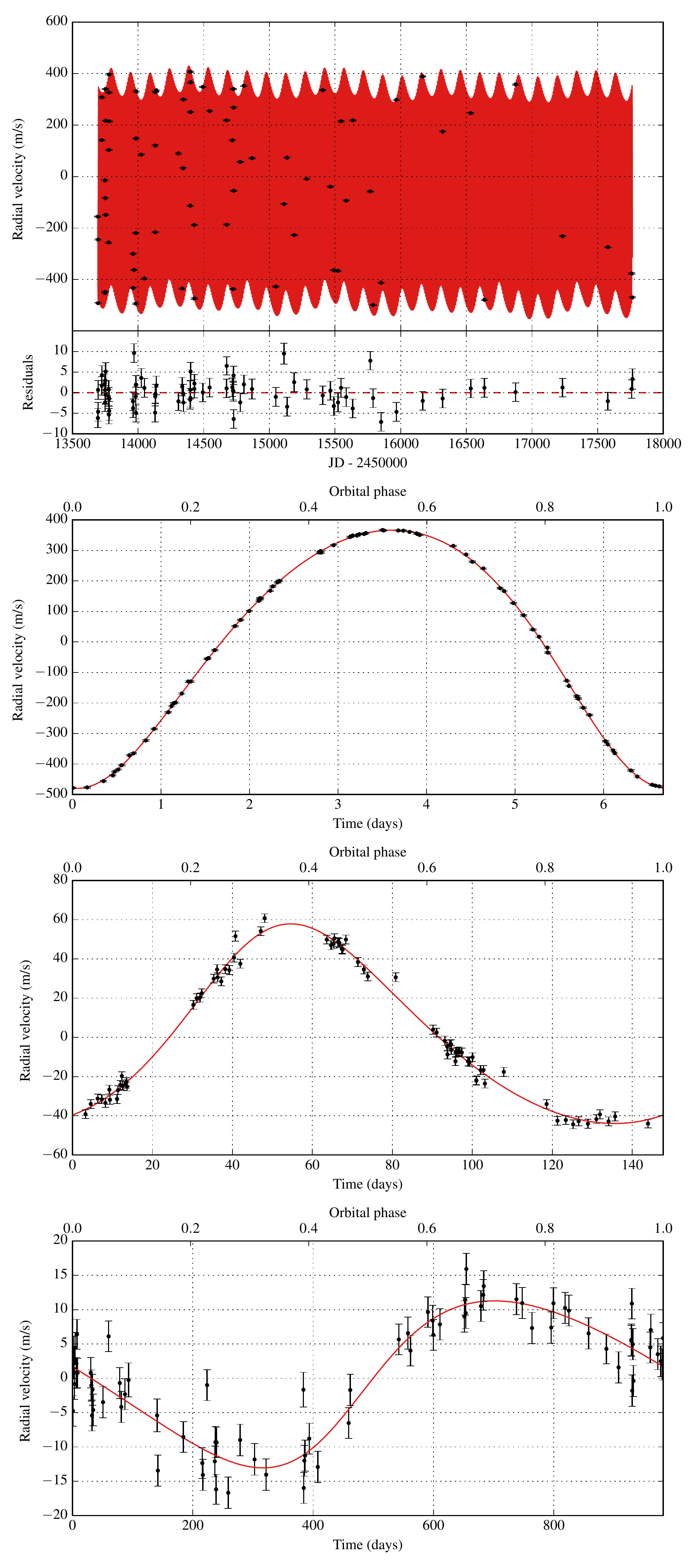}
	\caption{Keplerian model for HIP 14810 b, c, and d. The first plot shows the best Keplerian fit (in red) and the observed data points with an inset display of the residual velocities after fitting. The model includes a linear trend. The remaining plots show each of HIP 14810 b, c, and d folded by the orbital period after removing the other planets and the linear trend. The complete figure set (32 images) for all planetary systems in Section \ref{sec:results3} is available in the online journal.\label{figK10}}
\end{figure}

We also provide updated orbital parameters for the following planetary systems: the triple-planet systems HIP 14810 and HD 125612, the double-planet systems HD 1605, HD 5319, HD 11506, HD 37605, HD 75784, HD 219828, the eccentric transiting system HD 17156, the hot transiting planets orbiting HAT-P-1, HAT-P-3, HAT-P-4, HAT-P-12, HD 149026, XO-5, and the single-planet non-transiting systems HD 10442, HD 16760, HD 33283, HD 43691, HD 45652, HD 75354, HD 79498, HD 86081, HD 88133, HD 96167, HD 109749, HD 149143, HD 164509, HD 171238, HD 179079, HD 224693, HD 231701. In general, our results are in agreement with previously published values. However, we are able to quote smaller uncertainties due to the larger number of data points used in our modeling - in some cases, we are able to constrain the orbital period with a precision of ~500 ms (e.g. HIP 14810 b). For the transiting planets, we adopted the photometrically derived orbital periods for HAT-P-1 b \citep{Johnson2008}, HAT-P-3 b \citep{Chan2011}, HAT-P-12 b \citep{Sada2016}, HD 149026 b \citep{Carter2009}, and XO-5 b \citep{Smith2015}. For HAT-P-4 b and HD 17156 b, the radial velocities alone were sufficient to determine the orbital periods without any significant loss in precision. Please refer to Table \ref{t_starswithplanets} for information about the stars. Our best-fitted values for the Keplerian orbits are given in Table \ref{t_planets}. Figures for all of the Keplerian models are included in the online journal - please refer to Figure \ref{figK10} for an example. We also searched for additional detectable signals in the Lomb-Scargle periodograms of the residuals, but did not uncover anything significant.

We note that the HAT-P-4, HD 11506, HD 75784, HD 86081, and HD 164509 systems exhibit strong linear trends in the RVs which might be due to additional unseen companions or long-term stellar activity. In addition, an acceleration term of $0.58 \pm 0.11$ m/s/yr$^2$ is detectable in the RVs of HD 164509, pointing to the potential existence of an additional long-period companion whose orbital elements cannot be resolved at the current time baseline.

\figsetstart
\figsetnum{1}
\figsettitle{Keplerian models for planetary systems in Section \ref{sec:results3}}
 
\figsetgrpstart
\figsetgrpnum{1.1}
\figsetgrptitle{Keplerian model for HIP 14810 b, c, and d}
\figsetplot{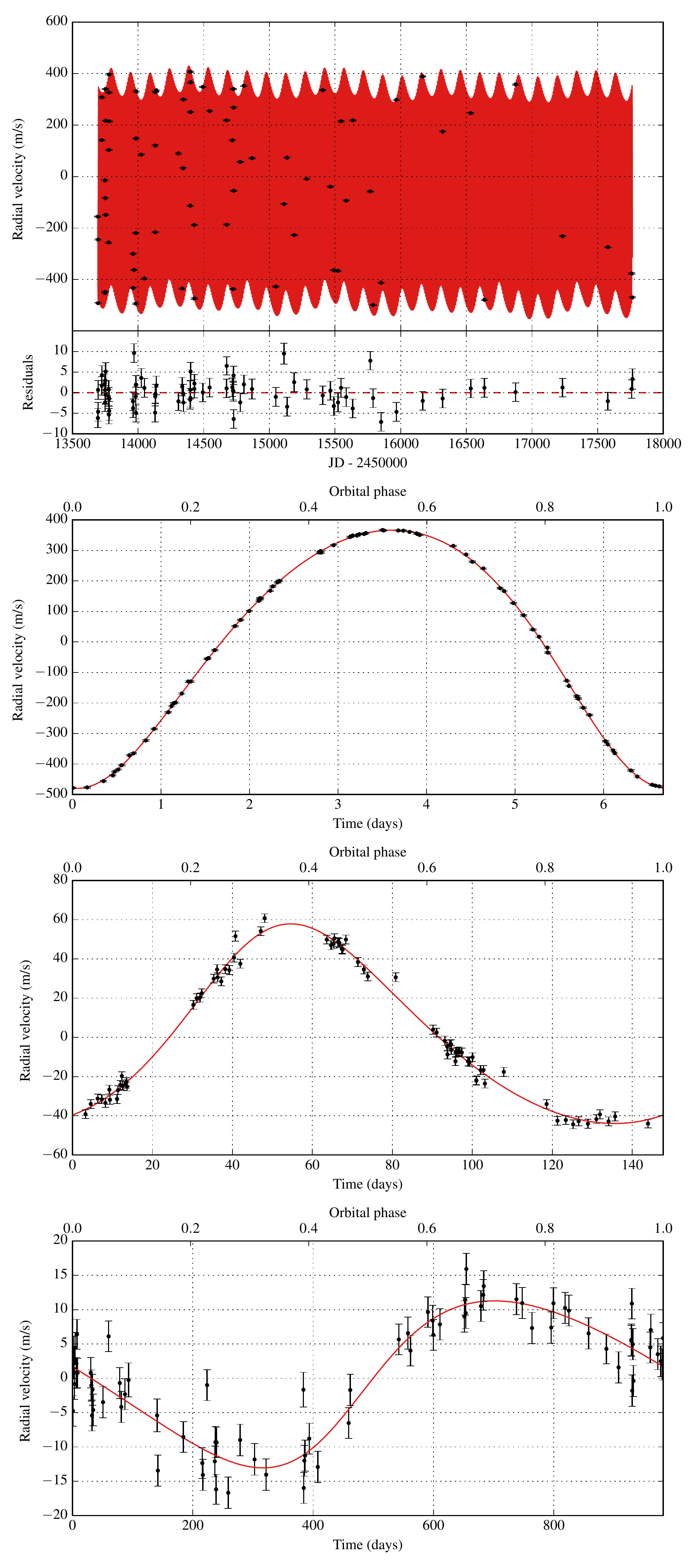}
\figsetgrpnote{The first plot shows the best Keplerian fit (in red) and the observed data points with an inset display of the residual velocities after fitting. The model includes a linear trend. The remaining plots show each of HIP 14810 b, c, and d folded by the orbital period after removing the other planets and the linear trend.}
\figsetgrpend

\figsetgrpstart
\figsetgrpnum{1.2}
\figsetgrptitle{Keplerian model for HD 219828 b and c}
\figsetplot{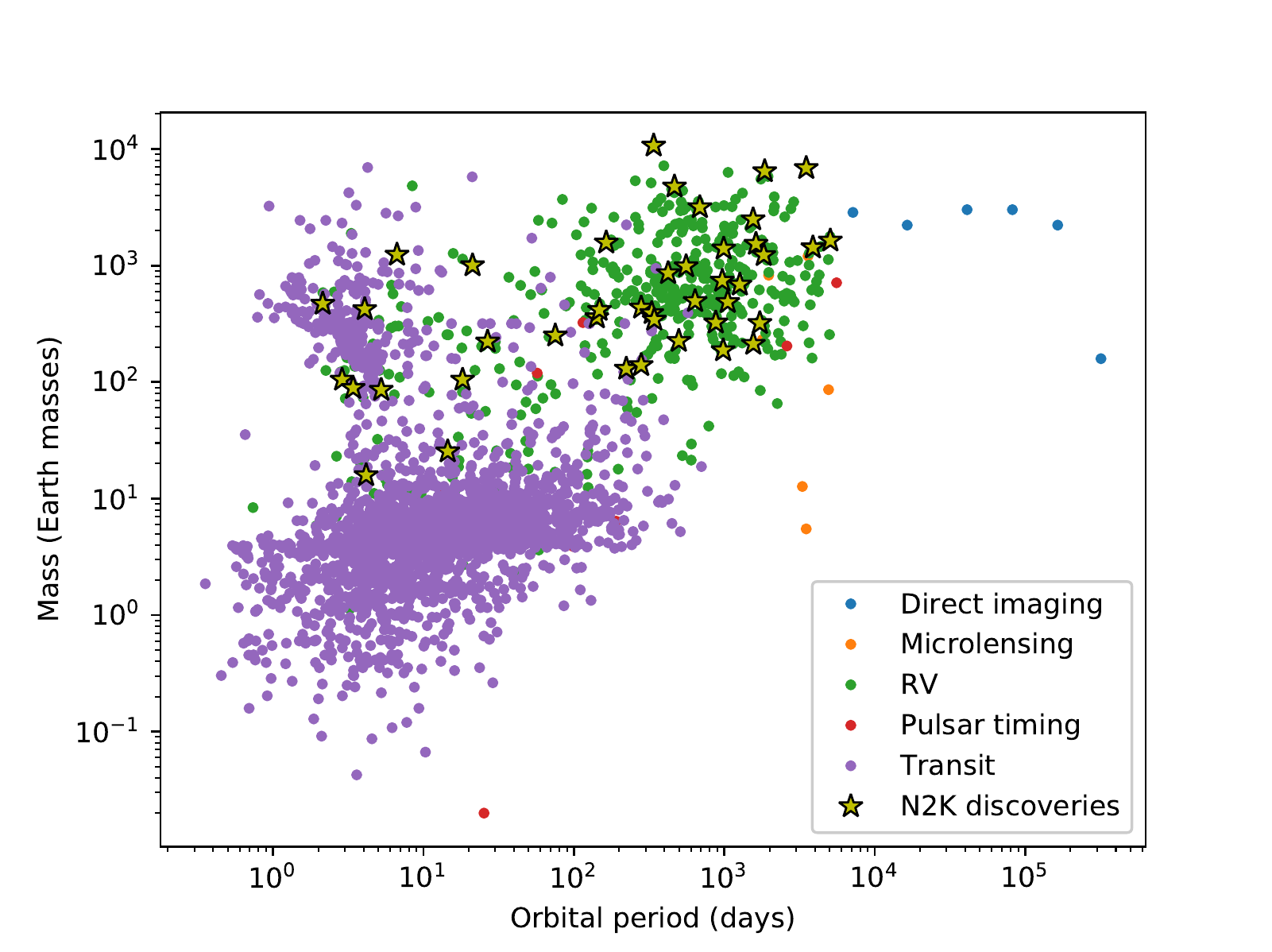}
\figsetgrpnote{The first plot shows the best Keplerian fit (in red) and the observed data points with an inset display of the residual velocities after fitting. The remaining plots show HD 219828 c and HD 219828 b, respectively, folded by the orbital period after removing the other planet.}
\figsetgrpend

\figsetend

\section{Discussion \label{sec:discussion}}

\subsection{Target list for direct imaging and astrometric surveys \label{sec:direct_astrometry}}

Accumulating over 13 years of radial velocity measurements has produced several gas giants on wide, long-period orbits from the N2K data set. Since direct imaging surveys are sensitive to such planets, we compiled a list of potential targets for state-of-the-art direct imaging surveys. The results are given in Table \ref{t_direct}, ordered by the maximal sky-projected angular separation (for a face-on orbit) between the star and its companion. While the largest separations correspond to stellar binary companions with $M \sin i > 13$ \mjup, we also recover several relatively young gas giant planets (HD 75898 b, HD 148164 b, HD 125612 d) with ages of 2-3 Gyr and potential angular separations between 75 and 100 milliarcseconds. In particular, the age of the HD 125612 system has an arguably significant uncertainty: according to \citet{Brewer2016}, it falls somewhere between 0.6 and 3.1 Gyr. Finally, we would like to point out that our data is consistent with an unseen long-period companion orbiting HD 163607 (refer to Section \ref{sec:K40} for additional information) with an angular separation of 100 mas or more.

We also compiled a similar list of targets for astrometric detection, presented in Table \ref{t_astrometry}. The results are ordered by the astrometric signature (semi-amplitude) for a circular orbit. While the largest signature amplitudes are produced by stellar and/or brown dwarf companions, nearly one half of our sample of discoveries induces a greater than 30 $\rm \mu$as astrometric shift in the host star, and thus would likely be detectable by Gaia \citep{Perryman2014}. The largest semi-amplitude caused by a companion with a plausible planetary origin is $469 \pm 108\ \rm \mu$as by HD 125612 d ($M \sin i = 7.3 \pm 0.9$ \mjup). In addition, at least five of the detectable hot Jupiters have orbital periods below 3 years.

\subsection{Planet occurrence rates based on stellar iron abundances}

\begin{figure}
	\plotone{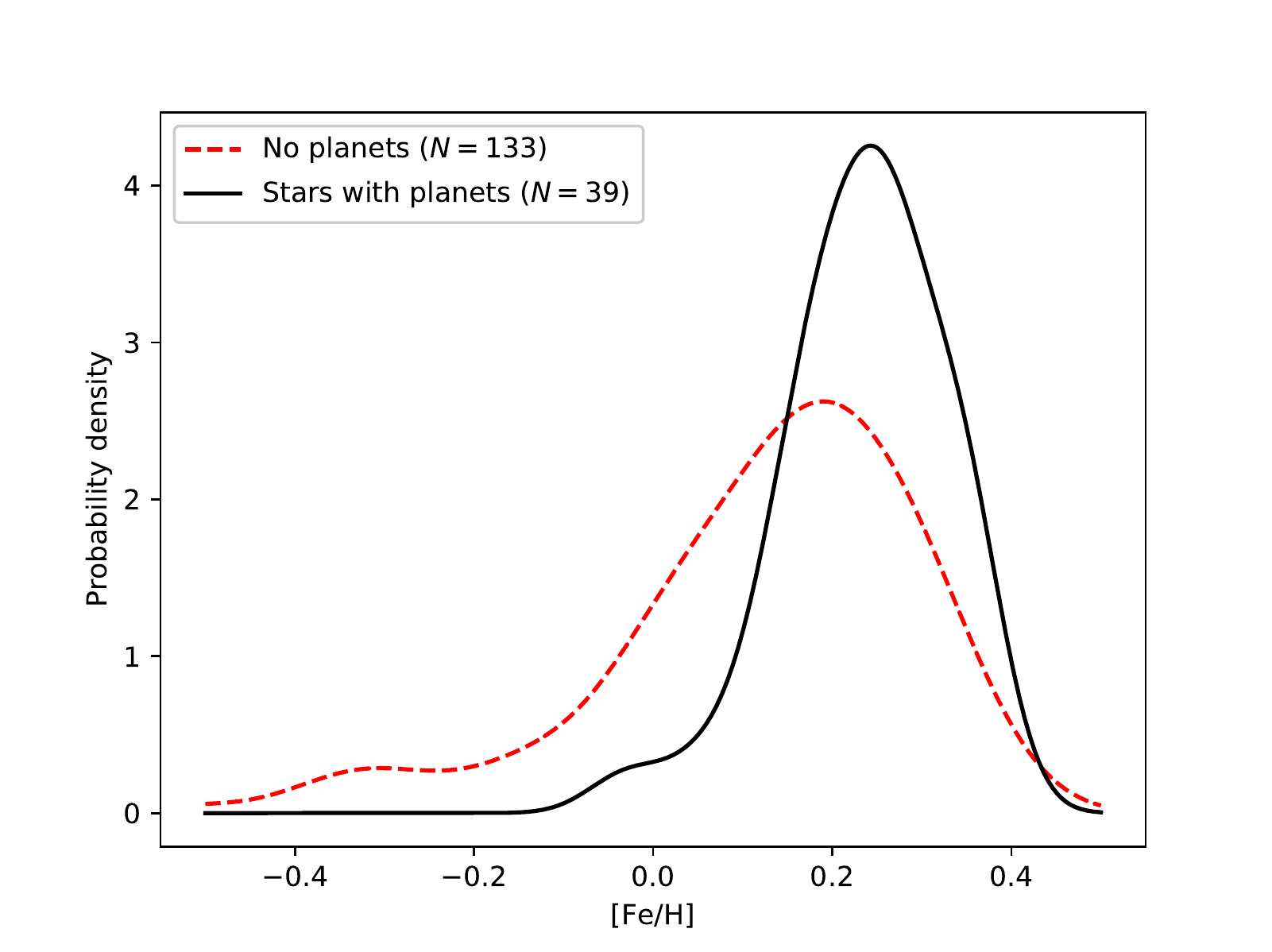}
	\caption{Gaussian Kernel density estimates (KDEs) of the [Fe/H] distribution of a subset of N2K stars, computed separately for stars with and without substellar companions. The number given by $N$ is the count of stars used in each data set.\label{fig:fehkde}}
\end{figure}

The N2K sample is intentionally biased towards metal-rich stars due to an observed correlation between stellar metallicity and the rate of occurrence of gas giant planets. In particular, \citet{FischerValenti2005} estimated that stars with [Fe/H] $> 0.2$ shelter at least 3 times as many exoplanets as solar metallicity stars. In order to further investigate this hypothesis, we computed a Gaussian kernel density estimate (KDE) of the [Fe/H] distribution of N2K stars with planets (including brown ddwarf companions) as well as a KDE of stars without known companions. In this analysis, we only included 39 planet-hosting stars and 133 N2K stars without planets for which updated iron abundances were available from \citet{Brewer2016}. KDE analysis yields a probability density function (PDF) which, integrated over finite intervals, is equivalent to the normalized histogram of [Fe/H] values. The results are plotted in Figure \ref{fig:fehkde}. As can be seen in the figure, planet-hosting stars are visibly concentrated towards the higher end of the [Fe/H] distribution. In order to test the significance of this discrepancy, we conducted a two-sided Welch's \textit{t}-test with the null hypothesis that the two distributions in Figure \ref{fig:fehkde} have equal means. We obtained a \textit{t}-statistic of 5.44 which corresponds to a p-value of $2.6 \times 10^{-7}$. Therefore, the planet occurrence bias towards higher metallicity stars is even greater than the selection bias of the N2K sample. Using the estimated PDFs, we find that the relative planet occurrence rate for stars with [Fe/H] $> 0.2$ is over 7 times greater than the rate for solar metallicity stars ([Fe/H] $= 0$) in our sample.

\section{Summary and conclusion \label{sec:summary}}

\begin{figure}
	\plotone{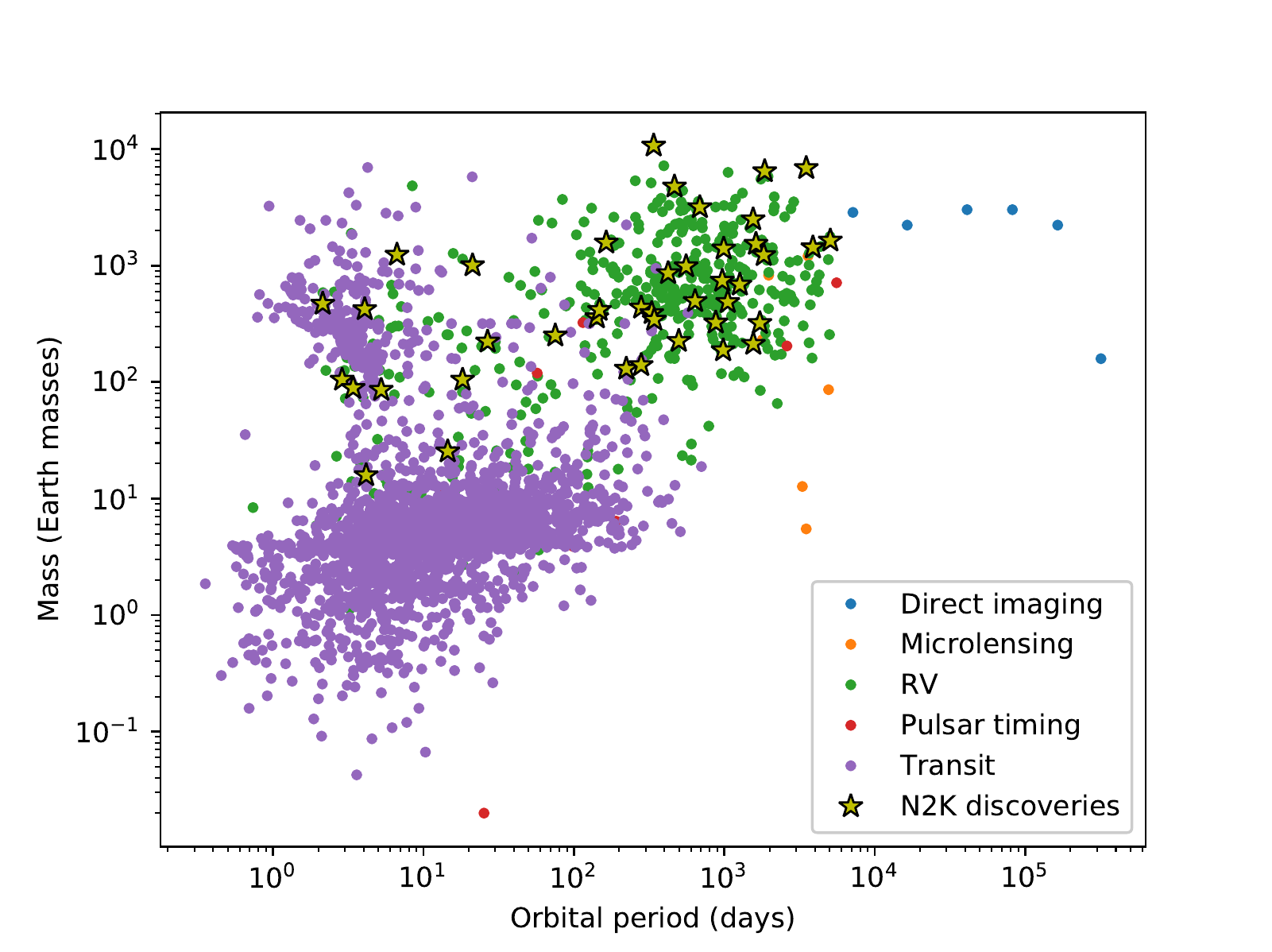}
	\caption{Masses and orbital periods of exoplanets known to date (February 2018), obtained from exoplanets.org. The N2K Consortium has contributed dozens of planets to this collection of alien worlds, spanning a wide spectrum of masses as well as orbital periods.\label{fig:massper}}
\end{figure}

Originally intended to detect hot Jupiters in short-period orbits, the N2K Consortium has now amassed over 13 years of high-precision radial velocity observations. This has allowed for the discoveries of dozens of hitherto unknown gas giants spanning a wide range of planetary masses and orbital periods, as illustrated in Figure \ref{fig:massper}. These discoveries include one of the shortest orbital periods ever found via the radial velocity technique, HD 86081 b ($P = 2.14$ days), as well as one of the longest orbital periods (HD 148164 b, $P = 5062$ days). Importantly, the parameter space inhabited by N2K planets has a significant overlap with transit discoveries. Combining the results from the two techniques yields precious information about the companions (e.g. the true mass, radius, and density of the object, orbital inclination) that cannot be deduced from RV data alone.

In this paper, we have present evidence for three previously undetected substellar companions HD 148284 B, HD 214823 B, and HD 217850 B, with masses greater than 20-34 \mjup. All three exhibit significant orbital eccentricities from $e = 0.16$ for HD 214823 B to $e = 0.76$ for HD 217850 B. We also announce the discovery of a double-planet system orbiting HD 148164 where the existence of both planets was previously unknown. Interestingly, the inner planet HD 148164 b is moving on a highly eccentric orbit with $e = 0.59$ and a semi-major axis of $a \sim 1$ AU which raises questions about the system's long-term stability. Nonetheless, a dynamical simulation shows that the system is stable over a time period of at least one million years. The outer planet HD 148164 c with an orbital period of $P = 5062$ days and a semi-major axis close to 6 AU is one of the longest-period RV discoveries known to date. The newly discovered cold planet HD 211810 b is a Jupiter analogue with a 4-year orbital period and a mass $m \sin i = 0.67$ \mjup. However, this orbit is strikingly eccentric as well with $e \sim 0.7$. This is similar to another discovery presented in this work: HD 55696 b, although the latter cold companion is somewhat more massive with $m \sin i \sim 4$ \mjup. We also find evidence for a 2.3-Jovian mass planet orbiting one of the stars in a previously known binary star system HD 98736. The orbital semi-major axis of this planet is close to 1.9 AU whereas the stellar binary companions are known to be at least 158 AU apart. Finally, we introduce a massive cold substellar companion on a 4.3-year orbit around HD 203473. Removing this companion from the radial velocity data leaves a significant RV acceleration term, suggesting the possible presence of yet another companion whose orbital signal cannot be resolved at the current time base line.

In Section \ref{sec:results2}, we list additional interesting finds from our modelling process that could not be classified as new substellar discoveries. HD 3404, HD 24505, HD 98360, and HD 103459 were found to host light stellar companions with lower mass boundaries between 0.13 and 0.27 Solar masses and orbital periods of 4.2 to 35.8 years. Interestingly, three of these four companions are moving on extremely eccentric orbits ($0.7 < e < 0.8$). We also present an alternative, two-planet model for HD 38801 where the existence of at least one companion has been previously established. However, more data needs to be accumulated before the second companion can be definitively confirmed. We investigated a residual linear trend and curvature in the radial velocities of HD 163607, and concluded that it might be due to a presently unresolvable third companion with an orbital semi-major axis of at least 7 AU. This translates into an angular separation of at least 100 mas as viewed from Earth, which makes it a good candidate for direct imaging surveys. Likewise, we find inconclusive evidence for a long-period companion HD 147506, or HAT-P-2, which has a known transiting planet on a 5.6-day orbit. Furthermore, our analysis suggests the presence of only one detectable companion around HD 207832 instead of two as reported in the literature. The residual RV signal is very likely due to stellar activity and disappears once the radial velocities are decorrelated from the \shk values. We also failed to recover a Keplerian signal of the putative 2.5-day companion HD 72356 b despite its large proposed RV semi-amplitude. We concluded that it is unlikely for HD 72356 to host a massive short-period companion based on our data. Finally, we propose that the plausible second companion in the HD 75898 system proposed by \citet{Robinson2007} is likely a manifestation of long-term magnetic activity.

Section \ref{sec:results3} provides up-to-date orbital parameters for 42 known substellar companions, including 24 systems with a single companion, 6 two-planet systems, and 2 with three companions. This includes several transiting planets from the HATNet survey (\citet{Bakos2004}): HAT-P-1 b, HAT-P-3 b, HAT-P-4 b, and HAT-P-12 b. In total, our data set encompasses 8 transiting companions. In addition to the four mentioned above, we give updated parameters for HD 147506 (a.k.a. HAT-P-2) and XO-5 as well as the two transiting planets that were originally discovered as RV planets by the N2K project: HD 17156 and HD 149026.

Finally, we produce target lists for direct imaging and astrometric surveys in Section \ref{sec:direct_astrometry} (Tables \ref{t_direct} and \ref{t_astrometry}). We find that many of the N2K discoveries could be detected by state-of-the-art direct imaging and astrometry equipment (e.g. Gaia).

\startlongtable
\begin{deluxetable}{lrrrrrrr}
	\tablewidth{0pt}
	\tabletypesize{\scriptsize}
	\tablecaption{Interesting Targets for Direct Imaging \label{t_direct}}
	\tablehead{
		\colhead{Planet}         &
		\colhead{$M \sin i$} & \colhead{a}          &
		\colhead{e} & \colhead{Age} & \multicolumn{3}{c}{Angle (mas)\tablenotemark{a}}     \\
		& \colhead{(\mjup)} & \colhead{(AU)} & & \colhead{(Gyr)} & \colhead{Min.} & \colhead{Max.} & \colhead{Err.}	}
	\startdata
	HD 24505 b& 221.86(29.64)& 10.82(0.61)& 0.798(0.001)& 7.3(0.8)& 93.0& 277.4& (15.6) \\
	HD 219828 b& 14.55(2.31)& 5.79(0.41)& 0.81(0.005)& 5.7(0.7)& 46.9& 144.9& (10.2) \\
	HD 217850 b& 21.61(2.61)& 4.56(0.24)& 0.762(0.002)& 7.6(1.3)& 48.7& 132.6& (6.9) \\
	HD 98630 b& 359.17(26.41)& 13.17(0.84)& 0.059(0.032)& \nodata& 114.8& 121.8& (8.6) \\
	HD 79498 b& 1.34(0.21)& 2.98(0.15)& 0.575(0.023)& 2.7(1.4)& 52.8& 101.8& (5.5) \\
	HD 148164 b& 5.16(0.82)& 6.15(0.5)& 0.125(0.017)& 2.4(0.8)& 84.7& 96.0& (7.9) \\
	HD 75898 c& 2.9(0.57)& 7.03(0.69)& 0.0(0.0)& 3.6(0.6)& 92.8& 92.8& (9.1) \\
	HD 103459 b& 139.77(20.14)& 3.21(0.16)& 0.699(0.005)& 6.0(0.6)& 38.3& 91.0& (4.7) \\
	HD 37605 c& 3.19(0.38)& 3.74(0.21)& 0.03(0.012)& 5.7(3.4)& 84.9& 87.5& (5.1) \\
	HD 125612 d& 7.28(0.93)& 4.06(0.25)& 0.117(0.006)& 1.8(1.2)& 74.4& 83.7& (5.2) \\
	HD 11506 b& 4.83(0.52)& 2.9(0.14)& 0.374(0.005)& 2.3(0.6)& 52.1& 77.2& (3.8) \\
	HD 75784 b& 4.5(1.18)& 5.22(0.58)& 0.266(0.04)& 5.4(1.4)& 58.0& 76.2& (8.8) \\
	HD 55696 b& 3.87(0.72)& 3.18(0.18)& 0.705(0.022)& 2.6(0.5)& 31.6& 76.0& (4.3) \\
	HD 211810 b& 0.67(0.44)& 2.66(0.04)& 0.675(0.143)& 0.0(0.0)& 32.5& 73.9& (6.4) \\
	HD 98736 b& 2.33(0.78)& 1.86(0.09)& 0.226(0.064)& 7.4(3.0)& 58.5& 73.6& (5.3) \\
	HD 3404 b& 145.05(27.96)& 2.86(0.16)& 0.738(0.004)& 6.7(1.3)& 27.0& 69.6& (4.0) \\
	HD 171238 b& 2.72(0.49)& 2.57(0.16)& 0.234(0.028)& 7.6(3.0)& 49.6& 63.0& (4.3) \\
	HD 203473 b& 7.84(1.15)& 2.73(0.17)& 0.289(0.01)& 5.2(1.0)& 40.6& 54.7& (3.5) \\
	HD 1605 b& 3.62(0.23)& 3.58(0.1)& 0.099(0.011)& 4.4(1.0)& 44.9& 49.6& (1.5) \\
	HIP 14810 d& 0.59(0.1)& 1.94(0.13)& 0.185(0.035)& 5.8(2.4)& 35.7& 43.0& (3.2) \\
	HD 10442 b& 1.53(0.86)& 2.03(0.55)& 0.09(0.019)& 10.7(3.9)& 39.1& 42.8& (11.5) \\
	HD 214823 b& 20.34(2.57)& 3.23(0.2)& 0.164(0.003)& 4.3(0.5)& 32.7& 38.6& (2.4) \\
	HD 163607 c& 2.16(0.27)& 2.38(0.12)& 0.076(0.023)& 7.8(0.7)& 34.5& 37.2& (2.1) \\
	HD 125612 b& 3.1(0.4)& 1.37(0.08)& 0.455(0.005)& 1.8(1.2)& 22.5& 36.8& (2.2) \\
	HD 73534 b& 1.01(0.21)& 2.95(0.22)& 0.0(0.0)& 7.0(1.4)& 36.4& 36.4& (2.7) \\
	HD 16760 b& 15.04(2.54)& 1.16(0.1)& 0.081(0.002)& 3.6(2.9)& 25.4& 27.6& (2.3) \\
	HD 96167 b& 0.71(0.18)& 1.33(0.09)& 0.681(0.033)& 4.8(0.6)& 11.2& 25.7& (1.9) \\
	HD 148164 c& 1.23(0.25)& 0.99(0.07)& 0.587(0.026)& 2.4(0.8)& 11.2& 21.9& (1.5) \\
	HD 1605 c& 0.93(0.08)& 1.49(0.04)& 0.095(0.057)& 4.4(1.0)& 18.7& 20.6& (1.2) \\
	HD 164509 b& 0.44(0.08)& 0.87(0.05)& 0.238(0.062)& 2.3(1.3)& 16.1& 20.5& (1.6) \\
	HD 38801 b& 9.97(1.43)& 1.66(0.11)& 0.057(0.006)& 4.9(1.0)& 18.3& 19.3& (1.3) \\
	HD 5319 c& 1.02(0.22)& 1.94(0.16)& 0.109(0.067)& 5.0(1.7)& 16.9& 18.8& (1.9) \\
	HD 11506 c& 0.41(0.06)& 0.77(0.04)& 0.193(0.038)& 2.3(0.6)& 14.7& 17.9& (1.0) \\
	HD 75898 b& 2.71(0.36)& 1.19(0.07)& 0.11(0.01)& 3.6(0.6)& 15.6& 17.5& (1.1) \\
	HD 148284 b& 33.73(5.52)& 0.97(0.08)& 0.389(0.001)& 8.7(1.2)& 9.5& 14.4& (1.2) \\
	HD 5319 b& 1.56(0.29)& 1.57(0.13)& 0.015(0.016)& 5.0(1.7)& 13.7& 13.9& (1.1) \\
	HD 75784 c& 1.08(0.35)& 1.03(0.07)& 0.142(0.078)& 5.4(1.4)& 11.8& 13.6& (1.3) \\
	HD 207832 b& 0.56(0.09)& 0.59(0.03)& 0.197(0.053)& 1.4(1.3)& 10.6& 12.9& (0.9) \\
	HIP 14810 c& 1.31(0.18)& 0.55(0.03)& 0.157(0.01)& 5.8(2.4)& 10.1& 11.9& (0.8) \\
	HD 45652 b& 0.43(0.08)& 0.24(0.01)& 0.607(0.026)& 8.4(3.2)& 5.5& 11.1& (0.6) \\
	HD 37605 b& 2.69(0.3)& 0.28(0.01)& 0.675(0.002)& 5.7(3.4)& 4.7& 10.6& (0.6) \\
	HD 163607 b& 0.79(0.11)& 0.36(0.02)& 0.744(0.012)& 7.8(0.7)& 3.5& 9.2& (0.4) \\
	HD 231701 b& 1.13(0.25)& 0.57(0.05)& 0.13(0.032)& 3.4(0.7)& 4.7& 5.4& (0.5) \\
	HD 17156 b& 3.16(0.42)& 0.16(0.01)& 0.675(0.005)& 3.5(0.7)& 1.6& 3.6& (0.2) \\
	HD 43691 b& 2.55(0.34)& 0.24(0.02)& 0.08(0.007)& 2.4(0.4)& 3.0& 3.2& (0.2) \\
	HD 33283 b& 0.33(0.07)& 0.15(0.01)& 0.399(0.056)& 3.6(0.5)& 1.5& 2.2& (0.2) \\
	HD 224693 b& 0.7(0.12)& 0.19(0.01)& 0.104(0.017)& 4.1(0.7)& 1.9& 2.1& (0.2) \\
	HD 179079 b& 0.08(0.02)& 0.12(0.01)& 0.049(0.087)& 7.0(0.7)& 1.9& 1.9& (0.2) \\
	HIP 14810 b& 3.9(0.49)& 0.07(0.0)& 0.144(0.001)& 5.8(2.4)& 1.3& 1.5& (0.1) \\
	HD 109749 b& 0.27(0.05)& 0.06(0.0)& 0.0(0.0)& 3.3(1.2)& 1.1& 1.1& (0.1) \\
	HD 125612 c& 0.05(0.01)& 0.05(0.0)& 0.049(0.038)& 1.8(1.2)& 1.0& 1.0& (0.1) \\
	HD 147506 b& 8.62(0.17)& 0.07(0.0)& 0.517(0.002)& 1.4(0.5)& 0.5& 0.9& (0.0) \\
	HD 149143 b& 1.33(0.15)& 0.05(0.0)& 0.017(0.004)& 4.3(0.7)& 0.9& 0.9& (0.0) \\
	HD 219828 c& 0.07(0.01)& 0.05(0.0)& 0.101(0.063)& 5.7(0.7)& 0.7& 0.8& (0.1) \\
	HD 88133 b& 0.28(0.05)& 0.05(0.0)& 0.0(0.0)& 5.4(0.9)& 0.6& 0.6& (0.0) \\
	HD 149026 b& 0.33(0.04)& 0.04(0.0)& 0.051(0.019)& 2.2(0.4)& 0.5& 0.6& (0.0) \\
	HAT-P-1 b& 0.53(0.05)& 0.06(0.0)& 0.0(0.0)& 3.6(0.0)& 0.4& 0.4& (0.0) \\
	HD 86081 b& 1.48(0.23)& 0.03(0.0)& 0.012(0.005)& 3.6(0.9)& 0.4& 0.4& (0.0) \\
	HAT-P-3 b& 0.59(0.03)& 0.04(0.0)& 0.0(0.0)& 1.6(2.1)& 0.3& 0.3& (0.0) \\
	HAT-P-12 b& 0.21(0.01)& 0.04(0.0)& 0.0(0.0)& 2.5(2.0)& 0.3& 0.3& (0.0) \\
	XO-5 b& 1.04(0.03)& 0.05(0.0)& 0.0(0.0)& 14.8(2.0)& 0.2& 0.2& (0.0) \\
	HAT-P-4 b& 0.65(0.04)& 0.04(0.0)& 0.084(0.014)& 4.2(1.6)& 0.1& 0.2& (0.0) \\
	\enddata
	\tablenotetext{a}{Depending on the exact orbital configuration, the maximum sky-projected separation falls within the range given here, with ``Min.'' corresponding to an edge-on orbit with the apastron passage occurring right behind the center of the star, and ``Max.'' representing a completely face-on orbit. The exact time of the largest separation depends on the inclination angle $i$ which cannot be determined from the RVs alone. The uncertainties (``Err.'') are given for the ``Max.'' angles.}
\end{deluxetable}


\startlongtable
\begin{deluxetable}{lrrrrr}
	\tablewidth{0pt}
	\tabletypesize{\scriptsize}
	\tablecaption{Interesting Targets for Astrometric Detection \label{t_astrometry}}
		\tablehead{
		\colhead{Planet}         &
		\colhead{$M \sin i$} & \colhead{P} & \colhead{a}          &
		\colhead{e} & \colhead{Angle\tablenotemark{a}}     \\
		& \colhead{(\mjup)} & \colhead{(days)} & \colhead{(AU)} & & \colhead{($\rm \mu$as)}	}
	\startdata
	HD 24505 b& 221.86(29.64)& 11315.0(91.7)& 10.82(0.61)& 0.798(0.001)& 29437.8(6808.2) \\
	HD 98630 b& 359.17(26.41)& 13074.0(982.1)& 13.17(0.84)& 0.059(0.032)& 27380.3(2982.1) \\
	HD 103459 b& 139.77(20.14)& 1831.9(0.9)& 3.21(0.16)& 0.699(0.005)& 6053.0(1381.9) \\
	HD 3404 b& 145.05(27.96)& 1540.8(1.9)& 2.86(0.16)& 0.738(0.004)& 4742.2(1305.7) \\
	HD 217850 b& 21.61(2.61)& 3501.3(2.1)& 4.56(0.24)& 0.762(0.002)& 1507.6(306.7) \\
	HD 219828 b& 14.55(2.31)& 4681.6(98.6)& 5.79(0.41)& 0.81(0.005)& 942.5(228.6) \\
	HD 214823 b& 20.34(2.57)& 1854.4(1.1)& 3.23(0.2)& 0.164(0.003)& 491.1(113.3) \\
	HD 125612 d& 7.28(0.93)& 2835.0(7.9)& 4.06(0.25)& 0.117(0.006)& 468.9(107.5) \\
	HD 16760 b& 15.04(2.54)& 466.0(0.1)& 1.16(0.1)& 0.081(0.002)& 381.8(119.4) \\
	HD 148164 b& 5.16(0.82)& 5061.7(113.8)& 6.15(0.5)& 0.125(0.017)& 347.2(92.5) \\
	HD 148284 b& 33.73(5.52)& 339.3(0.0)& 0.97(0.08)& 0.389(0.001)& 311.6(94.7) \\
	HD 203473 b& 7.84(1.15)& 1552.9(3.4)& 2.73(0.17)& 0.289(0.01)& 283.3(69.8) \\
	HD 37605 c& 3.19(0.38)& 2720.4(15.2)& 3.74(0.21)& 0.03(0.012)& 275.2(57.2) \\
	HD 11506 b& 4.83(0.52)& 1622.1(2.1)& 2.9(0.14)& 0.374(0.005)& 208.7(39.0) \\
	HD 75784 b& 4.5(1.18)& 3877.8(261.1)& 5.22(0.58)& 0.266(0.04)& 205.3(71.1) \\
	HD 75898 c& 2.9(0.57)& 6066.4(337.2)& 7.03(0.69)& 0.0(0.0)& 204.2(58.1) \\
	HD 98736 b& 2.33(0.78)& 968.8(2.2)& 1.86(0.09)& 0.226(0.064)& 144.9(53.4) \\
	HD 171238 b& 2.72(0.49)& 1531.6(11.7)& 2.57(0.16)& 0.234(0.028)& 137.9(35.9) \\
	HD 38801 b& 9.97(1.43)& 687.1(0.5)& 1.66(0.11)& 0.057(0.006)& 134.9(34.6) \\
	HD 55696 b& 3.87(0.72)& 1827.2(10.0)& 3.18(0.18)& 0.705(0.022)& 127.6(31.7) \\
	HD 1605 b& 3.62(0.23)& 2148.8(16.3)& 3.58(0.1)& 0.099(0.011)& 117.4(11.4) \\
	HD 79498 b& 1.34(0.21)& 1806.6(15.3)& 2.98(0.15)& 0.575(0.023)& 76.5(16.7) \\
	HD 125612 b& 3.1(0.4)& 557.0(0.4)& 1.37(0.08)& 0.455(0.005)& 67.6(15.6) \\
	HD 163607 c& 2.16(0.27)& 1267.4(7.2)& 2.38(0.12)& 0.076(0.023)& 63.6(12.5) \\
	HD 10442 b& 1.53(0.86)& 1053.2(3.4)& 2.03(0.55)& 0.09(0.019)& 56.7(57.5) \\
	HD 75898 b& 2.71(0.36)& 422.9(0.3)& 1.19(0.07)& 0.11(0.01)& 32.3(7.6) \\
	HD 73534 b& 1.01(0.21)& 1721.4(35.6)& 2.95(0.22)& 0.0(0.0)& 30.3(8.6) \\
	HD 211810 b& 0.67(0.44)& 1557.7(22.3)& 2.66(0.04)& 0.675(0.143)& 27.4(17.9) \\
	HIP 14810 d& 0.59(0.1)& 981.8(6.9)& 1.94(0.13)& 0.185(0.035)& 20.2(5.3) \\
	HD 37605 b& 2.69(0.3)& 55.0(0.0)& 0.28(0.01)& 0.675(0.002)& 17.2(3.5) \\
	HD 5319 b& 1.56(0.29)& 638.6(1.2)& 1.57(0.13)& 0.015(0.016)& 16.1(5.0) \\
	HD 148164 c& 1.23(0.25)& 328.5(0.4)& 0.99(0.07)& 0.587(0.026)& 13.4(3.9) \\
	HD 5319 c& 1.02(0.22)& 877.0(4.9)& 1.94(0.16)& 0.109(0.067)& 13.0(4.3) \\
	HIP 14810 c& 1.31(0.18)& 147.7(0.0)& 0.55(0.03)& 0.157(0.01)& 12.8(3.1) \\
	HD 1605 c& 0.93(0.08)& 577.2(2.5)& 1.49(0.04)& 0.095(0.057)& 12.6(1.4) \\
	HD 75784 c& 1.08(0.35)& 341.5(1.3)& 1.03(0.07)& 0.142(0.078)& 9.7(3.7) \\
	HD 96167 b& 0.71(0.18)& 498.1(0.8)& 1.33(0.09)& 0.681(0.033)& 8.1(2.7) \\
	HD 164509 b& 0.44(0.08)& 280.2(0.8)& 0.87(0.05)& 0.238(0.062)& 6.3(1.6) \\
	HD 207832 b& 0.56(0.09)& 160.1(0.2)& 0.59(0.03)& 0.197(0.053)& 5.5(1.3) \\
	HD 43691 b& 2.55(0.34)& 37.0(0.0)& 0.24(0.02)& 0.08(0.007)& 5.5(1.3) \\
	HD 17156 b& 3.16(0.42)& 21.2(0.0)& 0.16(0.01)& 0.675(0.005)& 5.3(1.2) \\
	HIP 14810 b& 3.9(0.49)& 6.7(0.0)& 0.07(0.0)& 0.144(0.001)& 4.8(1.1) \\
	HD 11506 c& 0.41(0.06)& 223.4(0.3)& 0.77(0.04)& 0.193(0.038)& 4.7(1.0) \\
	HD 231701 b& 1.13(0.25)& 141.6(0.1)& 0.57(0.05)& 0.13(0.032)& 4.3(1.6) \\
	HD 147506 b& 8.62(0.17)& 5.6(0.0)& 0.07(0.0)& 0.517(0.002)& 3.7(0.1) \\
	HD 163607 b& 0.79(0.11)& 75.2(0.0)& 0.36(0.02)& 0.744(0.012)& 3.5(0.7) \\
	HD 45652 b& 0.43(0.08)& 44.1(0.0)& 0.24(0.01)& 0.607(0.026)& 3.1(0.7) \\
	HD 224693 b& 0.7(0.12)& 26.7(0.0)& 0.19(0.01)& 0.104(0.017)& 1.0(0.3) \\
	HD 149143 b& 1.33(0.15)& 4.1(0.0)& 0.05(0.0)& 0.017(0.004)& 0.9(0.2) \\
	HD 86081 b& 1.48(0.23)& 2.1(0.0)& 0.03(0.0)& 0.012(0.005)& 0.4(0.1) \\
	HD 33283 b& 0.33(0.07)& 18.2(0.0)& 0.15(0.01)& 0.399(0.056)& 0.4(0.1) \\
	HD 109749 b& 0.27(0.05)& 5.2(0.0)& 0.06(0.0)& 0.0(0.0)& 0.2(0.1) \\
	XO-5 b& 1.04(0.03)& 4.2(0.0)& 0.05(0.0)& 0.0(0.0)& 0.2(0.0) \\
	HAT-P-3 b& 0.59(0.03)& 2.9(0.0)& 0.04(0.0)& 0.0(0.0)& 0.2(0.0) \\
	HAT-P-1 b& 0.53(0.05)& 4.5(0.0)& 0.06(0.0)& 0.0(0.0)& 0.2(0.0) \\
	HD 149026 b& 0.33(0.04)& 2.9(0.0)& 0.04(0.0)& 0.051(0.019)& 0.1(0.0) \\
	HD 179079 b& 0.08(0.02)& 14.5(0.0)& 0.12(0.01)& 0.049(0.087)& 0.1(0.0) \\
	HD 88133 b& 0.28(0.05)& 3.4(0.0)& 0.05(0.0)& 0.0(0.0)& 0.1(0.0) \\
	HAT-P-12 b& 0.21(0.01)& 3.2(0.0)& 0.04(0.0)& 0.0(0.0)& 0.1(0.0) \\
	HAT-P-4 b& 0.65(0.04)& 3.1(0.0)& 0.04(0.0)& 0.084(0.014)& 0.1(0.0) \\
	HD 125612 c& 0.05(0.01)& 4.2(0.0)& 0.05(0.0)& 0.049(0.038)& 0.0(0.0) \\
	HD 219828 c& 0.07(0.01)& 3.8(0.0)& 0.05(0.0)& 0.101(0.063)& 0.0(0.0) \\
	\enddata
    \tablenotetext{a}{This is the astrometric signature angle for a circular orbit. The detectability of orbits with $e \neq 0$ depends on the orbital angle.}
\end{deluxetable}

\acknowledgments
We thank the many observers who helped to obtain data for this project, including Geoff Marcy, R. Paul Butler, Steve Vogt, John Johnson, Jason Wright, Katie Peek, Julien Spronck, Matt Giguere, John Brewer, BJ Fulton, Evan Sinukoff, Erik Petigura, Lauren Weiss, Lea Hirsch, Joel Hartman.

DAF gratefully acknowledges support from NASA NNH11ZDA001. We thank Tom Blake and the PNNL EMSL for obtaining the FTS scan of our iodine cell.  We also thank contributors to Matplotlib, the Python Programming Language, and the free and open-source community. Simulations in this paper made use of the REBOUND code which can be downloaded freely at http://github.com/hannorein/rebound.

The data presented herein were obtained at the W. M. Keck Observatory, which is operated as a scientific partnership among the California Institute of Technology, the University of California and the National Aeronautics and Space Administration. The Observatory was made possible by the generous financial support of the W. M. Keck Foundation.

The authors wish to recognize and acknowledge the very significant cultural role and reverence that the summit of Maunakea has always had within the indigenous Hawaiian community.  We are most fortunate to have the opportunity to conduct observations from this mountain. 

\facilities{Keck:I (HIRES), Pacific Northwest National Labs (PNNL) Environmental Molecular Sciences Laboratory (EMSL)}

\bibliographystyle{aasjournal} 
\bibliography{n2k.bib} 

\end{document}